\documentclass[aps,amsmath,amssymb,reprint,prl]{revtex4-2}

\usepackage{graphicx}% Include figure files
\usepackage{dcolumn}% Align table columns on decimal point
\usepackage{bm}% bold math
\usepackage{multirow}% For the tabular environnement
\usepackage{array}% For the tabular environnement
\usepackage[colorlinks]{hyperref}% add hypertext capabilities
\usepackage[usenames,dvipsnames]{color}
\newcommand{\com}[1]{{\color{Green} #1}} % visible questions/comments

\begin{document}

\title{Upper bounds on broadband absorption}

%\date{}
\author{Stéphane Collin}
\email{stephane.collin@c2n.upsaclay.fr}
\affiliation{Centre de Nanosciences et de Nanotechnologies (C2N), CNRS, Université Paris-Saclay, Palaiseau, France}
\affiliation{Institut Photovolta{\"\i}que d'Ile-de-France (IPVF), Palaiseau, France}
\author{Maxime Giteau}
\affiliation{Centre de Nanosciences et de Nanotechnologies (C2N), CNRS, Université Paris-Saclay, Palaiseau, France}
\affiliation{Laboratoire PROcédés, Matériaux et Energie Solaire (PROMES), CNRS, Odeillo, France}

\begin{abstract}

We address the question of the optimal broadband absorption of waves in an open, dissipative system.
We develop a general framework for absorption induced by multiple overlapping resonances, based on quasi-normal modes and radiative and non-radiative decay rates.
Upper bounds on broadband absorption in a slab of thickness $d$ take the simple form: $\displaystyle A= 1-\exp(-F \alpha d)$, where $\alpha$ is the absorption coefficient and $F$ the path enhancement factor.
We apply these results to sunlight absorption in photovoltaics and answer the long-standing debate on the best light-trapping strategy in solar cells.
For angle-independent absorption, we derive the isotropic scattering upper bound $F = 4 n^2$ ($n$ the refractive index), extending the well-know Yablonovitch limit beyond the ray optics and weak absorption regimes.
For angle-restricted illumination, we show that $F$ can be further increased up to $8 \pi n^2 / \sqrt{3}$ using multi-resonant absorption induced by periodical patterning.
These results have a general scope in the field of wave physics and open new opportunities to maximize absorption, detection, and attenuation of electromagnetic or mechanical waves in ultrathin devices.
\end{abstract}

%\keywords{Suggested keywords}%Use showkeys class option if keyword
                              %display desired
\maketitle

\subsection*{Introduction}
Absorption of waves is an elemental process in physics. 
Maximizing absorption over a broad spectral range is crucial in many applications such as solar energy harvesting, optical telecommunications, radar stealth, or attenuation of acoustic and seismic waves.

In optics, the blackbody is an ideal picture of a perfect broadband absorber, but no actual object absorbs all incident light. In pictorial art, the tiny amount of light reflected by \emph{black} surfaces has inspired the vast artwork of Pierre Soulages, named \emph{Outrenoir} (beyond black). Actually, close to perfect absorption of waves can be achieved with bulk and strongly absorbing materials. In thin and weakly absorbing materials, absorption can be enhanced by leveraging scattering or resonance phenomena, but it remains constrained by the absorption coefficient $\alpha$ and the volume of the absorbing medium. The best strategy to optimize absorption is a fundamental but still open question that impacts many areas of physics.

Lambertian scattering is commonly used as a reference model for light-trapping in solar cells~\cite{yablonovitch_statistical_1982,green_lambertian_2002} but, contrary to common belief, it is not a fundamental limit~\cite{minano1990optical,rau_thermodynamics_2014,massiot_progress_2020}. 
More generally, energy conservation and the causality principle have been used to derive theoretical limits of the response of lossy systems in specific cases, leading for instance to upper bounds on absorption and scattering cross-sections~\cite{Gustafsson2020,Chao2021} and on the bandwidth-thickness absorption limit~\cite{Rozanov2000,Benzaouia2019,Acher2009,Yang2017c,Qu2022}, or to constraints on the (local) density of states~\cite{Scheel2008,Chao2022, Miller2016}. However, they do not provide general models, guidelines and limits for absorption optimization over a broad spectral range.

In the following, we first introduce the historical approaches to the absorption problem. Then, we develop a comprehensive theoretical model for absorption induced by multiple overlapping resonances in an open, dissipative system.
We derive analytical expressions for upper bounds on broadband absorption, with a wide domain of validity in wave physics.
We apply these results to sunlight absorption in photovoltaics and we discuss their implications. We show that our theoretical framework embraces and extends the whole scope of previous theories and provides a powerful tool to determine the optimal light trapping strategy for solar cells. 
Simulation results beyond the Lambertian light-trapping limit have raised questions recently~\cite{massiot_progress_2020}: our new multi-resonant upper bounds reconcile results from the literature with theoretical models.

%%%%%%%%%%%%

\subsection*{Historical approaches to the absorption problem}

In optics, absorption models are of particular importance for sunlight harvesting in solar cells~\cite{guillemoles_guide_2019}. The theoretical efficiency limit of an ideal solar cell, derived by Shockley and Queisser in 1961~\cite{shockley_detailed_1961}, is calculated by assuming perfect absorption for photon energy above the semiconductor bandgap $E_g$, and no absorption below.
However, this simplistic approximation does not take into account the material properties (refractive index $n$ and absorption coefficient $\alpha$) and finite thickness $d$ of actual solar cells~\cite{Wong2021}.
As a starting point, assuming no reflection at the front and back interfaces leads to single-pass absorption, expressed by the Beer-Lambert law: $\displaystyle A= 1-\exp(-\alpha d)$. The combination of a back mirror and surface patterning can be used to reflect, scatter, and trap light within the cell, resulting in an enhancement of the optical path by a factor $F$. What are the optimal light-trapping strategies and the resulting absorption upper bounds?

Previous light-trapping theories have been developed for ray-optics and isotropic illumination (Supplementary Section \com{S.1}). They are based on a statistical approach that considers the distribution of propagation directions in the absorber, and the probabilities for light to be trapped by total internal reflection or to escape in free space. It results in an average optical path $F d$.
Assuming weak absorption, E. Yablonovitch found the well-known maximum enhancement factor $F = 4 n^2$ in a seminal paper published in 1982~\cite{yablonovitch_statistical_1982}, and further extended to restricted acceptance angles $\theta$ as: $F = 4 n^2 / \sin^2{\theta}$~\cite{campbell_limiting_1986}.
He also proposed an absorption model for Lambertian scattering induced by surface texturation: $A_l = \alpha d / (\alpha d + 1/F)$~\cite{yablonovitch_statistical_1982,green_lambertian_2002}. This expression, often misleadingly named the Lambertian or Yablonovitch limit, is not a fundamental limit. It has been shown recently that it can be overcome, for example with carefully designed regularly patterned absorbers~\cite{massiot_progress_2020,eyderman_solar_2013}. An upper bound on absorption was derived by J. C. Mi{\~n}ano in 1990~\cite{minano1990optical} with the same factor $F$: $A = 1 - \exp{(-F \alpha d)}$. This expression is valid for both weak and strong absorption but it was obtained in the ray-optics regime.

In wave physics, open dissipative systems can be rigorously described in the framework of non-hermitian physics~\cite{Ashida2020}. In a one-port system (no transmission of incident waves), perfect absorption can be achieved at a given resonant frequency in the well-known critical coupling regime~\cite{Haus1984}. It has been extended to two-port systems as the so-called coherent perfect absorption~\cite{Chong2010}, and more recently to reflectionless multiport systems with the scattering matrix formalism~\cite{Sweeney2020,Stone2020,Guo2022a}. In contrast to previous approaches, we need to consider multiple strongly overlapping resonances to address the general question of broadband absorption of lossy systems.

%%%%%%%%%%%%

\subsection*{General multi-resonant model for absorption}

\begin{figure*}[t!]
    \centering
    \includegraphics[width=\textwidth]{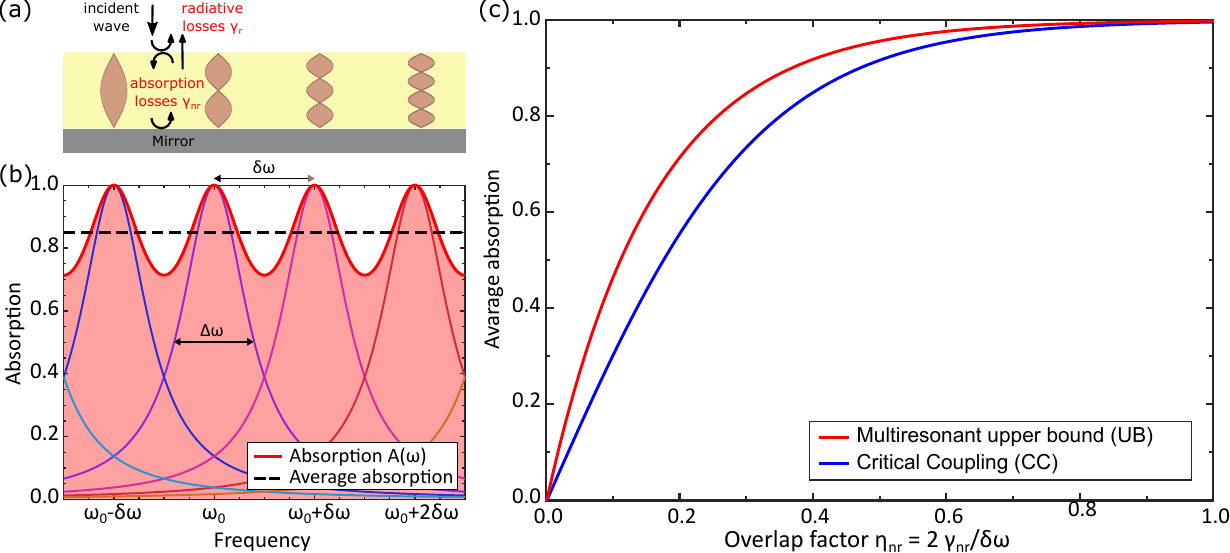}
    \caption{\textbf{General model for multi-resonant absorption.} (a) Sketch of an open system absorbing incident waves through multiple resonances and coupled through a single radiative channel. (b) Multiple evenly-distributed and independent overlapping resonances (colored Lorentzian curves) in the critical coupling regime, along with their total absorption envelop (red line and area) and the average absorption (dashed black line). (c) Average absorption as a function of the non-radiative overlap factor $\eta_{nr}$ in the critical coupling regime (radiative and non-radiative losses are equal) in blue (Eq.~\ref{eq:Abs_CC}), and upper bound in red (Eq.~\ref{eq:Abs_UB}).}
    \label{Fig1}
\end{figure*}

Here, we present a general framework based on the analysis of the response of a multi-resonant system in the frequency domain, described in detail in Supplementary Sections \com{S.2}, \com{S.3}, and \com{S.4}. We consider a slab of absorber with a back mirror, excited by a plane wave, as an illustrative example of a multi-resonant, open dissipative system (Fig.~\ref{Fig1}(a)). Interactions of incident waves with the system can be described by a set of resonant modes, the quasi-normal modes (QNMs)~\cite{Lalanne2018,Ching1998}, characterized by their complex frequencies $\omega_m^p = \omega_m + j \gamma$.

The absorption depends on the spectral distribution, the width, and the amplitude of resonances.
For given internal properties, i.e., absorption coefficient and thickness, the spectral response of the system can be tailored by optimizing its boundary conditions. For example, a frontside patterning allows tuning the density of resonant modes, and an anti-reflection coating their free-space coupling~\cite{battaglia_light_2012,eyderman_light-trapping_2015,massiot_progress_2020}.

The decay rate $\gamma$ represents the total losses. 
We use a phenomenological description of the system dynamics, the temporal coupled-mode theory~\cite{Fan2003,Collin2014}, to distinguish two different origins of losses: internal losses due to absorption are proportional to the non-radiative decay rate $\gamma_{nr}$, and external losses due to outcoupling are proportional to the radiative decay rate $\gamma_r$.
For a single resonance $m$, the absorption spectrum $A_m(\omega)$ is a simple Lorentzian function~\cite{Collin2014}:
\begin{equation}
    A_m(\omega) = \frac{4 \gamma_{r} \gamma_{nr}}{(\omega - \omega_m)^2 + (\gamma_{r} + \gamma_{nr})^2} \label{eq:Lor}.
\end{equation}
Considering a narrow spectral range relative to the resonance bandwidth $\Delta \omega = 2\gamma = 2 (\gamma_r + \gamma_{nr})$, absorption is maximized at the critical coupling condition $\gamma_{r}=\gamma_{nr}$, with perfect absorption at the resonance frequency $A_m(\omega_m) = 1$. 

Achieving broadband absorption requires considering a series of overlapping resonances, as illustrated in Fig. \ref{Fig1}(b). 
When the overlap between resonances is negligible, total absorption can be approximated as a sum of Lorentzian functions~\cite{yu_fundamental_2010}.
In contrast, the general formalism developed here accounts for the overlap of multiple resonances. We introduce $R_m(\omega) = 1 - A_m(\omega)$, the probability of reflection, by the mode $m$, of the energy carried by an incident wave. Intuitively, the total probability of reflection may be expressed as the product of the independent probabilities of reflection $R(\omega) = \prod\limits_m R_m(\omega)$, which leads to
\begin{equation}
	R(\omega) = \prod\limits_m \left| \frac{\omega - \omega_m^z}{\omega - \omega_m^p} \right|^2.
	\label{eq:A_tot}
\end{equation}
with
\begin{align} 
\omega_m^p &= \omega_m + j (\gamma_r + \gamma_{nr}) \label{eq:pole} \\ 
\omega_m^z &= \omega_m + j (\gamma_{nr} - \gamma_r). \label{eq:zero}
\end{align}
Equation~\ref{eq:A_tot} is an expansion of $R(\omega)$ over the poles and zeros of the system, which can be rigorously deduced from the Weierstrass factorization theorem~\cite{Grigoriev2013, Krasnok2019}, see the Supplementary Section \com{S.3}.

We now derive general reference models and upper bounds on absorption from Equation~\ref{eq:A_tot} by optimizing the spectral response of the system. First, the pole and zero expressed in Eqs.~\ref{eq:pole} and \ref{eq:zero} are exact for isolated resonances and provide the optimal solution to maximize the average absorption.
Second, we construct an infinite set of resonances evenly distributed in the frequency domain around a frequency $\omega_0$: $\omega_m = \omega_0 + m \delta \omega$, where $m$ is an integer and $\rho = 1/ \delta \omega$ is the spectral density of modes in the system (see Fig.~\ref{Fig1}(b)). The resulting absorption $A(\omega) = 1 - R(\omega)$ is a periodical function of $\omega$ with period $1/\rho$.

We define the \emph{radiative and non-radiative overlap factors}, $\eta_r= 2 \gamma_r / \delta \omega$ and $\eta_{nr}= 2 \gamma_{nr} / \delta \omega$, respectively. The total overlap factor $\eta = \eta_r + \eta_{nr} = \Delta\omega / \delta \omega$ quantifies the degree of overlap between neighboring resonances. We obtain a simple expression for the average absorption:
\begin{equation}
	\langle A \rangle = \frac{2\sinh(\pi\eta_r)\sinh(\pi\eta_{nr})}
	{\sinh\left(\pi(\eta_r+\eta_{nr})\right)}.
	\label{eq:A_av}
\end{equation}

Eq.~\ref{eq:A_av} is a key result of this work. The average absorption of a general multi-resonant system is a simple function of the dimensionless radiative and non-radiative overlap factors. While $\eta_{nr}$ depends on the material properties of the absorber, $\eta_r$ can be engineered by tailoring the coupling between the system interior and free space. Hence, $\eta_r$ can be considered an adjustable parameter to investigate reference models and upper bounds for absorption.

We propose the \emph{multi-resonant critical coupling} (CC) regime as a first reference model: the CC condition $\eta_{r}=\eta_{nr}$ is fulfilled for every resonance, leading to an average absorption:

\begin{equation}
    \langle A \rangle_{cc} = \tanh(\pi\eta_{nr}).
    \label{eq:Abs_CC}
\end{equation}

Although the CC regime leads to perfect absorption at each resonance frequency, it does not provide the highest average absorption. For a given $\eta_{nr}$, $\langle A \rangle$ is an increasing function of $\eta_{r}$ (Eq.~\ref{eq:A_av}). The \emph{upper bound} on multi-resonant absorption is reached as $\eta_r \to \infty$ (overcoupling regime):
\begin{equation}
    \langle A \rangle_{ub} = 1 - e^{-2\pi\eta_{nr}}.
    \label{eq:Abs_UB}
\end{equation}

Equations~\ref{eq:Abs_CC} and~\ref{eq:Abs_UB} are simple yet effective formulae. Multi-resonant absorption limits are expressed as a function of a single parameter, the non-radiative overlap factor $\eta_{nr}=2\gamma_{nr}\rho$, which accounts for the intrinsic properties of the system.
The CC model and the upper bound are compared in Fig.~\ref{Fig1}(c), where the absorption is plotted as a function of the non-radiative overlap factor $\eta_{nr}$. The average absorption approaches unity for both models when $\eta_{nr}$ is close to 1 or above. The difference becomes significant as $\eta_{nr}$ decreases, and the ratio between the upper bound and the CC absorption tends to a factor of two in the weak absorption regime ($\eta_{nr} \to 0$).

Assuming the resonant system is a slab of thickness $d$ filled with a linear, homogeneous, and isotropic material, the attenuation (or absorption) coefficient $\alpha$ of a propagating wave intensity can be linked to its energy decay time $2\gamma_{nr}$ and group velocity $\nu_g$ through $2\gamma_{nr} = \alpha \nu_g$, and the non-radiative overlap factor is expressed as $\eta_{nr} = \alpha \nu_g \rho$.

The absorption upper bound can be written as a function of an effective path length $F d$, where $F$ is defined as the wave path enhancement factor:
\begin{equation}
    \langle A \rangle_{ub} = 1 - e^{-F \alpha d},
    \label{eq:Abs_UBF}
\end{equation}
\begin{equation}
    F = \frac{2\pi \eta_{nr}}{\alpha d} = 2\pi \nu_g \rho/d.
    \label{eq:F}
\end{equation}

This expression allows for a direct comparison with the usual Beer-Lambert law obtained for single-pass absorption, $F=1$. It is worth noting that $F$ has no physical meaning at the resonance frequency in the case of critical coupling ($F \to +\infty$) but is a useful figure of merit for spectrally-averaged absorption enhancement.

The above derivation assumes a single radiative channel, which is appropriate to describe a planar slab textured with a sub-wavelength periodical pattern, i.e., with no diffraction in free space.
This theoretical framework can be extended to the case of $N$ identical radiative channels (see Supplementary Section \com{S.4}). The absorption upper bound takes the same form by replacing $\eta_{nr}$ by $\eta_{nr}/N$. Eqs. \ref{eq:Abs_UBF} and \ref{eq:F} will be used in the following to model isotropic scattering induced by random structures with $N \to +\infty$.

The assumptions used in the above derivation are detailed in the Supplementary Section \com{S.3}.
The main requirement is a slow variation of the overlap factors over the spectral ranges relative to the resonance spacing $\delta \omega$ and the resonance width $\Delta \omega$. The average absorption is not significantly affected by other assumptions. For instance, replacing the infinite distribution of evenly-spaced modes by a finite number of resonances or a non-uniform distribution has little impact on $\langle A \rangle$.
There is an apparent contradiction between the overcoupling regime of the upper bound and the limited losses required to apply the Lorentzian approximation. Actually, we show that radiative overlap factors as low as 0.5 are sufficient to reach 95\% of the upper bound absorption. We also note that the closed-form expressions of Eqs.~\ref{eq:A_av}, \ref{eq:Abs_CC}, and \ref{eq:Abs_UB} are intrinsically bounded between 0 and 1, ensuring that absorption tends to 1 for large internal losses, even beyond the validity of the Lorentzian approximation.

%%%%%%%%%%%%%
\subsection*{Upper bounds on sunlight absorption}

\begin{figure*}[t!]
    \centering
    \includegraphics[width=\textwidth]{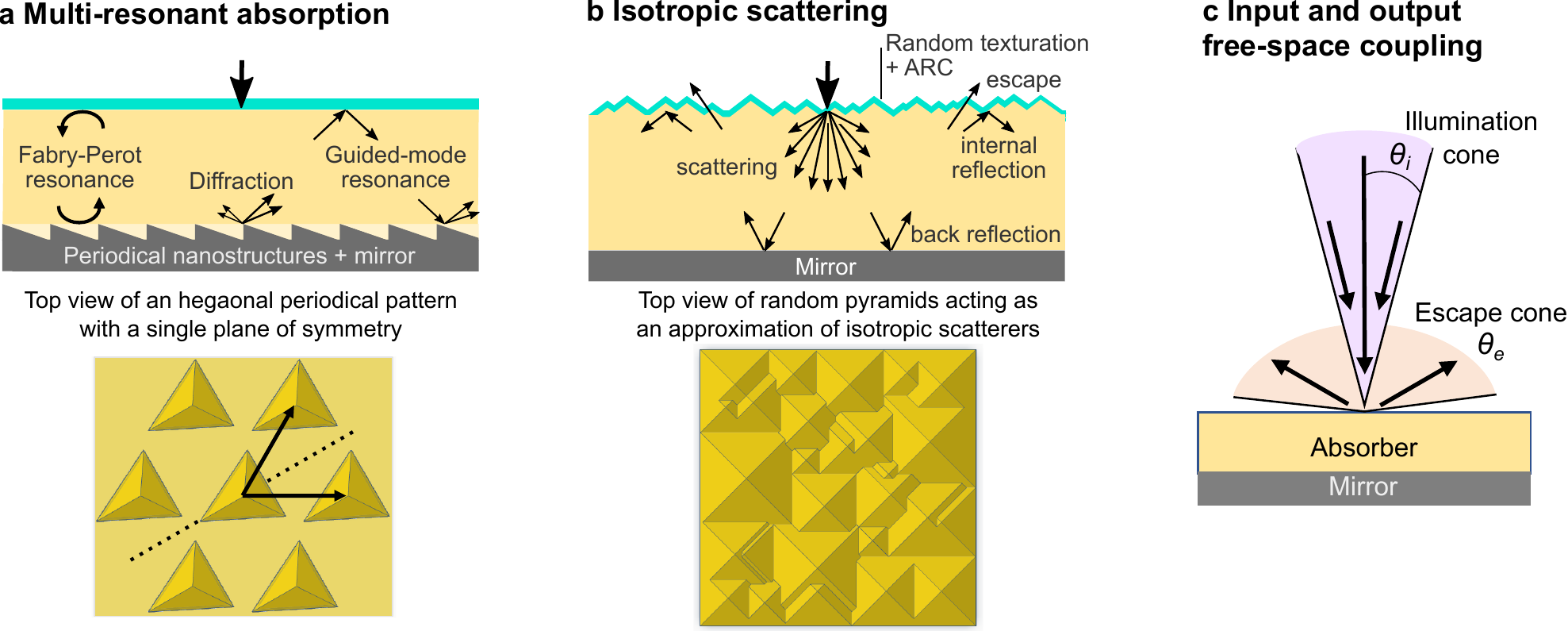}
    \caption{\textbf{Light-trapping structures based on periodic and random patterning, and free-space coupling schemes.} (a-b) Sketches of the two reference light-trapping strategies. (a) Multi-resonant absorption induced by a regular pattern, here an hexagonal lattice of tetrahedra on the back side. (b) Isotropic scattering induced by random texturation, here pyramids of random size and position on the front side, as obtained by chemical etching of silicon wafers and typically used in solar cells. (c) Illumination and escape cones constrain the free-space coupling.
    }
    \label{fig:patterning}
\end{figure*}

\paragraph{Multi-resonant absorption.}
In the following, we apply this multi-resonant theoretical framework to optics, and we address the important question of the upper bounds on sunlight absorption in a solar cell modeled by a slab of semiconductor (see details in the Supplementary Section \com{S.5}).
A perfect back reflector is used to avoid transmission through the backside of the absorber.
Multiple well-defined resonances can be induced by a two-dimensional regular pattern at the front or back side, leading to diffraction effects and guided-modes resonances in the absorber~\cite{mellor_upper_2011,yu_fundamentalgrating_2010,Chen2019}, see Fig.~\ref{fig:patterning}(a).  The spectral density of modes depends on the period $p$.
Their coupling to free space $\gamma_{r}$ can be controlled by tuning the shape of the sub-wavelength patterning, and by using an anti-reflection coating. Since most optical modes are well confined in the absorber, the group velocity can be expressed as $\nu_g = c/n$, and $\displaystyle F = \frac{2 \pi c \rho}{n d}$.

Sunlight incident on the absorber can be decomposed as an ensemble of independent plane waves linearly polarized along two orthogonal directions~\cite{Divitt2015}.
In the most general case, polarization rotation can be induced by the resonant system, such that two radiative channels should be considered.
However, if the 2D pattern has a plane of symmetry perpendicular to the slab, then the two polarization states parallel and orthogonal to the mirror plane do not couple for normal incidence~\cite{menzel_advanced_2010}.
The total density of modes $2\rho$ is then divided in two halves, each of which is coupled to a single radiative channel. For each of the two polarization states, the previous derivations (Eqs.~\ref{eq:Abs_CC}, \ref{eq:Abs_UB}, \ref{eq:Abs_UBF}, \ref{eq:F}) apply.

Optimizing sunlight absorption requires maximizing the density of modes $\rho$ in the system while avoiding diffraction in free space. Considering the reciprocal space of the diffraction pattern, the problem is equivalent to the optimization of circle packing in the plane. The optimal lattice is hexagonal, and the optical path enhancement factor is
\begin{equation}
    F = 2 \sqrt{3} \pi n^2 \left( \frac{p}{\lambda} \right)^2.
    \label{eq:Fub_hex_p}
\end{equation}
\noindent where $p$ is the period of the lattice.
The upper bound is achieved for $p \to 2 \lambda / \sqrt{3}$:
\begin{equation}
    F_{ub} = \frac{8}{\sqrt{3}} \pi n^2.
    \label{eq:Fub}
\end{equation}

The plane of symmetry simplifies the design and requires less computationally intensive simulations. It also enables the use of the multi-resonant critical coupling model valid only with a single radiative channel. This symmetry can be broken by design or with oblique incidence. In this case, every mode (density $2 \rho$) is coupled to both radiative channels and we find the same path enhancement factor and upper bound. On the contrary, higher orders of symmetry would result in mode degeneracies and lower average absorption. As a result, the optimal pattern has no more than one plane of symmetry~\cite{menzel_advanced_2010}.

The multi-resonant upper bound given by Eq.~\ref{eq:Fub} is derived for normal-incident light ($\theta_i = 0$) and no angular restriction for photon escape (no diffraction allowed in the whole hemisphere above the absorber, $\theta_e = \pi/2$). It can be generalized by imposing the single-channel condition to an illumination cone with half-angle $\theta_i$ and an escape cone restricted to a half-angle $\theta_e \geq \theta_i$ (Fig.~\ref{fig:patterning}(c)):

\begin{equation}
    F_{ub}(\theta_i,\theta_e) = \frac{F_{ub}}{[\sin{\theta_i}+\sin{\theta_e}]^2}.
    \label{eq:FubTheta}
\end{equation}

The path enhancement factors apply to both the critical coupling reference model and the upper bound. The formulas are summarized in Table~\ref{table:expressions}.

\newcolumntype{M}[1]{>{\centering\arraybackslash}m{#1}}
\begin{table*}[t!]
\centering
 \begin{tabular}{m{0.23\textwidth}|M{0.3\textwidth} M{0.3\textwidth}} 
 & \textbf{Multi-resonant absorption} & \textbf{Isotropic scattering}\\
 \hline
 Upper bound & \multicolumn{2}{c}{$\displaystyle \langle A \rangle_{ub} = 1 - e^{-F \alpha d}$} \\
 \hline
 \multirow{2}{*}{Reference models} & Critical coupling & Lambertian scattering\\
 & $\displaystyle \langle A \rangle_{cc} = \tanh\left(\frac{F \alpha d}{2} \right)$ & $\displaystyle A_l = \frac{\alpha d}{\alpha d + 1/F}$\\
 \hline
 Maximum path enhancement factor $F$ & $\displaystyle F = \frac{8}{\sqrt{3}} \pi n^2$ & $\displaystyle F = 4 n^2$\\
 \hline
 Restricted cones for illumination ($\theta_i$) and escape ($\theta_e$) & $\displaystyle F(\theta_i,\theta_e) = \frac{8}{\sqrt{3}} \frac{\pi n^2}{[\sin{\theta_i}+\sin{\theta_e}]^2}$ & $\displaystyle F(\theta_i=\theta_e) = \frac{4 n^2}{\sin^2{\theta_e}}$\\
 \end{tabular}
\caption{\textbf{Summary of the reference absorption models for light-trapping in solar cells}: the upper bounds and the maximum path enhancement factors derived for multi-resonant absorption and isotropic scattering, and two reference models (critical coupling absorption and Lambertian scattering).}
\label{table:expressions}
\end{table*}

%%%%%%%%%%%%%%%%%%%%%%%%%%%%%%%%%%
\paragraph{Isotropic scattering.}
As an alternative to multi-resonant absorption, isotropic scattering can be used to trap sunlight in the absorber (Fig.~\ref{fig:patterning}(b)). Oblique propagation and total internal reflection leads to light path enhancement that has been widely studied in the ray-optics regime. 
Our multi-resonant theoretical framework provides a way to generalize previous results: isotropic scatterers are modeled as large-period structures that allow free-space coupling to a large number $N$ of identical channels with the same coupling rate $\gamma_r$~\cite{yu_fundamental_2010}. The light path enhancement is then (see Supplementary Section \com{S.6}):
\begin{equation}
    F = \frac{2 \pi c}{n d}\frac{\rho}{N},
\end{equation}
and in the limit of large periods ($p \gg \lambda$,) both $\rho$ and $N$ scale as $p^2$, leading to:
\begin{equation}
    F_{s} = 4 n^2.
    \label{eq:Fll}
\end{equation}

Assuming restricted illumination and escape cones with the same half-angle $\theta_e = \theta_i$, the maximum path enhancement factor for isotropic scattering becomes:
\begin{equation}
    F_{s}(\theta_i) = \frac{F_{s}}{\sin^2{\theta_i}}.
    \label{eq:Fis}
\end{equation}

Equations~\ref{eq:Abs_UBF}, \ref{eq:Fll} and \ref{eq:Fis} (see also Table~\ref{table:expressions}) recover the well-known $4 n^2$ factor and the upper bound absorption previously derived for isotropic scattering in the ray-optics regime~\cite{yablonovitch_statistical_1982,minano1990optical,campbell_limiting_1986}. We show that they are also consistent with previous results obtained for wave optics in the weak absorption regime~\cite{yu_fundamental_2010,yu_fundamentalgrating_2010} (Supplementary Section \com{S.7}) and that the ``thermodynamic upper bound" derived in reference~\cite{yu_thermodynamic_2012} does not hold (Supplementary Section \com{S.8}). Here, these seminal results are generalized to a wide domain of validity that includes both thick and ultrathin solar cells in the weak and strong absorption regimes.

\begin{figure}[t!]
    \centering
    \includegraphics[width=\columnwidth]{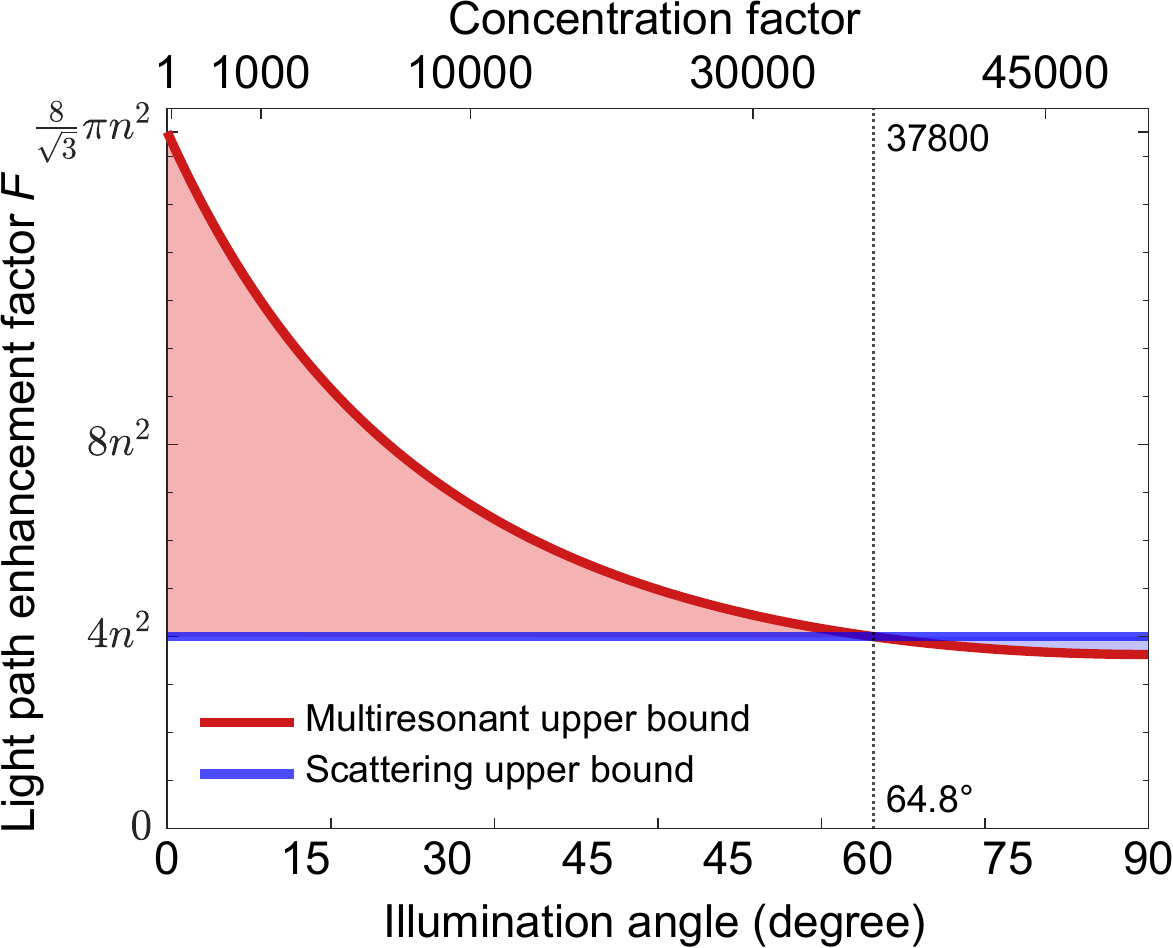}
    \caption{\textbf{Upper bounds on light absorption with multiple resonances and isotropic scattering.} Maximum light path enhancement factor $F$ as a function of the half-angle of the illumination cone for multi-resonant absorption with no diffraction in free space, and for isotropic scattering. Photon escape is considered here without restriction (full hemisphere, $\theta_e = \pi/2$). The half-angle $\theta_i$ can be translated into concentration factors $\displaystyle C = \sin^2(\theta_i) / \sin^2(\theta_{1sun})$, where $\theta_{1sun} = 0.27^{\circ}$ is the half-angle of the sun as seen from the earth.
    }
    \label{fig:comparison_models}
\end{figure}

\paragraph{Optimal light-trapping strategy.}
We compare the absorption enhancement achieved by multi-resonant absorption and isotropic scattering as a function of the half-angle of the illumination cone $\theta_i$, assuming no restriction for photon escape ($\theta_e = \pi/2$), see Fig.~\ref{fig:comparison_models}. At normal incidence, multi-resonant absorption is the most efficient strategy by far, and its enhancement factor $F$ stays beyond the $4 n^2$ limit of isotropic scattering for illumination half-angles up to $64.8^\circ$. As an illustration, concentration photovoltaics is currently limited to factors up to about 1000 suns, corresponding to narrow illumination cones close to normal incidence (half-angle of $8.5^\circ$). For wider angles and in the case of angular restriction on photon escape with $\theta_e = \theta_i$, isotropic scattering leads to higher $F$ factors, with a gain over multi-resonant absorption limited to $2\sqrt{3}/\pi \approx 1.10$ maximum. These results provide an answer to the long-debated question of the best light-trapping strategy for solar cells~\cite{battaglia_light_2012, van_lare_optimized_2015,burresi_complex_2015}.

\subsection*{Application to silicon photovoltaics}

\begin{figure*}[t!]
    \centering
    \includegraphics[width=\textwidth]{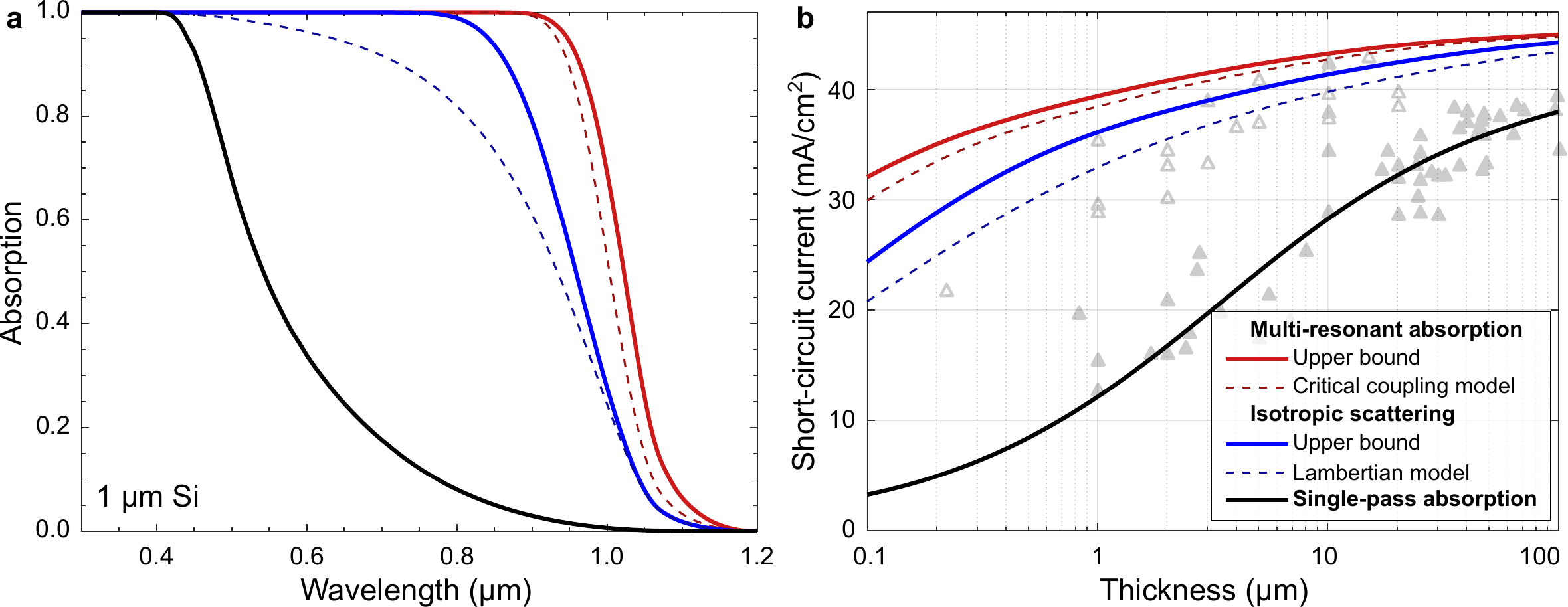}
    \caption{\textbf{Upper bounds on light-trapping in silicon solar cells.} Comparison of the different light-trapping models for silicon solar cells modeled as a simple, textured slab with a back mirror. (a) Absorption spectra for a $1~\rm{\mu m}$-thick silicon slab. (b) Short-circuit current density $J_{sc}$ calculated for the reference models and upper bounds for one-sun illumination. The grey symbols show results of short-circuit current measurements (filled triangles) and simulation (open triangles) taken from the literature, see reference~\cite{massiot_progress_2020}.
    }
    \label{Fig3}
\end{figure*}

We apply the previous results to silicon (Si) solar cells, the most common photovoltaic technology.
Reducing the thickness of Si wafers with respect to the current industry standards (about 150~µm), with no performance degradation, is an important pathway for material, energy, and cost savings~\cite{liu_revisiting_2020,haegel_photovoltaics_2023}.
In the following, we assume normal incidence and no restriction on the escape cone. The refractive index of Si is taken from reference~\cite{green_improved_2022}. The model assumptions are satisfied over the whole range of thicknesses and wavelengths investigated (see the Supplementary Section \com{S.9}).

As an illustration, we first consider a 1~µm-thick silicon layer, about two orders of magnitude thinner than commercial solar cells.  
Absorption spectra plotted in Figure~\ref{Fig3}(a) show the large gap between the two main light-trapping strategies: light scattering and multi-resonant absorption.
Lambertian scattering, the reference light-trapping scheme for state-of-the-art solar cells, provides a huge improvement over single-pass absorption. However, it remains far from the isotropic scattering upper bound. Critical-coupling absorption is close to the multi-resonant upper bound and exhibits a significant gain over light-scattering strategies.

The short-circuit current $J_{sc}$ represents the figure of merit for the overall absorption in a solar cell under one sun illumination. 
Assuming each photon absorbed in the semiconductor contributes to one electron-hole pair, $J_{sc}$ is calculated by integrating the absorption spectrum weighted by the spectral density of photon flux (AM1.5G spectral irradiance) between 300~nm and 1200~nm. The results are plotted as a function of the slab thickness in Fig.~\ref{Fig3}(b).
They confirm that there is plenty of room for improvement beyond the Lambertian light-trapping strategy, with a ten-fold reduction in thickness enabled by approaching to the multi-resonant upper bound.

The derivation of reference models and upper bounds provide guidelines for the design of ultrathin solar cells \cite{massiot_progress_2020}. In practice, the spectral position and free-space coupling of resonant modes are determined by the shape, size, and arrangement of nanostructures.
We have shown that the best multi-resonant absorption scheme requires a pattern with an hexagonal arrangement, no more than one plane of symmetry, and a maximum interspacing of $2 \lambda_0 / \sqrt{3}$. If the pattern is placed at the backside, diffraction in free space may only occur after a double-pass in the absorber and for $\lambda < \lambda_0$. Hence, $\lambda_0$ defines the long-wavelength region where light-trapping is needed.

Experimental and numerical results taken from the literature are also shown in Fig.~\ref{Fig3}(b)~\cite{massiot_progress_2020}. Importantly, several data points lie above the Lambertian model and the isotropic upper bound.
It is interesting to point out that these results were obtained with a periodical texture that complies with most of the criteria required for optimal multi-resonant absorption: the pattern is made of slanted cones with a single plane of symmetry, and a period of 0.85~µm\cite{eyderman_light-trapping_2015}.
Therefore, our theoretical framework reconciles the results found in the literature with reference models and upper bounds, and sets the upper limits for each light-trapping scheme.
Still, intensive numerical work and efficient optimization strategies are needed to account for the vast geometrical parameter space and reach absorption close to the upper bound with realistic designs for ultrathin solar cells~\cite{massiot_progress_2020}.

The fundamental efficiency limit of a single-junction silicon solar cell is the result of a trade-off between photon absorption and intrinsic charge carrier recombination. Assuming Lambertian scattering, an optimal thickness of 100~µm leads to a theoretical efficiency of 29.4\%, with $J_{sc} = 43.4~\ \mathrm{mA/cm^2}$~\cite{Niewelt2022}. Actually, the same short-circuit current is obtained with thicknesses of only 12~µm for the multi-resonant upper bound, and 42~µm with isotropic scattering. Such thickness reduction results in decreased Auger recombination and a predicted voltage gain of more than 20~mV~\cite{Niewelt2022}. Hence, the maximum theoretical efficiency of silicon solar cells should be reassessed and is expected to be above 30~\% for thicknesses of a few tens of µm.

\subsection*{Discussion}

In the previous section, our theoretical models were compared to the experimental and numerical results published in the literature for thin and ultrathin silicon solar cells. The upper bounds on absorption have a wide domain of validity, and can be applied to other materials, including direct bandgap semiconductors (see the example of GaAs in the Supplementary Section \com{S.10}).

It is interesting to note that with a regular patterning, optical absorption beyond the multi-resonant upper bound can be achieved in specific cases. First, the density of modes used in the above derivation assumes bulk materials. Absorption can be further enhanced by increasing the local density of states close to sub-wavelength structures such as plasmonic particles~\cite{Balian1970,Callahan2012,Atwater2010,Massiot2014}. 
Such effects are limited to very thin absorbers, or inhomogeneous, nanostructured media.
Second, we can break the requirement of a single radiative channel. Larger periods lead to an increase of the density of modes, but also of free-space losses due to diffraction. Sweet spots can be found for normal incidence illumination by canceling the coupling to diffracted waves in free-space, but at the expense of the acceptance angle~\cite{Bhattacharya2019}.

The absorption models provide guidelines to design optimal light-trapping structures. However, significant efforts still need to be made to optimize numerically the shape and size of patterning to get close to the upper bounds. For instance, it raises the challenge of designing texturation that meets the requirement of large $\gamma_r$ over a broad spectral range.
Moreover, our theoretical framework was applied to perfectly periodical patterning for multi-resonant absorption, and to fully random structures for isotropic scattering. Beyond these two limiting cases, the physics of correlated disordered media have emerged recently and open new routes to control finely both free-space coupling and light propagation in absorbers~\cite{Vynck2021}. In particular, hyperuniform structures offer a promising route to favor diffraction at high angles in the absorber~\cite{Bigourdan2019,buencuerpo_efficient_2022,camarillo_abad_transparent_2022,tavakoli_over_2022}. The extension of our multi-resonant framework to the different classes of correlated disordered patterning is an intriguing research direction.

In conclusion, we have proposed a general theoretical framework for multi-resonant absorption in an open, dissipative system. Analytical expressions for the absorption upper bounds where derived for a slab with either random or periodical patterning, and applied to light-trapping in solar cells. Our results show the route towards high-efficiency, ultrathin solar cells, enabling manufacturing with lower material usage and carbon footprint, and the design of flexible devices~\cite{massiot_progress_2020}.
Beyond the realm of optics and photovoltaics, our general formalism can be applied to any kind of electromagnetic and mechanical wave, and should guide the development of ultrathin, broadband detectors and absorbers for radar waves~\cite{li_ultrathin_2012}, compact acoustic sinks~\cite{Leroy2015,huang_compact_2020}, the attenuation of seismic waves~\cite{Brule2014,Brule2020}, or energy harvesting of ocean water waves~\cite{Benzaouia2019}.

\subsection*{Acknowledgements}

The authors acknowledge Philippe Lalanne and Andrea Cattoni for early discussions, and Daniel Suchet, Jean-François Guillemoles, Sylvain Finot, Jerónimo Buencuerpo, Mariano Pascale, and Raffaele Colombelli for helpful discussions and suggestions on the manuscript. Part of this work was performed within the framework of LIA NextPV, currently IRP NextPV-II, with support from CNRS  (France) and RCAST/University of Tokyo (Japan). The work of S.C. was supported by the French Government in the frame of the ''Programme d'Investissement d'Avenir'' (ANR-IEED-002-01), and the ANR projects NANOCELL (ANR-15-CE05-0026) and ICEMAN (ANR-19-CE05-0019).

\subsection*{Additional information}
See Supplemental Material for the full derivation of the theoretical framework, additional discussions on the historical perspective on light-trapping, critical comments on previous works, and additional results of absorption models applied to silicon and GaAs slabs.

\bibliography{MRABib.bib}

\end{document}

\title{Upper bounds on broadband absorption: Supplemental Material}
\author[1,2]{Stéphane Collin\thanks{Corresponding author: \texttt{stephane.collin@c2n.upsaclay.fr}}}
\author[1,3]{Maxime Giteau}
 
\affil[1]{Centre de Nanosciences et de Nanotechnologies (C2N), CNRS, Université Paris-Saclay, Palaiseau, France}
\affil[2]{Institut Photovolta{\"\i}que d'Ile-de-France (IPVF), Palaiseau, France}
\affil[3]{Laboratoire PROcédés, Matériaux et Energie Solaire (PROMES), CNRS, Odeillo, France}

\date{}

\maketitle

\clearpage
\tableofcontents

%%%%%%%%%%%%%%%%%%%%%%%%
\clearpage
\section{A historical perspective on light-trapping models and upper bounds in the ray-optics regime}
\label{sec:lit}

\subsubsection{Statistical approach and the $F=4n^2$ factor: E. Yablonovitch, 1982}

In 1982, Yablonovitch proposed a \emph{statistical approach} based on ray optics and valid in the regime of \emph{weak absorption}~\cite{yablonovitch_statistical_1982,yablonovitch_intensity_1982}.
The main assumption is that \emph{the behavior of light rays is ergodic}, which means that the angular distribution of light in the absorber is randomized (the distribution is uniform in the phase space). In this framework, isotropic and collimated illuminations are equivalent.

Using simple arguments based on the equilibrium between external and internal radiation, it is shown that the maximum light intensity enhancement obtained with a back mirror is $2 n^2$. Light absorption can be approximated by $A = F \alpha d$, where the optical path enhancement factor is $F=4n^2$.

Using geometrical optics, it is also shown that optical absorption in an slab textured to create Lambertian (i.e. isotropic) scattering can be modeled as: $A = \frac{\alpha d}{\alpha d + 1/F}$ (without parasitic losses).

\subsubsection{Restricted angle: P. Campbell and M. A. Green, 1986}

In 1986, Campbell and Green extended this approach to the case of a \emph{restricted angle} $\theta_e$ for both illumination and light escape~\cite{campbell_limiting_1986}. Keeping the assumptions of angular randomization of rays in a weakly absorbing medium, the light path enhancement $F$ is increased by a factor $1/\sin^2 \theta_e$.

\subsubsection{Upper bound: J. C. Miñano, 1990}

The work published by Miñano in 1990 deserves a great consideration since it provides \emph{an upper bound for absorption valid for ray optics which is not restricted to the weak absorption regime}~\cite{minano1990optical,luque_optical_1991}.

He introduces the distribution of optical path lengths and the average light path $\langle l \rangle$, and demonstrates that the maximum absorption is given by $A = 1-\exp(-\alpha \langle l \rangle)$. This limit is reached when all path lengths are equal (the distribution of path lengths is a Dirac delta function). Consistently with previous works, assuming an isotropic optical response within a restricted angle $\theta_e$ leads to the maximum average path length $\langle l \rangle = 4 n^2 d / \sin^2 \theta_e = F d$. Overall, this derivation provides the upper bound absorption valid in the regime of ray optics: $A = 1-\exp{(-F \alpha d)}$.

In the weak absorption regime, the impact of the distribution of path lengths vanishes, and the absorption is approximated by $A \simeq \alpha \langle l \rangle$, in agreement with the previous papers.

\subsubsection{Generalization of the Lambertian model: M. A. Green, 2002}

In 2002, Green showed numerically that the expression derived by Yablonovitch to model Lambertian scattering in weakly absorbing media remains a good approximation for any absorption $\alpha d$~\cite{green_lambertian_2002}.

\subsubsection{Comparison of the low-absorption limit, the Lambertian model and the upper bound}

\begin{figure}[ht]
    \centering
    \includegraphics[width=0.6\textwidth]{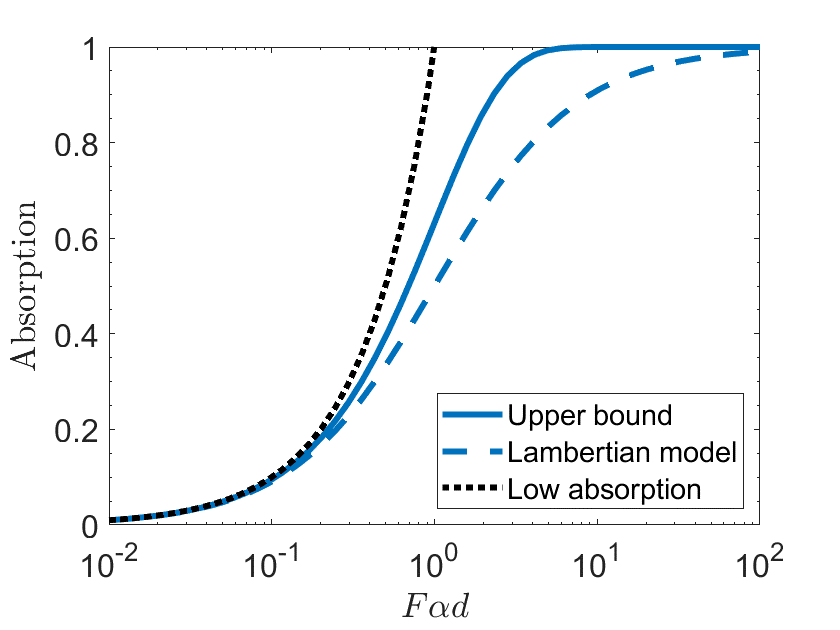}
    \caption{Absorption as a function of $F \alpha d$ for different models: the upper bound model (Eq.~\ref{eq:ray_optics_ub}, blue solid line), the Lambertian model (Eq.~\ref{eq:ray_optics_lamb}, blue dashed line) and the low-absorption approximation $ A = F \alpha d$ (black dotted line). 
    }
    \label{fig:A_models_comparison}
\end{figure}

Although the same light path enhancement factor $F$ is found for the Lambertian model and the upper bound, different expressions for the absorption were found. They are compared in Fig.~\ref{fig:A_models_comparison}, along with the low-absorption limit $A = F \alpha d$. We note that this linear approximation is only valid up to $A \approx 0.1$. The upper bound curve is well above the Lambertian model and converges much more rapidly towards 1.
\newline

\definecolor{shadecolor}{gray}{0.9}
\begin{shaded}
In summary, it was shown that the \emph{upper bound absorption} is expressed by
\begin{equation}
    A = 1-\exp{(-F \alpha d)}
    \label{eq:ray_optics_ub}
\end{equation}
and \emph{Lambertian scattering} can be modeled by
\begin{equation}
    A = \frac{\alpha d}{\alpha d + 1/F}.
    \label{eq:ray_optics_lamb}
\end{equation}
For an isotropic response restricted to an illumination and escape cone of half-angle $\theta_e$, the maximum optical path length enhancement is given by
\begin{equation}
    F = \frac{4 n^2}{\sin^2 \theta_e}.
\end{equation}
\textbf{These expressions are not restricted to weak absorption, but they were only demonstrated in the ray optics regime.}
\end{shaded}

\subsubsection{Generalization of the light path enhancement: I.M. Peters, 2014~\cite{peters_phase_2014} and U. Rau, 2014~\cite{rau_thermodynamics_2014}}

In reference~\cite{peters_phase_2014}, Peters further investigates the maximum optical path length. He shows that for an incident illumination through a solid angle $\Omega$, the maximum light path enhancement is bounded by $4n^2/\Omega$. This bound is reached when all the available states in the phase space are filled. This condition can be achieved in a single scattering event by Lambertian scatterers, or by successive events based on well-defined redirection of light rays.

In reference~\cite{rau_thermodynamics_2014}, Rau discusses the importance of the optical path length distribution to reach the upper bound. The reason why Lambertian scattering is significantly lower than the upper bound is due to the dispersion of the optical path lengths. He introduces $\langle 1/l \rangle$ as a figure of merit to assess the scattering efficiency. He also presents a gedanken experiment for a deterministic light trapping scheme that approaches the absorption upper bound.

\subsubsection{Extension of ray optics to diffraction-based light trapping: I. Tobías, 2008~\cite{tobias_light_2008} and A. Mellor, 2011~\cite{mellor_upper_2011}}

Still in the ray optics formalism, light-trapping theories have been extended to the use of diffraction gratings. In particular, Tobias \textit{et al.} have shown that light path enhancements $F>4n^2$ can be achieved at normal incidence without angular restriction, but with a limited angle of acceptance and spectral bandwidth~\cite{tobias_light_2008}.
Mellor \textit{et al.} emphasized that Lambertian scattering light-trapping can be overcome by diffraction gratings, approaching the upper bound derived by Miñano under specific conditions~\cite{mellor_upper_2011}.

%%%%%%%%%%%%%%%%%%%%%%%%%%%%%%%%%%%%
\clearpage
\section{A model for multi-resonant absorption with a single radiative channel}
\label{sec:MRAsinglechannel}

We first derive an analytical model for the absorption of an incident plane wave by a \textbf{multi-resonant structure coupled to a single radiative channel}. It will be generalized to multiple radiative channels in section~\ref{sec:GeneralFormalism}. A slab of absorber with a back reflector and a two-dimensional periodical pattern with a sub-wavelength period is an example of such a structure. The pattern provides a way to create multiple resonances in the system, and the sub-wavelength period prevents the propagation of diffracted waves in free-space, except from the zero order (specular reflection).

In the following, we model the optimal absorption spectrum as the result of an infinite series of Lorentzian resonances regularly spaced in frequency, and we derive closed-form expressions for the absorption limits in two reference cases. The validity of the assumptions of the model will be further discussed in section~\ref{sec:GeneralHypotheses}.

%%%%%%%%%%%%%%%%%%%%
\subsection{Pole-zero decomposition}
\label{ssec:polezero}

Open resonant systems can be described by their natural resonant modes, named quasinormal modes (QNMs)~\cite{Ching1998,Lalanne2018}, and in particular their complex eigenfrequencies.

We start with a single resonant mode $m$ characterized by a resonance frequency $\omega_m$, a non-radiative decay rate $\gamma_{nr}$ (absorption, internal losses), and a radiative decay rate $\gamma_r$ (free-space coupling).
The total decay rate is $\gamma=\gamma_r+\gamma_{nr}$.
Following the temporal coupled-mode theory~\cite{Fan2003,Collin2014}, the absorption spectrum $A_m(\omega)$ of the system takes the form of a Lorentzian function:

\begin{equation}
	A_m(\omega)=\frac{4\gamma_r\gamma_{nr}}{(\omega-\omega_m)^2+(\gamma_r+\gamma_{nr})^2},
	\label{eq:Lorentzian_m}
\end{equation}

\noindent with a full width at half maximum $\Delta \omega = 2 \gamma$.
The reflection is 

\begin{equation}
    R_m(\omega) = 1 - A_m(\omega) =  \left| \frac{\omega - \omega_m^z}{\omega - \omega_m^p} \right|^2,
\end{equation}

\noindent where the right-hand side of the equation is a decomposition over the pole $\omega_m^p = \omega_m + j \gamma$ and zero $\omega_m^z = \omega_m + j \left(\gamma_{nr} - \gamma_{r}\right)$ of the resonant mode. This decomposition is consistent with the phenomenological theory of total resonant absorption in diffraction gratings~\cite{Hutley1976,Maystre2013}. 
At the resonance frequency, perfect absorption $A_m(\omega_m)=1$ is achieved for the critical coupling condition $\gamma_r=\gamma_{nr}$.

We consider now a series of overlapping resonances. The absorption is derived from a Weierstrass factorization of the amplitude of reflected waves $r(\omega)$~\cite{Grigoriev2013,Krasnok2019}. The total reflection $R(\omega)=|r(\omega)|^2$ is written:
\begin{equation}
	R(\omega) = \prod\limits_m R_m(\omega) = \prod\limits_m \left| \frac{\omega - \omega_m^z}{\omega - \omega_m^p} \right|^2.
	\label{eq:R_tot}
\end{equation}
The pole and zero of each mode are poles and zeros of the total reflection. In the case of incident light waves, Eq.~\ref{eq:R_tot} can be interpreted as follows: the total probability of reflection of an incident photon is the product of the independent probabilities of reflection of every mode $m$. More details on the Weierstrass factorization are given in sections \ref{sec:Weierstass} and \ref{ssec:MultipleResonances}.
The total absorption is then:

\begin{equation}
	A(\omega) = 1 - \prod_{m=-\infty}^{+\infty}(1-A_m(\omega)).
	\label{eq:A_tot}
\end{equation}

As an optimal scheme to maximize absorption, we assume a series of resonances regularly spaced in frequency every $\delta\omega$, in the frequency range of interest around a frequency $\omega_0$: $\omega_m= \omega_0 + m \cdot \delta\omega$ with $m\in\mathbb{Z}$. We also assume that the system properties, i.e. the density of modes and the decay rates, vary slowly over spectral ranges given by the period $\delta\omega$ and the resonance width $\Delta\omega$. These assumptions are further discussed in section~\ref{sec:GeneralHypotheses}. Finally we observe that for large values of $m$, resonances have no impact on the absorption around $\omega_0$ since $m \delta\omega \gg \Delta\omega$, so that the series of modes can be extended from $m=-\infty$ to $m=+\infty$.

The absorption spectrum $A(\omega)$ is a periodical function of $\omega$ (period: $\delta\omega$) which is derived analytically in the following, and the averaged absorption can be expressed as:

\begin{equation}
	\langle A \rangle = \frac{1}{\delta\omega} \int_{\omega_m-\delta\omega/2}^{\omega_m+\delta\omega/2} A(\omega) d\omega.
	\label{eq:Average_abs1}
\end{equation}

The closed-form expression of the averaged absorption $\langle A \rangle$ will provide a simple way to evaluate the upper bound for multi-resonant absorption. In the following, we first express total absorption in the critical coupling condition, and then we derive a more general formula valid for any radiative coupling.

%%%%%%%%%%%%%%%%%%%%
\subsection{Critical coupling}

In the case of critical coupling $\gamma_r=\gamma_{nr}$, perfect absorption is achieved at each resonance frequency: $A(\omega_m)=1$.

Equation \ref{eq:Lorentzian_m} leads to:

\begin{equation}
\begin{aligned}
	1- A_m(\omega) &= 1- \frac{\gamma^2}{({\omega-m\delta\omega})^2 + \gamma^2}\\
	&= \frac{1}{1+\left(\frac{\gamma/\delta\omega}{\omega/\delta\omega-m}\right)^2}.
\end{aligned}
	\label{eq:Lorentzian_m2}
\end{equation}

Using Eqs.~\ref{eq:A_tot} and \ref{eq:Lorentzian_m2} and the identity

\begin{equation}
	\prod_{m=-\infty}^{+\infty}\left(1+\left(\frac{a}{b-m}\right)^2\right)=1+\sinh^2(\pi a)\csc^2(\pi b),
	\label{eq:identity}
\end{equation}

\noindent where $\csc = 1/\sin$, valid for $b\in\big]-\frac{1}{2},\frac{1}{2}\big[$, an analytical formula is found for the total absorption:

\begin{equation}
	A(\omega) = 1 - \frac{1}{1+\sinh^2\left(\frac{\pi\gamma}{\delta\omega}\right)\csc^2\left(\frac{\pi\omega}{\delta\omega}\right)}.
	\label{eq:infinite_CC}
\end{equation}

The average absorption is obtained by integration over a period $\delta\omega$:

\begin{equation}
    \langle A \rangle=\frac{1}{\delta\omega} \left[ \omega-\frac{\delta\omega}{\pi} \tanh\left(\frac{\pi\gamma}{\delta\omega}\right)\arctan
    \left(\frac{\tan(\frac{\pi\omega}{\delta\omega})}{\tanh(\frac{\pi\gamma}{\delta\omega})}\right) \right]_{\omega \to -\delta \omega/2}^{\omega \to +\delta \omega/2}.
\label{eq:A_av_integral}
\end{equation}

\noindent It results in a simple expression for the average absorption:

\begin{equation}
	\langle A \rangle_{cc}=\tanh\left(\frac{\pi\gamma}{\delta\omega}\right).
\end{equation}

\definecolor{shadecolor}{gray}{0.9}
\begin{shaded}
\textbf{At critical coupling, the average absorption $\langle A \rangle_{cc}$ depends only on an \emph{overlap factor} $\eta$ defined as the ratio between the resonance width $\Delta \omega = 2\gamma$ and the spectral spacing between two resonances $\delta\omega$}:

\begin{equation}
	\boxed{
	\eta=\frac{2\gamma}{\delta\omega}
	}
	\label{eq:OverlapFactor}
\end{equation}

\begin{equation}
	\boxed{
	\langle A \rangle_{cc} = \tanh\left(\frac{\pi\eta}{2}\right)
	}
	\label{eq:A_av_CC}
\end{equation}
\end{shaded}

%%%%%%%%%%%%%%%%%%%%
\subsection{General expression for the average absorption}

A similar derivation can be performed in a more general case, for any radiative and non-radiative decay rates:

\begin{equation}
	1-A_m(\omega)=\frac{1+\left(\frac{(\gamma_r-\gamma_{nr})/\delta\omega}{\omega/\delta\omega-m}\right)^2}{1+\left(\frac{(\gamma_r+\gamma_{nr})/\delta\omega}{\omega/\delta\omega-m}\right)^2},
 \label{eq:1-Am_gen}
\end{equation}

\begin{equation}
	\prod_{m=-\infty}^{+\infty}(1-A_m(\omega))=\frac{
\sin^2\left(\frac{\pi\omega}{\delta\omega}\right)+\sinh^2\left(\frac{\pi(\gamma_r-\gamma_{nr})}{\delta\omega}\right)
}{
\sin^2\left(\frac{\pi\omega}{\delta\omega}\right)+\sinh^2\left(\frac{\pi(\gamma_r+\gamma_{nr})}{\delta\omega}\right)
}.
\end{equation}

The average absorption becomes:

\begin{equation}
\langle A \rangle=\frac{1}{\delta\omega} \left[\omega-\frac{\delta\omega}{\pi} 
\frac{
\left[\sinh^2\left(\frac{\pi(\gamma_r+\gamma_{nr})}{\delta\omega}\right)-
\sinh^2\left(\frac{\pi(\gamma_r-\gamma_{nr})}{\delta\omega}\right)\right]
\arctan\left(\frac{\tan\left(\frac{\pi\omega}{\delta\omega}\right)}{\tanh\left(\frac{\pi(\gamma_r+\gamma_{nr})}{\delta\omega}\right)} \right)
}{
\cosh\left(\frac{\pi(\gamma_r+\gamma_{nr})}{\delta\omega}\right)
\sinh\left(\frac{\pi(\gamma_r+\gamma_{nr})}{\delta\omega}\right)
} \right]_{\omega \to -\delta \omega/2}^{\omega \to +\delta \omega/2},
\end{equation}

\noindent leading to

\begin{equation}
	\langle A \rangle =\frac{
\sinh^2\left(\frac{\pi(\gamma_r+\gamma_{nr})}{\delta\omega}\right)-\sinh^2\left(\frac{\pi(\gamma_r-\gamma_{nr})}{\delta\omega}\right)
}{
\cosh\left(\frac{\pi(\gamma_r+\gamma_{nr})}{\delta\omega}\right)
\sinh\left(\frac{\pi(\gamma_r+\gamma_{nr})}{\delta\omega}\right)
}.
\label{eq:1chanAbs}
\end{equation}

\definecolor{shadecolor}{gray}{0.9}
\begin{shaded}
\textbf{We introduce the \emph{radiative} and \emph{non-radiative overlap factors} $\eta_r$ and $\eta_{nr}$, respectively:}
\begin{equation}
	\boxed{
	\eta_r=\frac{2\gamma_r}{\delta\omega}
	}
	\hspace{2mm}
	\boxed{
	\eta_{nr}=\frac{2\gamma_{nr}}{\delta\omega}
	},
	\label{eq:OverlapFactorRNR}
\end{equation}

\noindent with $\eta = \eta_r + \eta_{nr}$ and we find a simple, general expression for the average absorption:
\begin{equation}
	\boxed{
	\langle A \rangle %(\eta_r,\eta_{nr})
	=\frac{2\sinh(\pi\eta_r)\sinh(\pi\eta_{nr})
}{
\sinh\left(\pi(\eta_r+\eta_{nr})\right)
}
	}.
	\label{eq:A_av_noCC}
\end{equation}
\end{shaded}

$\langle A \rangle$ is a function of $\eta_r$ and $\eta_{nr}$. At critical coupling ($\eta_r=\eta_{nr}$), equation (\ref{eq:A_av_noCC}) leads to the expression (\ref{eq:A_av_CC}).

%%%%%%%%%%%%%%%%%%%%
\subsection{Upper bound}
\label{ssec:UBOneChannel}

We now address the question of the maximum absorption that can be achieved by tuning the radiative decay rate.
$\eta_r$ depends on the coupling between the modes in the absorber and the free space, and it can be adjusted by design. 
In contrast, non-radiative dissipation depends on the intrinsic properties of the absorber: $\eta_{nr}$ is considered as a constant independent of the radiative coupling. 
Using equation \ref{eq:A_av_noCC}, the partial derivative of $\langle A \rangle$ with respect to $\eta_{r}$ yields

\begin{equation}
\frac{\partial \langle A \rangle}{\partial\eta_{r}}
= 2\pi \frac{\sinh^2(\pi \eta_{nr})}{\sinh^2\left(\pi(\eta_r+\eta_{nr})\right)},
\end{equation}

\noindent which is a positive function of $\eta_r$ over $[0,+\infty[$. The limit

\begin{equation}
\max\left(\langle A \rangle \right)_{\eta_{nr}}
=\lim_{\eta_r \to \infty} \left(\langle A \rangle \right)_{\eta_{nr}}
\end{equation}

\noindent can be approximated using $\sinh{x}\underset{x \to \infty}{\sim} e^x/2$.

\definecolor{shadecolor}{gray}{0.9}
\begin{shaded}
\textbf{The absorption maximum, or \emph{upper bound}, is reached when $\eta_r \to \infty$:}

\begin{equation}
\boxed{
\langle A \rangle_{ub}=1-e^{-2\pi\eta_{nr}}
}.
\label{eq:Abs_OC}
\end{equation}
\end{shaded}

The ratio between the upper-bound and the critical coupling absorption can be written
\begin{equation}
    \frac{\langle A \rangle_{ub}}{\langle A \rangle_{cc}} = 1+e^{-2\pi\eta_{nr}},
    \label{eq:A_UB_CC_ratio}
\end{equation}

\noindent and takes values between 1 (strong absorption, $\eta_{nr} \to \infty$) and 2 (weak absorption, $\eta_{nr} \to 0$).

%%%%%%%%%%%%%%%%%%%%
\subsection{Path enhancement factor}
\label{ssec:PathF}

We consider a slab of absorber of thickness $d$ made of a linear, homogeneous and isotropic material with absorption coefficient $\alpha$, which supports resonant modes.
By analogy with the relations derived for a pulse traveling in such a material at group velocity $\nu_g$, using Fourier transforms within the slow varying envelop approximation and neglecting the group velocity dispersion as a first-order approximation~\cite{Saleh2019}, the energy time decay $2\gamma_{nr}$ can be linked to the spatial attenuation of the wave intensity $\alpha$ through $2\gamma_{nr} = \alpha \nu_g$. With a spectral density of states (DOS) $\rho = 1 / \delta \omega$, the non-radiative overlap factor is simply expressed as:

\begin{equation}
\boxed{
\eta_{nr} = \alpha \nu_g \rho
}.
\label{eq:etanr_nu_DOS}
\end{equation}

In a bulk material, it is convenient to describe the attenuation of a plane wave with its effective path $d_\emph{eff}$. In case of resonances, a plane wave entering the absorber may travel an effective distance $d_\emph{eff}$ much longer than the actual slab thickness $d$. The path enhancement factor $F$ is defined as $d_\emph{eff}=F d$. $F=1$ for single-pass absorption. It is worth noting that $F$ has no physical meaning at the resonance frequency in case of critical coupling ($F \to +\infty$), but is a useful figure of merit for spectrally-averaged absorption enhancement.

\definecolor{shadecolor}{gray}{0.9}
\begin{shaded}
The exponential decay of absorption (the Beer-Lambert law in optics) can be re-written with this effective thickness $F d$ as:
\begin{equation}
\boxed{
	A = 1- e^{-F \alpha d}}.
\label{eq:EnhancFactor}
\end{equation}

The maximal path enhancement factor obtained for the absorption upper bound is deduced from Eqs.~(\ref{eq:Abs_OC}) and (\ref{eq:EnhancFactor}), and can also be expressed as a function of the group velocity $\nu_g$ and DOS $\rho$ using Eq.~\ref{eq:etanr_nu_DOS}:

\begin{equation}
\boxed{
F_{ub}=\frac{2 \pi \eta_{nr}}{\alpha d} = \frac{2\pi \nu_g \rho}{d}
}.
\label{eq:Fub}
\end{equation}
\end{shaded}

%%%%%%%%%%%%%%%%%%%%%%%%%%%%%%%%%%%%
\clearpage
\section{Discussion of model assumptions}
\label{sec:GeneralHypotheses}

In this section, we discuss in detail the different assumptions made in the previous section.

%%%%%%%%%%%%%%%%%%%%%%%
\subsection{Temporal coupled-mode theory}
\label{sec:TCMT}

The temporal coupled-mode theory (TCMT) relies on the following general assumptions~\cite{Joannopoulos2008}: the linearity and time-invariance of the system response, the conservation of energy, the reciprocity, and low losses. The last assumption is the more demanding. It is assumed that the rate of the energy loss is proportional to the energy stored in the system. In absence of input wave, the energy stored in the system decays exponentially with time, with a decay that should be slow with respect to the period: $\gamma \ll \omega/(2 \pi)$, or $Q \gg \pi$, with the quality factor $Q=\omega/(2 \gamma)$. In case of critical coupling, the condition translates into $\gamma_{nr} \ll \omega/(4 \pi)$. In practice, we are interested in weakly absorbing systems which fulfill this requirement. In highly-absorbent systems, this condition is not fulfilled but the average absorption is close to 1. The overcoupling regime is discussed in section~\ref{sec:Infinite_eta_r}.

%%%%%%%%%%%%%%%%%%%%%%%
\subsection{Weierstrass factorization (single radiative channel)}
\label{sec:Weierstass}

The reflection coefficient of a plane wave on the system, $r(\omega)$, can be decomposed using the Weierstrass factorization valid for meromorphic functions~\cite{Grigoriev2013}:
\begin{equation}
    r(\omega) = \mathbb{A} \cdot \exp(j\omega\mathbb{B}) \cdot \prod_m \frac{\omega-\omega_m^z}{\omega-\omega_m^p},
    \label{eq:Weierstrass_general}
\end{equation}
\noindent where $\mathbb{A}$ and $\mathbb{B}$ are two complex numbers, and $\omega_m^p$ and $\omega_m^z$ are the poles and zeros of the system, respectively.
Assuming absorption vanishes when $\omega \rightarrow 0$, $\mathbb{A}$ and $\mathbb{B}$ can be expressed as~\cite{Grigoriev2013}: 
\begin{align}
    \mathbb{A} &= \prod_m \frac{\omega_m^p}{\omega_m^z}\\
    j\mathbb{B} &= \sum_m \left( \frac{1}{\omega_m^z}-\frac{1}{\omega_m^p} \right)
\end{align}

To derive the optimal absorption in section~\ref{sec:MRAsinglechannel}, the expression of the poles and zeros of the multi-resonant system takes the same form as in the case of a single resonant mode $m$ (isolated modes). We note that the spectrum of $(\omega-\omega_m^z)/(\omega-\omega_m^p)$ is symmetrical with respect to $\omega_m$, and that this assumption is consistent with the requirement that $\omega_m^p$ and $\omega_m^z$ are complex conjugates when $\gamma_{nr} \rightarrow 0$ (time-reversal symmetry). For sufficiently high quality factors $(\omega_m \gg \gamma)$ or in the overcoupling regime ($\gamma_r \gg \gamma_{nr}$), $\omega_m^z \to \omega_m^p$ so $\left|\mathbb{A}\right| \rightarrow 1$ and $\mathbb{B} \rightarrow 0$ and the Weierstrass factorization \ref{eq:Weierstrass_general} leads to the simple pole-zero decomposition for $R(\omega)$ given in Equation~\ref{eq:R_tot}. It may be required to keep a prefactor in the pole-zero decomposition of $R(\omega)$ in some intermediate situations which go beyond the scope of this work.

%%%%%%%%%%%%%%%%%%%%%%%
\subsection{Slow variation approximation}
\label{sec:SlowVaryingEnvelop}

As stated in section~\ref{sec:MRAsinglechannel}, we assume that the system properties, i.e. the density of modes and the decay rates, or at least the overlap factor $\eta_r$ and $\eta_{nr}$, vary slowly over spectral 
ranges corresponding to the period $\delta\omega$ and the resonance width $\Delta\omega$.
It can be observed that for slow variations of the decay rate, the absorption spectrum becomes slightly asymmetric around the resonance frequency, but this does not affect the average absorption in first-order approximation.
This approximation will be further discussed and illustrated in Section~\ref{sec:validity}.

%%%%%%%%%%%%%%%%%%%%%%%
\subsection{Overcoupling regime}
\label{sec:Infinite_eta_r}

The upper bound for multi-resonant absorption is obtained for $\eta_{r} \to \infty$. In practice, large values of $\eta_r$ should be avoided in order to ensure a large quality factor ($Q \gg \pi$). Can we get close to the upper bound with finite, low enough radiative overlap factors?

\begin{figure}[ht]
\centering
\includegraphics[width=0.5\textwidth]{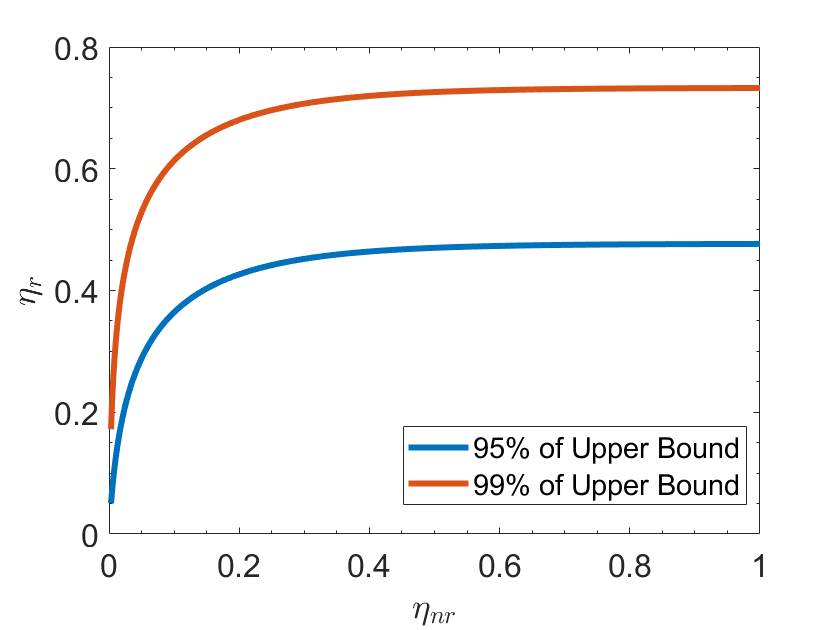}
\caption{Radiative overlap factor required for the average absorption (Eq.~\ref{eq:A_av_noCC}) to reach 95\% (blue) and 99\% (red) of the upper bound absorption (Eq.~\ref{eq:Abs_OC}), as a function of the non-radiative overlap factor.}
\label{fig:finite_eta_r}
\end{figure}

In Figure~\ref{fig:finite_eta_r}, we plot the radiative overlap factors $\eta_r$ required to reach an absorption equal to $95\%$ (blue) and $99\%$ (red) of the upper bound absorption, respectively, as a function of $\eta_{nr}$. The results show that limited values of $\eta_r$ are sufficient to get close to the upper bound, which appears as a reasonable approximation for optimized realistic systems.

%%%%%%%%%%%%%%%%%%%
\subsection{Decomposition over an infinite number of modes}
\label{sec:Infinite_number}

The pole-zero decomposition of the absorption spectrum has been extended to an infinite number of modes in order to derive analytical formulas. In particular, we question the addition of modes with negative frequencies. In the following, we show that only a small number of resonances play a role around a frequency of interest $\omega_0$, and that the infinite extension of the series of resonances has a negligible impact on the average absorption.

\subsubsection{Critical coupling}

\begin{figure}[h]
\centering
\includegraphics[width=0.9\textwidth]{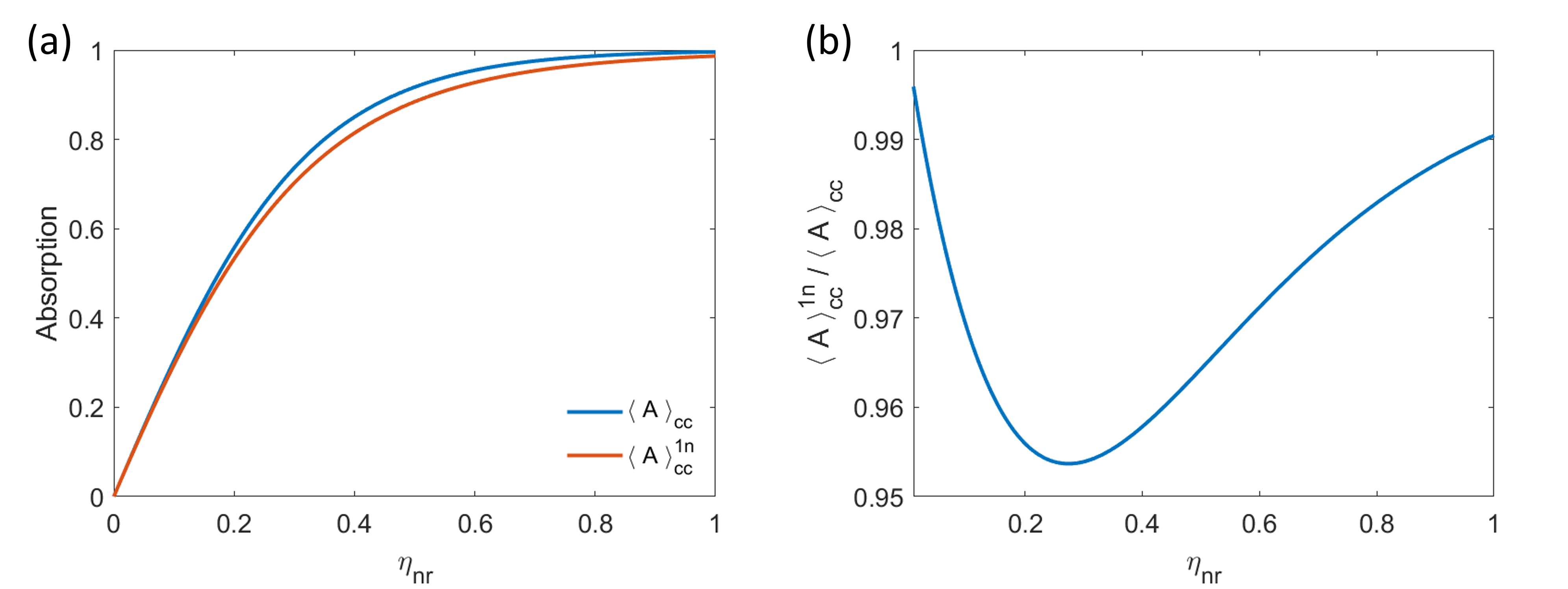}
\caption{(a) Average absorption in the critical coupling as a function of the non-radiative overlap factor $\eta_{nr}$, obtained by considering an infinite series of modes ($\langle A \rangle_{cc}$ in blue, Eq. \ref{eq:A_av_CC}), and a resonant mode with a single neighbor on each side ($\langle A \rangle_{cc}^{1n}$ in red). (b) Ratio between the two absorption models as a function of $\eta_{nr}$.}
\label{fig:one_neighbor_CC}
\end{figure}

We first illustrate this effect in the critical coupling regime.
In Figure \ref{fig:one_neighbor_CC}, we compare the average absorption $\langle A \rangle_{cc}$ derived with infinite series (Eq. \ref{eq:A_av_CC}) with the absorption $\langle A \rangle_{cc}^{1n}$ obtained by taking into account a single neighbor on each side of a resonance centered at $\omega_0$. In this way, $\langle A \rangle_{cc}^{1n}$ is the average, between $\omega_0 - \delta \omega/2$ and $\omega_0 + \delta \omega/2$, of the absorption spectrum formed by three resonances placed at $\omega_0 - \delta \omega$, $\omega_0$ and $\omega_0 + \delta \omega$.

As expected, $\langle A \rangle_{cc}^{1n}$ is lower than the absorption obtained with an infinite series $\langle A \rangle_{cc}$ (Figure \ref{fig:one_neighbor_CC}(a)). However, it is remarkable that the relative difference between the two models is less than $5\%$ for all overlap factors $\eta_{nr}$ (Figure \ref{fig:one_neighbor_CC}(b)). For small overlap factors, it is obvious that the neighboring resonances have little impact: $\langle A \rangle_{cc}^{1n} / \langle A \rangle_{cc} \rightarrow 1$ for $\eta_{nr} \rightarrow 0$. For large overlap factors, absorption is already close to one, so the relative impact of additional resonances is weak and we also have $\langle A \rangle_{cc}^{1n} / \langle A \rangle_{cc} \rightarrow 1$. The minimum is observed around $\eta_{nr}=0.25$. It corresponds to the transition between those two regimes, where the spacing between modes is about twice their width ($\eta=0.5$). Overall, only the closest resonances contribute significantly to the average absorption, and the extension to an infinite series provides a convenient and accurate analytical solution.

\subsubsection{Overcoupling}

We now turn to the absorption upper bound, demonstrating that it can be approached even when considering both a finite number of modes and a finite radiative overlap factor.

We consider the single-neighbor model described above, this time without assuming critical coupling ($\langle A \rangle_{ub}^{1n}$).
We also consider a zero-neighbor model accounting only for the average absorption, between $\omega_0 - \delta\omega/2$ and $\omega_0 + \delta\omega/2$, of a single resonance centered at $\omega = \omega_0$ ($\langle A \rangle_{ub}^{0n}$).

\begin{figure}[h]
\centering
\includegraphics[width=0.9\textwidth]{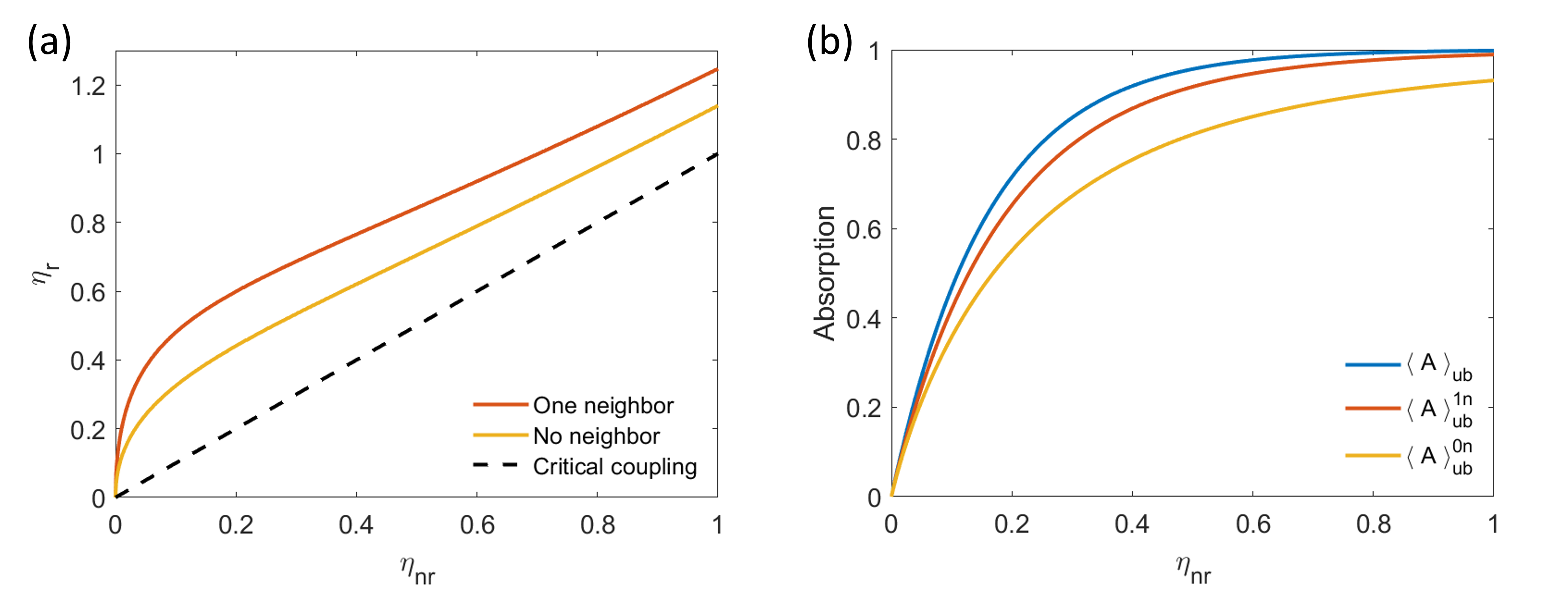}
\caption{(a) Optimal radiative overlap factor $\eta_r$ that maximizes the average absorption as a function of the non-radiative overlap factor $\eta_{nr}$, when considering only one mode (yellow line) or one mode and its first neighbors (red line). The critical coupling condition $\eta_r = \eta_{nr}$ (dashed black line) is shown as a reference.
(b) Average absorption as a function of the non-radiative overlap factor $\eta_{nr}$ for the upper bound (blue line), and for the one-neighbor (red line) and the zero-neighbor (yellow line) models.
}
\label{fig:one_neighbor_UB}
\end{figure}

When considering an infinite series of modes, the average absorption increases monotonically with $\eta_r$ and the upper bound $\langle A \rangle_{ub}$) is reached for $\eta_r \to \infty$. In contrast, for a finite number of neighbors, the optimal $\eta_r$ becomes finite and decreases as fewer neighbors are considered, see Fig.~\ref{fig:one_neighbor_UB}(a). Using these optimal values for $\eta_r$, the upper bounds are plotted in Fig.~\ref{fig:one_neighbor_UB}(b) for zero, one and an infinite number of neighbors. Similarly to the critical coupling scheme, it is found that the absorption of the one-neighbor model is already close to the upper bound modeled with an infinite series of modes and infinite radiative overlap factors.

\subsubsection{A criterion for the use a infinite series of modes}
\label{sec:criterion_modes}

We question the impact of the modes $m \leq 0$: they are non physical but they have been added to derive simple analytical formulas.

The total reflection is derived as the product of the reflections $R_m(\omega)$ (see Eq.~\ref{eq:R_tot}). At a given frequency $\omega$, the relative contribution of modes $m \leq 0$ is negligible if $\prod_{m=-\infty}^{0} R_m(\omega)$ is close to 1. If we consider the mode $M$ at a frequency $\omega = M \times \delta \omega$, the contribution of negative modes can be expressed by $\prod_{m=M}^{+\infty} R_m(0)$. If $\eta \ll 2M$, then for any $m \geq M$, $R_m(0)$ can be approximated by (Eq.~\ref{eq:1-Am_gen}):
\begin{equation}
	R_m(0) \simeq 1 - \left[1 - \delta^2 \right] \left( \frac{\eta}{2m} \right)^2.
	\label{eq:R_m_a}
\end{equation}
Using $1 - \delta^2 = 4 \eta_r \eta_{nr}/\eta^2$, we obtain:
\begin{equation}
	\prod_{m=M}^{+\infty} R_m(0) = \prod_{m=M}^{+\infty} \left[ 1 - \frac{\eta_r \eta_{nr}}{m^2} \right] = \frac{\Gamma(M)^2}{\Gamma(M-\sqrt{\eta_r \eta_{nr}}) \Gamma(M+\sqrt{\eta_r \eta_{nr}})},
	\label{eq:R_m_b}
\end{equation}
where $\Gamma$ is the gamma function. The contribution of negative modes increases with the product $\eta_r \eta_{nr}$, but observe that $\eta_r$ and $\eta_{nr}$ are smaller than 1 in most cases of interest (overlap factors close to unity are sufficient to reach nearly perfect absorption). Even in the worst-case scenario $\eta_r \eta_{nr} = 1$, the infinite product $\prod_{m=M}^{+\infty} R_m(0)$ converges quickly to 1 as $M$ increases, reaching 0.96 for $M=5$ and 0.99 for $M=10$ (relative impact of $1\%$), in accordance with the assumption $\eta \ll 2M$.

In conclusion, the addition of non-physical modes $m \leq 0$ provides a convenient way to derive simple analytical formulas for the average absorption. Their contribution is negligible in most practical cases, except for low frequencies close to the fundamental modes of the structures.

%%%%%%%%%%%%%%%%%%%%%%%
\subsection{Uniform distribution of the resonances}
\label{sec:Uniform_distrib}

We have assumed that the resonances are regularly spaced in frequency every $\delta \omega$. We can intuitively expect that this uniform mode distribution is the optimal scheme. In the following, we evaluate the impact of a non-uniform distribution.

\begin{figure}[h]
\centering
\includegraphics[width=0.5\textwidth]{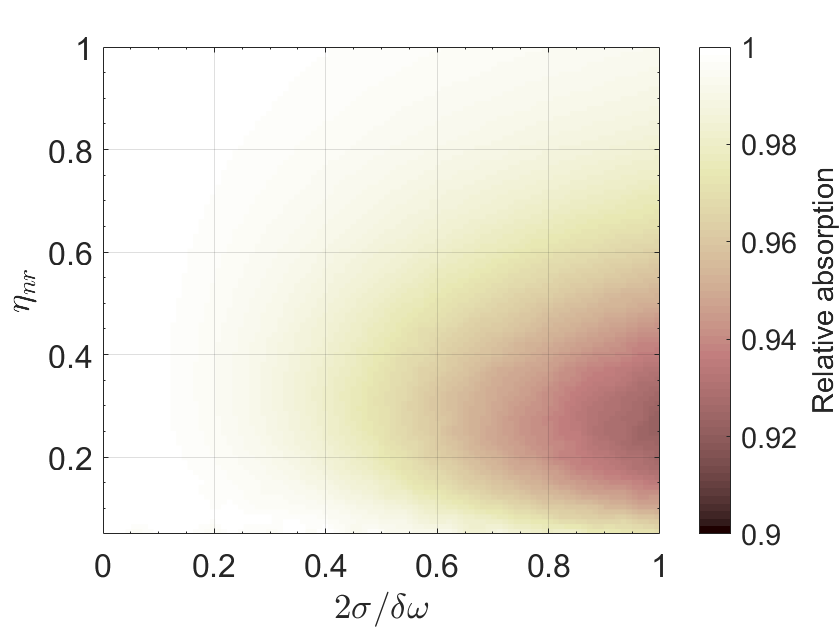}
\caption{Absorption as a function of $\eta_{nr}$ and the normalized standard deviation of the mode position $2\sigma / \delta \omega$, relative to the uniform distribution case. In this range, the absorption decreases by less than 8\% relative to the ideal case.}
\label{fig:normal_distrib_modes}
\end{figure}

We consider the critical coupling regime. We start with a uniform distribution of a large number of modes, every $\delta \omega$. Then, we create a new series of modes by shifting randomly the position of each resonance according to a normal law $N(0,\sigma^2)$. The standard deviation $\sigma$ is varied between 0 (uniform distribution) and $\delta \omega /2$ (corresponding to $68\%$ of the generated resonance peaks located between $-\delta \omega/2$ and $\delta \omega/2$).
The average absorption is calculated numerically for various $\eta_{nr}$ and $\sigma$, and the simulation is repeated many times to get statistically averaged results. The resulting absorption is normalized to the uniform case ($\sigma=0$) for each $\eta_{nr}$, and the result is shown in Figure \ref{fig:normal_distrib_modes}.

As expected, the absorption decreases for non-uniform distributions of modes. Still, in this range, even for the highest values of $\sigma = \delta \omega/2$, the absorption loss is no more than $8\%$ relative to the uniform case.
Therefore, the non uniformity of the mode distribution in real systems has a limited impact on the overall average absorption, confirming the relevance of the multi-resonant absorption model developed in this work.
We note that the highest relative efficiency drop is observed around $\eta_{nr}=0.25$, for the same reason as discussed in section~\ref{sec:Infinite_number}.

In the overcoupling regime, the impact of non-uniform distributions is still maximal around $\eta_r=0.5$.
In this regime, the resonance broadening reduces the impact of the mode position, and non-uniform distributions are even closer to the model.

%%%%%%%%%%%%%%%%%%%%%%%%%%%%%%%%%%%%
\clearpage
\section{General formalism for multi-resonant absorption}
\label{sec:GeneralFormalism}

In this section, we propose a more general model for multi-resonant absorption by taking into account several distinct radiative channels. This model extents the multi-resonant absorption model described in section~\ref{sec:MRAsinglechannel}, and it will be used to derive a general absorption upper bound valid for both scattering and diffraction phenomena.

%%%%%%%%%%%%%%%%%%%%
\subsection{Absorption by a single resonant mode}
\label{ssec:abssinglemode}

We consider an absorber slab with a back mirror and a periodical pattern as a resonant system coupled to $N$ input/output channels (diffracted plane waves). In contrast with section~\ref{sec:MRAsinglechannel}, the period is not restricted to the sub-wavelength range. We first assume a single resonance $m$, with a resonance width much smaller than the resonance frequency.

\textbf{Temporal coupled-mode theory.} Using the temporal coupled-mode theory~\cite{Fan2003,Collin2014}, the temporal variation of the mode amplitude is written as a function of the energy dissipation (total decay rate $\gamma$) and the input wave amplitudes $I_n$ (equation~\ref{eq:coupled-eq1}). The output wave amplitude $O_q$ in channel $q$ is the result of the direct coupling of input waves, and the radiative dissipation of the system (equation~\ref{eq:coupled-eq2}). The amplitudes are defined such that $| a |^2$ is the energy of the mode in the resonator, $| I_n |^2$ the input power in channel $n$ and $| O_q |^2$ is the output power in channel $q$:
\begin{align}
\frac{da}{dt} = \left( j \omega_m - \gamma \right) a + \sum\limits_{n=1}^{N} \alpha_n^i . I_n \label{eq:coupled-eq1}\\
O_q = \sum\limits_{n=1}^{N} C_{nq} . I_n + \alpha_q^o a \label{eq:coupled-eq2}
\end{align}
where $[C_{nq}]$ is the scattering matrix for the direct pathways between the incoming and outgoing waves, and $\alpha_n^i$ ($\alpha_n^o$) are the input (output) coupling coefficients of the mode for channel $n$. Reciprocity imposes that the input and output coupling coefficients are equal ($\alpha_n^i=\alpha_n^o=\alpha_n$) and constrains the direct scattering coefficients $C_{nq}$ and the coupling coefficients $\alpha_n$~\cite{Wang2018,Zhao2019}. Using the conservation of energy and the radiative decay rate through channel $n$, $\gamma_{r}^n = | \alpha_n |^2 / 2$, the total radiative decay rate of the resonant mode is $\gamma_{r} = \sum\limits_{n=1}^{N} \gamma_{r}^n$. The non-radiative decay rate (internal dissipation) is $\gamma_{nr} = \gamma - \gamma_{r}$. Note that with this notation, the decay rate of the \emph{energy} stored in the system is $2 \gamma$ (lifetime: $\tau /2$).

\textbf{Absorption of a plane wave.} The mode amplitude of the resonance $m$ (complex frequency $\omega_{m} + j \gamma$) excited by a single monochromatic input plane wave through channel $n$ (time-dependance $e^{j \omega t}$) is simply given by:
\begin{equation}
a (\omega)= \frac{\alpha_n I_n}{j (\omega - \omega_m)+ \gamma},
\label{eq:a}
\end{equation}
and the non-radiative loss power is $2 \gamma_{nr} | a |^2$.
For the resonant mode $m$, the frequency-dependent absorption of the wave intensity incident through channel $n$ can be written as:
\begin{equation}
A_{m}^n (\omega)= \frac{4 \gamma_{r}^n \gamma_{nr}}{(\omega - \omega_{m})^2 + \gamma^2}.
\label{eq:amn}
\end{equation}

In the following, we consider an incident plane wave through a single channel $n$.
The total reflection, sum of the normalized intensities scattered in all channels, can be decomposed over the pole and zero of the resonant mode:
\begin{equation}
    R_m(\omega) = 1 - A_m(\omega) =  \left| \frac{\omega - \omega_m^z}{\omega - \omega_m^p} \right|^2,
\label{eq:rmn}
\end{equation}
with $\omega_m^p = \omega_m + j \gamma$ and $\omega_m^z = \omega_m + j \left(\gamma^2 - 4 \gamma_r^n \gamma_{nr}\right)^{1/2}$. The labels $n$ are skipped in $R_m$ and $A_m$ for simplicity.

%%%%%%%%%%%%%%%%%%%%
\subsection{Multiple resonances}
\label{ssec:MultipleResonances}

\textbf{Phenomenological approach.}
In the case of multiple resonant modes, the phenomenological approach introduced in section~\ref{ssec:polezero} can be used to describe total reflection: the probability of reflection of a quantum of energy, i.e. a photon in the case of light waves, is the product of the independent probabilities of reflection of every mode $m$. It leads to the same decomposition over poles and zeros:
\definecolor{shadecolor}{gray}{0.9}
\begin{shaded}
\begin{equation}
	R(\omega) = \prod\limits_m R_m(\omega) = \prod\limits_m \left| \frac{\omega - \omega_m^z}{\omega - \omega_m^p} \right|^2.
	\label{eq:R_totmulti}
\end{equation}
\end{shaded}
A more rigorous derivation is proposed below.

\textbf{Scattering matrix.}
The scattering matrix relates the amplitude of the output waves $O_q$ as a function of the input waves. 
Considering a single input wave $I_n$, the diffraction coefficients are $r_q (\omega) = O_q/I_n$, and the total reflection can be decomposed as:
\begin{equation}
    R(\omega) = \sum_{q} \left| r_q(\omega) \right|^2 = \sum_{q} \left| \frac{O_q}{I_n} \right|^2.
    \label{eq:R_totdiffeff}
\end{equation}

Inserting Eq.~\ref{eq:a} into Eq.~\ref{eq:coupled-eq2} provides expressions for the elements of the scattering matrix in the case of a single resonance. In the case of multiple resonances, a general approach would need to take into account the coupling between the resonant modes.
Contrary to the theory developed in section~\ref{sec:MRAsinglechannel} for a single radiative channel, Eq.~\ref{eq:R_totmulti} can not be readily linked to a Weierstrass factorization of a reflection coefficient, since $R$ is not derived from a single component of the scattering matrix, but from the sum of the diffraction efficiencies (Eq.~\ref{eq:R_totdiffeff}).
In the following, we discuss the conditions that enable to derive a simple pole-zero decomposition of $R(\omega)$.

\textbf{Properties of $R(\omega)$.}
The diffraction coefficients $r_q$ all have the same poles $\omega_m^p$, which are the complex eigenfrequencies of the quasinormal modes~\cite{Ching1998,Lalanne2018}. They are the singularities of the scattering matrix and the poles of $R(\omega)$. In general, the zeros of the diffraction coefficients $r_q$ and of $R(\omega)$ are different from each other. A general theory of reflection-zeros ($R$-zeros) has been developed recently for arbitrary scattering structures~\cite{Sweeney2020,Krasnok2019} in the context of coherent perfect absorption (CPA) phenomena and parity-time ($PT$) symmetric structures~\cite{Chong2010,Chong2011}. It extends the Weierstrass factorization to the determinant of the scattering matrix and of the so-called generalized reflection matrix, which are analytically continued in the complex frequency plane.

In the case of lossless systems, time-reversal symmetry implies that the poles and zeros are complex conjugates. If absorption is progressively added in the system, the poles and zeros are no longer complex conjugates, but they are still expected to go by pairs. They can be tuned by design~\cite{Sweeney2020}, allowing for example reflectionless scattering modes (RSMs) which are $R$-zeros occurring on the real frequency axis (they are a generalization to multiple channel scattering of the perfect absorption case obtained in the critical coupling regime for systems with a single radiative channel). This framework opens a lot of possibilities which go beyond the scope of this article. Eq.~\ref{eq:R_totmulti} with $\omega_m^p$ and $\omega_m^z$ given in section~\ref{ssec:abssinglemode} further imposes that each zero has the same real part as the associated pole, consistently with the properties of lossless (or low-loss) systems. This condition results in the symmetric shape of $R_m(\omega)$ around each pole, resulting in a regular and optimal distribution of poles and zeros. The prefactors that appear in the Weierstrass factorization of meromorphic functions are discussed in section~\ref{sec:Weierstass}.

\definecolor{shadecolor}{gray}{0.9}
\begin{shaded}
\textbf{Angular independence.}
If we further impose that the reflectivity is independent from the incidence angle, the diffraction coefficients $r_q$ should be identical.
As for the specular reflection $r(\omega)$ in section~\ref{ssec:polezero}, the diffraction coefficients $r_q(\omega)$ are components of the scattering matrix and can be decomposed over the Weierstrass factorization. Therefore, it leads to a pole-zero decomposition of the total reflection $R(\omega)$, resulting in the form of Eq.~\ref{eq:R_totmulti}.
\end{shaded}

%%%%%%%%%%%%%%%%%%%%
\subsection{Average absorption}

In the following, we use the pole-zero decomposition of Eq.~\ref{eq:R_totmulti} and we proceed in a similar way as in section~\ref{sec:MRAsinglechannel}. The optimal absorption scheme is modeled by assuming a uniform spectral distribution of the resonant modes around a frequency of interest $\omega_0$, and the series of modes is extended over $]-\infty,+\infty[$: $\omega_{m}= \omega_0 + m \cdot \delta\omega$ with $m \in \mathbb{Z}$. $\delta\omega$ is the frequency spacing between modes. To simplify the expressions, we define the reduced frequency:
\begin{equation}
    x=\frac{\omega - \omega_0}{\delta \omega}.
\end{equation}

Using Eqs.~\ref{eq:amn} and \ref{eq:rmn}, we can write the reflection $R_{m}$ for an incident for channel $n$ as:

\begin{align}
    & R_{m} (x)= 1 - \frac{\eta_{r}^n \eta_{nr}}{(x - m )^2 + (\eta/2)^2} \label{eq:r1}\\
	& R_{m} (x) = \frac{1+\left(\frac{\eta \delta/2}{x-m}\right)^2}{1+\left(\frac{\eta/2}{x-m}\right)^2}
	\label{eq:r2}
\end{align}

\noindent with the radiative overlap factor for the input channel $n$

\begin{equation}
    \eta_{r}^n = \frac{2 \gamma_r^n}{\delta\omega}
\label{eq:etarn}
\end{equation}

\noindent and
\begin{equation}
	\boxed{
	\delta = \sqrt{1-4\frac{\eta_{nr} \eta_{r}^n}{\eta^2}}.
	}
\label{eq:delta}
\end{equation}

The total energy reflected by the system $R(x) = \prod_{m=-\infty}^{+\infty} R_m(x)$ can be derived using Eq.~\ref{eq:r2} and applying again (Eq.~\ref{eq:identity}) for both the numerator and denominator:

\begin{align}
	R(x) & = \frac{\sin^2(\pi x)+\sinh^2(\pi \eta \delta/2)}{\sin^2(\pi x)+\sinh^2(\pi \eta/2)}\\
	& = 1 - \frac{\sinh^2(\pi \eta/2)-\sinh^2(\pi \eta \delta/2)}{\sin^2(\pi x)+\sinh^2(\pi \eta/2)}.
\end{align}

It leads to a closed-form expression for the total absorption $A(x) = 1 - R(x)$:

\begin{equation}
	A (x) = \left(1-\frac{\sinh^2(\pi \eta \delta/2)}{\sinh^2(\pi \eta/2)}\right) \frac{1}{1+\sin^2(\pi x)/\sinh^2(\pi \eta/2)}.
\end{equation}

$A (x)$ is a periodical function of $x$ (period: 1), and the averaged absorption is calculated as:
\begin{equation}
	\langle A \rangle = \int_{-1/2}^{1/2} A(x) dx,
	\label{eq:Average_abs}
\end{equation}

\noindent We use:
\begin{equation}
    \int_{-1/2}^{1/2}  \frac{1}{1+c \sin^2(\pi x)} dx = \frac{1}{\sqrt{1+c}},
\end{equation}

\noindent valid for $c > -1$, resulting in
\begin{align}
		\langle A \rangle & = \left(1-\frac{\sinh^2(\pi \eta \delta/2)}{\sinh^2(\pi \eta/2)}\right) \frac{1}{\sqrt{1+1/\sinh^2(\pi \eta/2)}}\\
	& = \left(1-\frac{\sinh^2(\pi \eta \delta/2)}{\sinh^2(\pi \eta/2)}\right) \frac{\sinh(\pi \eta/2)}{\cosh(\pi \eta/2)}\\ 
	& = \frac{\sinh^2(\pi \eta/2) - \sinh^2(\pi \eta \delta/2)}{\sinh(\pi \eta/2) \cosh(\pi \eta/2)} \label{eq:Aavgen}
\end{align}

Eq.~(\ref{eq:Aavgen}) can be rearranged into a form similar to Eq.~(\ref{eq:A_av_noCC}):

\begin{equation}
	\boxed{
	\langle A \rangle=\frac{
2\sinh{\left(\frac{\pi\eta}{2}(1+\delta)\right)} \sinh{\left(\frac{\pi\eta}{2}(1-\delta)\right)}
}{
\sinh\left(\pi\eta\right)
}
	}
	\label{eq:An_any}
\end{equation}
or
\definecolor{shadecolor}{gray}{0.9}
\begin{shaded}
\begin{equation}
	\boxed{
	\langle A \rangle=\frac{
\cosh{(\pi\eta)} - \cosh{(\pi\eta\delta)}
}{
\sinh\left(\pi\eta\right)
}
}
	\label{eq:An_any_ch}
\end{equation}
\end{shaded}

In the case of a single radiative channel, $\delta=\frac{\eta_r-\eta_{nr}}{\eta}$, and we recover Eq. \ref{eq:A_av_noCC}.

%%%%%%%%%%%%%%%%%%%%
\subsection{Discussion and approximate formula}

The parameter $\delta$ represents the balance between the decay rates: it is equal to 0 at critical coupling $\eta_{nr}=\eta_{r}^n$, while $\delta \to 1$ is achieved as soon as any radiative or non-radiative decay rate dominates the total decay rate ($\eta_{nr}/\eta \to 0$ or $\eta_{r}^n/\eta \to 0$).

Using the asymptotic development $\sinh(x) \approx \frac{1}{2}\exp(x)$ for which the error is smaller than 10\% for $x\geq1.15$, we can write, as long as $\eta$ is not too small:

\definecolor{shadecolor}{gray}{0.9}
\begin{shaded}
\begin{equation}
\boxed{
\langle A \rangle \simeq 1 - e^{- \pi \eta (1 - \delta)}
}
\label{eq:Abs_n_approx}
\end{equation}
\end{shaded}

In practice, Eq.~\ref{eq:Abs_n_approx} is a good approximation of Eq.~\ref{eq:An_any_ch} when $\eta \geq 0.5$ and $\delta \geq 0.5$. Actually, equation (\ref{eq:Abs_n_approx}) is even an excellent approximation of (\ref{eq:An_any_ch}) for any $\eta$ when $\delta \to 1$, with a relative error $\epsilon \leq 5\%$ for $\delta \geq 0.9$ and $\epsilon \leq 2.5\%$ for $\delta \geq 0.95$.

%%%%%%%%%%%%%%%%%%%%
\subsection{Upper bound absorption}

As a reminder, the total overlap factor is decomposed into radiative and non radiative contributions $\eta = \eta_r + \eta_{nr}$, and the radiative overlap factor is the sum over the $N$ diffraction channels: $\eta_r = \sum\limits_{i=1}^{N} \eta_{r}^i$.
In the following, we investigate how the average absorption $\langle A \rangle$ can be maximized by tuning the radiative decay rates.

\subsubsection{Derivative of the absorption as a function of the radiative decay rates}

We calculate the partial derivative of $\langle A \rangle$ with respect to the radiative overlap factor of channel $i$, which is not necessarily the incident channel $i=n$ (the non radiative overlap factor $\eta_{nr}$ is kept constant):
\begin{equation}
		\frac{\partial \langle A \rangle}{\partial \eta_r^i} =
		\pi \alpha \frac{-1+\cosh(\pi\eta)\cosh(\pi\eta\delta)-\frac{\beta}{\alpha}\sinh(\pi\eta)\sinh(\pi\eta\delta)}{\sinh^2(\pi\eta)},
\end{equation}

\noindent which can be rewritten
\begin{equation}
		\frac{\partial \langle A \rangle}{\partial \eta_r^i} =
		\frac{\pi \alpha}{\sinh^2(\pi\eta)}
		\left( -1 + \frac{1}{2}
	\left[ \left(1-\frac{\beta}{\alpha} \right) \cosh \big(\pi\eta(1+\delta)\big) 
	+ \left(1+\frac{\beta}{\alpha} \right) \cosh\big(\pi\eta(1-\delta)\big)
	\right] \right),
	\label{eq:dAeta}
\end{equation}

\noindent with
\begin{align}
	\alpha &= \frac{\partial \eta}{\partial \eta_r^i}=\frac{\partial \eta_r}{\partial \eta_r^i}\\
	\beta &= \frac{\partial (\eta\delta)}{\partial \eta_r^i} = \frac{\eta \alpha - 2 \eta_{nr} \epsilon}{\eta\delta}\\
	\epsilon &= \frac{\partial \eta_r^n}{\partial \eta_r^i}	
\end{align}

\subsubsection{Impact of free-space radiative losses ($\epsilon = 0$)}

If we vary the radiative coupling of channels $i \neq n$ independently from the input channel $i=n$ ($\eta_r^n$ is constant), we have $\epsilon = 0$ and $\beta/\alpha = 1/\delta$. The derivative of $\langle A \rangle$ simplifies into:

\begin{align}
		\frac{\partial \langle A \rangle}{\partial \eta_r^i} = 
		\frac{\pi \alpha}{\sinh^2(\pi\eta)} \Omega
\end{align}

\noindent with

\begin{equation}
    \Omega = -1+\frac{1}{2 \delta}\left[ (\delta-1)\cosh(\pi\eta (1+\delta)) + (\delta+1)\cosh(\pi\eta (1-\delta)) \right]
\end{equation}

We verify that $\Omega$ is always negative, since $\Omega(\eta = 0) = 0$ and $\frac{\partial\Omega}{\partial \eta} \leq 0$.
Therefore, $\frac{\partial \langle A \rangle}{\partial \eta_r^i}$ has the opposite sign of $\alpha$ for all $\eta > 0$.
This translates into the following intuitive result: \textbf{the absorption increases as we decrease the radiative decay rate of channels which act only as outputs, i.e. free-space losses}.

 As an illustration, diffraction losses can be reduced by the use of an angular filter. Reflecting or cancelling all radiative channels except the incident channel $i=n$ brings us back to the single radiative channel case (see Section~\ref{ssec:UBF}).

\definecolor{shadecolor}{gray}{0.9}
\begin{shaded}
If the radiative coupling rates can be tuned independently, the average absorption $\langle A \rangle$ increases with the radiative coupling of incident waves $\eta_r^n$ (overcoupling regime), and it is maximized when the coupling rate to other channels $i \neq n$ is reduced. \textbf{The upper bound calculated for the single radiative channel case (Section \ref{sec:MRAsinglechannel}) is therefore the overall absorption maximum.}
\end{shaded}

\subsubsection{Impact of the radiative coupling of incident waves ($\epsilon = 1$)}

With $\alpha \geq 0$ and $\forall x$, $\cosh(x)\geq 1$, a sufficient condition for $\frac{\partial \langle A \rangle}{\partial \eta_r^i} \geq 0$ is
\begin{equation}
    \left(\frac{\beta}{\alpha}\right)^2\leq 1.
\end{equation}

We have:
\begin{equation}
    \left(\frac{\beta}{\alpha}\right)^2 = 1 - \frac{4 \eta_{nr}}{\delta^2 \eta^2} \left[ \eta_{nr} \frac{\epsilon}{\alpha} \left(1 - \frac{\epsilon}{\alpha} \right) + \eta_r \frac{\epsilon}{\alpha} - \eta_r^n \right],
\end{equation}

\noindent such that the sign of the partial derivative of $\langle A \rangle$ (Eq.~\ref{eq:dAeta}) follows the sign of $\sigma_d$:

\begin{equation}
    \sigma_d = \eta_{nr} \frac{\epsilon}{\alpha} \left(1 - \frac{\epsilon}{\alpha} \right) + \eta_r \frac{\epsilon}{\alpha} - \eta_r^n.
\end{equation}

If we vary the incident radiative coupling $i = n$ independently from other coupling rates $\eta_r^{i \neq n}$, we have $\alpha = 1$ and $\epsilon = 1$, leading to $\sigma_d = \eta_r - \eta_r^n \geq 0$: $\langle A \rangle$ increases with the incident radiative overlap factor $\eta_r^n$.
This is also true if we assume all $N$ radiative channels have the same coupling rates: $\epsilon = 1$, $\alpha = N$, and the derivative $\frac{\partial \langle A \rangle}{\partial \eta_r^i}$ is always positive for $N \geq 1$.
More generally, $\sigma_d \geq 0$ when $1 \leq \alpha \leq \frac{\eta_r}{\eta_r^n}$, which includes the previous cases.

\definecolor{shadecolor}{gray}{0.9}
\begin{shaded}
\textbf{If all radiative channels have the same coupling rates $\eta_r^n$, as it is the case when the absorption is independent from the incidence angle, the absorption upper bound is achieved in the overcoupling limit: $\eta_r^n \to \infty$.}
\end{shaded}

\subsubsection{Analytical expression of the absorption upper bound}

In the overcoupling limit, $\eta_r^n \gg \eta_{nr}$ and a first-order approximation of $\delta$ leads to:
\begin{equation}
	\delta \approx 1- \frac{2\eta_r^n \eta_{nr}}{\eta^2}.
	\label{eq:delta_approx}
\end{equation}

The upper bound absorption is written:
\begin{equation}
    \langle A \rangle_{ub} = 1 - \exp{\left(-2\pi \frac{ \eta_r^n \eta_{nr}}{\eta} \right)}
\end{equation}

When the coupling rates of the $N$ radiative channels are equal, the expression is simplified into:

\begin{equation}
    \langle A \rangle_{ub} = 1 - \exp{\left(-\frac{2\pi  \eta_{nr}}{N} \right)}
    \label{eq:upper_bound_N_identical_channels}
\end{equation}

\subsubsection{Absorption upper bound for optical systems}

Similarly to section~\ref{ssec:PathF}, we can define the maximal path enhancement factor $F_{ub}$ in the case of multiple radiative channels:
\begin{align}
    \langle A \rangle_{ub} &= 1 - \exp{\left(- F_{ub} \alpha d \right)}\\
    F_{ub} &= \frac{2\pi \eta_{nr}}{\alpha d} \frac{\eta_r^n}{\eta}.
\end{align}

\definecolor{shadecolor}{gray}{0.9}
\begin{shaded}
For $N$ identical radiative coupling rates:
\begin{equation}
    F_{ub} = \frac{2\pi \eta_{nr}}{N \alpha d}.
    \label{eq:F_ub_N_channels}
\end{equation}
\end{shaded}

%%%%%%%%%%%%%%%%%%%%%%%%%%%%%%%%%%%%
\clearpage
\section{Application to broadband light absorption with sub-wavelength patterning}
\label{sec:MultiresonantAbsLight}

In this section, we apply the previous derivation of the optimal multi-resonant absorption to electromagnetic plane waves. We are particularly interested in the broadband absorption of sunlight in a thin slab of semiconductor for photovoltaic applications, although the following arguments could be applied to other applications involving any broadband light sources. The coupling of incident light with a large density of modes in the system is achieved with a two-dimensional (2D) diffraction grating. For the sake of simplicity, we assume that the incident light impinges the absorber slab at normal incidence, but the generalization to off-normal illumination is straightforward.
In the following, we consider short-period lattices to ensure that no diffracted wave propagates in free space. We express the optical path enhancement factor as a function of the optical properties of the absorber (refractive index) and the geometry of the pattern. We discuss the impact of both the lattice arrangement and the symmetries, and we derive the upper bound for multi-resonant absorption of light.

\subsection{Light path enhancement}

The formula obtained for the multi-resonant absorption upper bound with a single radiative channel (Eq.~\ref{eq:Abs_OC}) can be expressed, in the context of light absorption, as a generalized form of the Beer-Lambert law (see section~\ref{ssec:PathF}),

\begin{equation}
    \langle A \rangle_{ub} = 1 - \exp (-F \alpha d),
\end{equation}

\noindent where $d$ is the thickness of the absorber slab and $\alpha$ the absorption coefficient of the material. For a complex refractive index $n+i\kappa$, $\alpha = 2 \kappa \omega / c = 4 \pi \kappa/\lambda$, with $c$ the speed of light in vacuum and $\lambda$ the wavelength.
The light path enhancement factor $F$ is expressed as (Eq.~\ref{eq:Fub})
\begin{equation}
    F=\frac{2\pi \eta_{nr}}{\alpha d}.
\end{equation}

We can also express the critical coupling limit as a function of $F$:

\begin{equation}
    \langle A \rangle_{cc}=\tanh\left(\frac{F \alpha d}{2} \right)
\end{equation}

Using the spectral density of modes coupled to free-space $\rho = 1/\delta\omega$, and $\eta_{nr} = 2 \gamma_{nr} \rho$ (Eq.~\ref{eq:OverlapFactorRNR}), $F$ is written

\begin{equation}
\boxed{
    F=\frac{4\pi \gamma_{nr} \rho}{\alpha d}.}
    \label{F_gamma_rho_N}
\end{equation}

In the following, we derive $\gamma_{nr}$ and $\rho$ according to the material properties and the geometry of the patterned slab, and we investigate the conditions to maximize $F$ and the overall absorption of sunlight.

%%%%%%%%%%%
\subsection{Non-radiative decay rate}

The non-radiative decay rate $\gamma_{nr}$ (in $\mathrm{s^{-1}}$) corresponds to the temporal decay rate of the mode amplitude due to absorption in the material. Let us describe the mode as a resonant plane wave. Its intensity decays temporally with a rate $2 \gamma_{nr}$, and spatially with a rate $\alpha$.
The ratio between both decay rates is the group velocity $c/n$, leading to:
\begin{equation}
\gamma_{nr}=\frac{\alpha c}{2 n}=\frac{\kappa}{n}\omega
\label{eq:gamma_nr}
\end{equation}

In a thin absorber, resonant modes can extend partially outside the slab, which may lower the non-radiative decay rate.
This effect can be accounted for by considering a confinement factor $\Gamma$, such that:
\begin{equation}
\gamma_{nr}=\Gamma \frac{\alpha c}{2 n}
\end{equation}

In the following, we assume $\Gamma=1$. We can rewrite $F$ (Eq.~\ref{F_gamma_rho_N}) as
\begin{equation}
\boxed{
    F=\frac{2\pi c \rho}{n d}.}
    \label{eq:F_rho_1}
\end{equation}

We emphasize that $F$ is independent of alpha. For a given absorber (real part of the refractive index $n$ and thickness $d$), maximizing $F$ is equivalent to maximizing the density of modes. It will done by optimizing the periodical pattern while ensuring no diffraction channel.

%%%%%%%%%%%%%%%%%%%%%%%%%%5
\subsection{Spectral density of states}

\subsubsection{Wave vectors coupled to free space}

A plane wave of frequency $\omega$ inside the medium can couple to free space if its in-plane wave vector is smaller than $k_0 = \omega/c$. In the reciprocal space, it should lie in a disk of area:
\begin{align}
    S_{rad} = \pi k_0^2.
\end{align}

\subsubsection{Surface and volume of a mode in the reciprocal space}

We consider a periodic pattern that diffracts light inside the slab. Its reciprocal lattice is described by two primitive vectors $\mathbf{b_1}$ and $\mathbf{b_2}$, where $\mathbf{b_1}$ links the two closest lattice points. The mode surface is given by (Fig.~\ref{fig:optimal_grating_compacity}(a)):

\begin{equation}
    S_{mode} = |\mathbf{b_1} \wedge \mathbf{b_2}|.
\end{equation}

Assuming a periodical (or perfect reflection) boundary condition at the top and bottom interfaces of the slab of thickness $d$, the $z$-components of the mode wavevectors are multiples of $2\pi/d$. Overall, the modes lie at the nodes of a three-dimensional reciprocal lattice, with a unit cell volume
\begin{equation}
    V_{mode} = \frac{2 \pi}{d} S_{mode}.
\end{equation}

\subsubsection{Density of states}

Considering the frequency range $\big[\omega,\omega+d\omega\big[$, the resonant modes are in a three-dimensional shell of radius $k=\frac{n \omega}{c}$ and thickness $dk=\frac{n }{c}d\omega$. Its volume $V_{shell} = 4 \pi k^2 dk$ is expressed
\begin{align}
	V_{shell} &= 4 \pi \frac{{n}^3 \omega^2}{c^3} d\omega \\
	          &= \frac{4 n^3}{c} S_{rad}  d\omega.
\end{align}

If the dimensions of the mode volume are much smaller than the shell radius, the total spectral density of modes in the system is given by
\begin{equation}
    \rho_t(\omega) = 2 \frac{V_{shell}}{V_{mode}} \frac{1}{d\omega}
\end{equation}

\begin{equation}
    \boxed{
    \rho_t(\omega) = \frac{4 n^3 d}{\pi c} \frac{S_{rad}}{S_{mode}}}
\label{eq:rhoSradSmode}
\end{equation}
\noindent where the factor 2 accounts for the two polarizations of light.
In the case of a single radiative channel, we will see that the actual density of modes $\rho$ coupled to a linearly polarized plane wave is limited by additional constrains ($\rho \leq \rho_t$).

%%%%%%%%%%%
\subsection{Light polarization and pattern symmetries}

Sunlight is an incoherent source that can be decomposed as a superposition of independent plane waves linearly polarized in two orthogonal directions, with the same intensity. These two sets of plane waves impinge the absorber at normal incidence and can be treated separately. We are looking for the optimal structure that maximizes the light path enhancement factor $F$ for both polarization states.
In the following, we assume no diffraction channels and we discuss the impact of the symmetries on the density of modes.

%%%%%%%%%%%
\subsubsection{Symmetries and radiative channels}

We investigate here the additional constrains on the pattern symmetries to ensure a single radiative channel.

Using the Jones matrix formalism, we write the frequency-dependent reflection matrix of the system in the basis formed by two orthogonal linear polarizations in the directions $x$ and $y$ as:

\begin{equation}
	r=
	\begin{pmatrix}
		r_{xx} & r_{xy} \\
		r_{yx} & r_{yy}
	\end{pmatrix}
	\label{eq:Jones}
\end{equation}
\noindent where the matrix elements are complex reflection amplitudes.
In the general case, polarization rotation is induced by the anti-diagonal terms $r_{xy}$ and $r_{yx}$: an incident plane wave linearly polarized along the $x$ direction results in two radiative channels.

To restrict radiative losses to a single channel, polarization rotation needs to be suppressed. To fulfill this condition and to comply with the broadband nature of sunlight, there should exist a basis of orthogonal polarization eigenvectors independent of the frequency, in which the Jones matrix takes the form
\begin{equation}
	r=
	\begin{pmatrix}
		A & 0 \\
		0 & D
	\end{pmatrix}
\end{equation}
\noindent where $A$ and $D$ are the eigenvalues of $r$.

Menzel and al. published a complete classification of the Jones matrices of material slabs with two-dimensional periodical patterns, according to their symmetry properties~\cite{menzel_advanced_2010}. A plane of symmetry perpendicular to the slab plane is the simplest way to ensure a diagonal form of the Jones matrix independently of the frequency. The eigenvectors are two linear polarization states parallel and orthogonal to the mirror plane. We note that the same optical reflectivity is obtained for incident plane waves linearly polarized at $+45^{\degree}$ and $-45^{\degree}$ with respect to the mirror plane.

\subsubsection{Symmetries and spectral density of optical modes}

The optical modes are defined by their position in the reciprocal lattice. The plane of symmetry contains points of the reciprocal lattice, and the corresponding modes are coupled to a single polarization eigenstate. Their contribution becomes negligible for a large number of modes. The modes outside the plane of symmetry are degenerate of degree 2.
Coupling between degenerate modes removes degeneracy and results in two coupled modes, a symmetric mode and an anti-symmetric mode with respect to the mirror plane. Due to their symmetry, each mode couples to a single polarization state parallel or perpendicular to the mirror plane. Finally, the spectral density of modes available for each polarization state is the total density of modes $\rho_t$ divided by a factor of two.

\definecolor{shadecolor}{gray}{0.9}
\begin{shaded}
\begin{equation}
	\rho = \frac{\rho_{t}}{2}
    \label{eq:rhoeff}
\end{equation}

\textbf{To ensure a single radiative channel, each optical mode should be coupled to a single polarization eigenstate. For optimal absorption of sunlight, the total density of modes must be split evenly between both polarizations.
It can be achieved with a single mirror plane perpendicular to the slab.}
\end{shaded}

\begin{figure}[h]
\centering
\includegraphics[width=\textwidth]{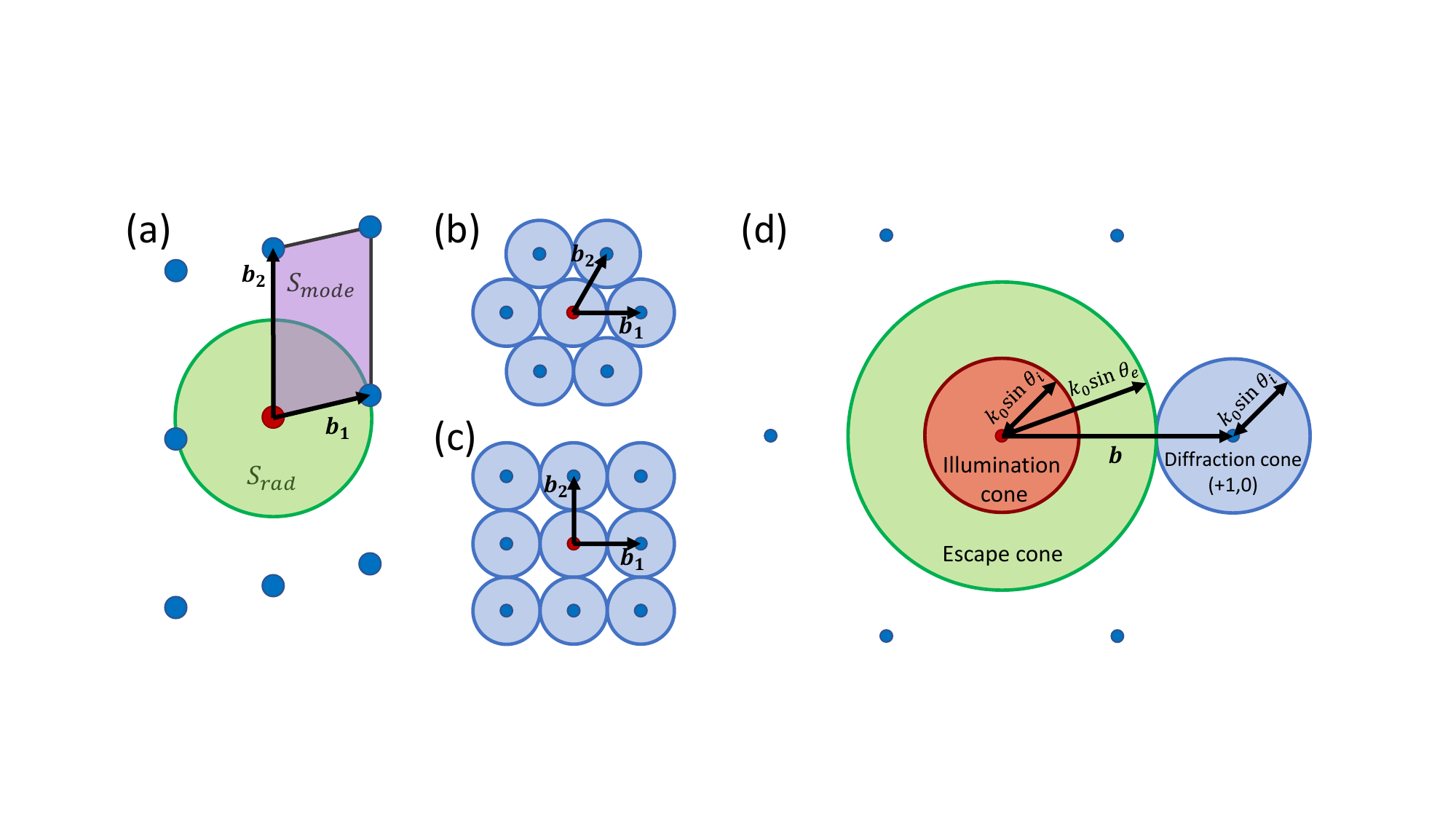}
\caption{
Representations in the reciprocal space of the incident (red dot) and diffracted (blue dots) wavevectors, for a two-dimensional periodic pattern and normal incidence.
(a) In the general case, the reciprocal lattice is characterized by two wavevectors $\mathbf{b_1}$ and $\mathbf{b_2}$ (with $\mathbf{b_1}$ connecting the closest lattice points) which define the surface occupied by a mode $S_{mode} = ||\mathbf{b_1} \wedge \mathbf{b_2}||$. To ensure a single radiative channel per polarization state, the diffracted wavevectors (blue dots) should be outside of $S_{rad}$ (here we assume no angle restriction: $\theta_e = 90^{\circ}$). The sketch illustrates the limiting case $||\mathbf{b_1}|| = k_0$.
(b,c) Illustration of circle packing in an hexagonal (b) and a square (c) arrangement. The hexagonal lattice provides the highest packing density, and thus the best solution to maximize the ratio $S_{rad}/S_{mode}$ and the $F$ factor (see the text).
(d) Sketch of an hexagonal reciprocal lattice (unit vector $b$) with angular restrictions for both illumination and photon escape. The orange disk represents the \emph{illumination cone} of possible incident wavevectors (half-angle $\theta_i$). Diffraction results in \emph{diffraction cones}, depicted by translating the illumination disk onto each lattice point (the blue disk shows the diffraction cone of order (+1,0)). Light escape can be restricted by an angular filter, defining an \emph{escape cone} of half-angle $\theta_e \geq \theta_i$ (green disk). The sketch shows the limiting case that ensures no overlap between the blue and green circles, i.e. no diffraction in free space and a single radiative channel for each polarization state: $||\mathbf{b_1}|| = k_0 (\sin \theta_e + \sin \theta_i)$.
}
\label{fig:optimal_grating_compacity}
\end{figure}

\subsection{Upper bound for the light path enhancement factor}
\label{ssec:UBF}

In the case of a single radiative channel enabled by a single mirror symmetry, the light path enhancement is obtained by combining Eqs.~\ref{eq:F_rho_1}, \ref{eq:rhoSradSmode} and \ref{eq:rhoeff}:

\begin{equation}
    F = 4 n^2 \frac{S_{rad}}{S_{mode}}.
    \label{eq:F_Srad_N_Smode}
\end{equation}

In the following, we investigate the impact of the periodical arrangement on the light path enhancement factor $F$, and we derive its upper bound.

\subsubsection{Optimal lattice: hexagonal arrangement}

The highest light path enhancement factor is obtained by maximizing the ratio $S_{rad}/S_{mode}$. For a given frequency $\omega$, the upper limit is reached for the smallest unit cell area $S_{mode}$, constrained by the condition to avoid diffraction in free space (the in-plane component of diffracted wavevectors must be larger that $k_0$):

\begin{equation}
    ||\mathbf{b_1}|| \geq k_0.
    \label{eq:k0max}
\end{equation}

This optimization problem is illustrated in Fig.\ref{fig:optimal_grating_compacity}.

\definecolor{shadecolor}{gray}{0.9}
\begin{shaded}
The maximization of $S_{rad}/S_{mode}$ is equivalent to the well-known problem of circle packing in the plane. The solution providing the highest density, and therefore the highest $F$ (Eq.~\ref{eq:F_Srad_N_Smode}), is the hexagonal arrangement (Fig.\ref{fig:optimal_grating_compacity}(b)).
\end{shaded}

\subsubsection{Hexagonal lattice}

The hexagonal reciprocal lattice is defined by a period
\begin{equation}
    b = \frac{4 \pi}{p \sqrt{3}}
\end{equation}
\noindent and an area
\begin{equation}
S_{mode} = \frac{\sqrt{3}}{2} b^2,
\end{equation}

\noindent leading to a light path enhancement

\begin{equation}
    \boxed{F_{hex} = 2 \sqrt{3} \pi n^2 \left( \frac{p}{\lambda} \right)^2}.
    \label{eq:Fub-hex-p}
\end{equation}

\definecolor{shadecolor}{gray}{0.9}
\begin{shaded}
The maximal $F$ factor that fulfills the condition~\ref{eq:k0max} is obtained when $b \rightarrow k_0$ ($p \rightarrow 2 \lambda / \sqrt{3} $):
\begin{equation}
    \boxed{F_{ub} = \frac{8 \pi}{\sqrt{3}} n^2}
    \label{Fub}
\end{equation}

\noindent and the upper-bound absorption takes the form (see Eq.~\ref{eq:EnhancFactor}):

\begin{equation}
    \boxed{\langle A \rangle_{ub} = 1-\exp(-F_{ub} \ \alpha d)}.
    \label{eq:A_ub}
\end{equation}
\end{shaded}

\subsubsection{Square lattice}

The important case of a square pattern is also derived straightforwardly: $S_{mode} = (2 \pi / p)^2$, 

\begin{equation}
    F_{sq} = 4 \pi n^2 \left( \frac{p}{\lambda} \right)^2.
    \label{eq:Fsquare-p}
\end{equation}

\noindent and the maximal $F$ is obtained for $p = \lambda$:
\begin{equation}
    F_{sq-ub} = 4 \pi n^2.
    \label{eq:Fsquaremax}
\end{equation}

\subsection{General condition for oblique illumination and a restricted escape cone}

In the following, we extend the previous derivations of the upper bounds (\ref{Fub} and \ref{eq:Fsquaremax}) obtained for normal incidence to the general cases of oblique illumination and angular restriction of the escape cone.

Oblique illumination further constrains the period to avoid diffraction in free space. We first define the \emph{illumination cone} of possible incident wavevectors (half-angle $\theta_i$), represented by a disk of radius $k_0 \sin \theta_i$ in the phase space (Fig.~\ref{fig:optimal_grating_compacity}(d)). Diffraction occurs in \emph{diffraction cones} represented by disks of the same radius, centered at each point of the reciprocal lattice.

By adding an angular filter above the absorber, it is possible to define a restricted \emph{escape cone} (half-angle $\theta_e$) and to cancel free-space losses through diffraction channels, allowing for larger periods and $F$ factors. The escape cone is represented by a disk of radius $k_0 \sin \theta_e$ in the reciprocal space. The single radiative channel condition is fulfilled if there is no overlap between the escape cone and the diffraction cones, see Fig.~\ref{fig:optimal_grating_compacity}(d).
The condition to avoid diffraction in free-space is expressed as

\begin{equation}
    ||\mathbf{b_1}|| \geq k_0 (\sin \theta_e + \sin \theta_i).
\end{equation}

\definecolor{shadecolor}{gray}{0.9}
\begin{shaded}
Considering an illumination cone of half-angle $\theta_i$ and an escape cone of half-angle $\theta_e$, the maximum light path enhancement is

\begin{equation}
\boxed{
    F(\theta_i,\theta_e) = \frac{F}{[\sin(\theta_e)+\sin(\theta_i)]^2}.
}
\end{equation}
\end{shaded}

\subsubsection{Angular restriction of the escape cone with normal-incidence illumination $(\theta_i=0^{\circ})$}

In the case of angular restriction of the escape cone defined by the half-angle $\theta_e$ and normal-incidence light, the optical path enhancement factor is given by:
\begin{equation}
    F(\theta_e) = \frac{F}{\sin^2{\theta_e}}.
    \label{eq:F-omega}
\end{equation}
The same angular dependence is found in the case of optical scattering, see section~\ref{sec:OpticalScattering}.

\subsubsection{Oblique incidence without angular restriction ($\theta_e = 90^\circ$)}

If we consider an illumination cone of half-angle $\theta_i$ and no angular restriction for light escape ($\theta_e = 90^\circ$), the condition to ensure a single radiative channel for any $\theta \leq \theta_i$ further constrains the period and leads to an optical path enhancement factor
\begin{equation}
    F(\theta_i) = \frac{F}{[1+\sin(\theta_i)]^2}.
\end{equation}

To avoid diffraction for any angle of incidence, the upper bound is divided by a factor 4 as compared to the reference case of normal incidence (Eq.~\ref{Fub}): $F = \frac{2 \pi}{\sqrt{3}} n^2 \simeq 3.63 n^2$, which is smaller than the optical path enhancement factor of isotropic scattering $4 n^2$ (see section~\ref{sec:lit}).

It is worth noting that the upper bound obtained with an hexagonal lattice is higher than the scattering limit for illumination half-angles $\theta_i$ lower than $65^{\circ}$.

\subsection{Impact of symmetries}

In the literature, most periodic patterns used for light trapping are made of square or hexagonal lattices with additional symmetries~\cite{Zhu2010,Chen2019}. It results in multiple diffracted waves with the same wavevectors and amplitudes, leading to degenerate modes and a lower effective density of modes. Coupling between modes may remove these degeneracies. In the following, we detail the impact of symmetries on the effective density of modes and on the maximal light path enhancement factor.

\subsubsection{Square lattice}

\begin{figure}[h!]
\centering
\includegraphics[width=0.8\textwidth]{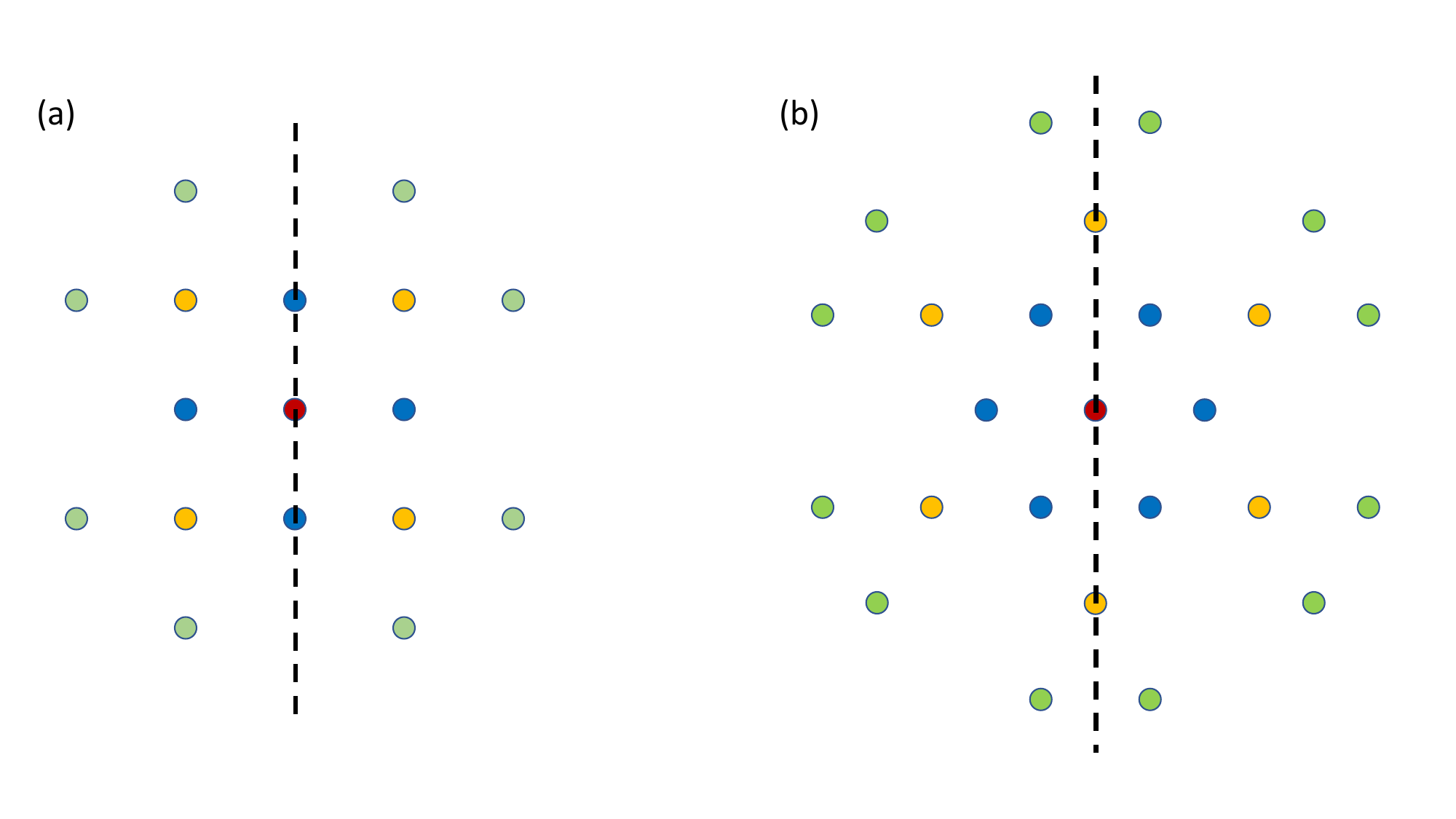}
\caption{
Representation of the lattice points in the reciprocal space. Points at an equal distance from the origin (red dot) correspond to wave vectors with the same magnitude, and are represented with the same color. 
The dashed black line shows a possible plane of symmetry to ensure that most of the modes (except for the lattice points located on the plane) are coupled to a single radiative channel (one polarization state). If the patterns share all the symmetries of the periodic lattice, wave vectors with the same magnitude lead to degenerate modes.
(a) For a square lattice, the mode degeneracy is 4 (blue and yellow  dots) or 8 (green dots).
(b) For an hexagonal lattice, the mode degeneracy is 6 (blue and yellow dots) or 12 (green dots).
}
\label{fig:mode_degeneracies}
\end{figure}

Let us first consider a square pattern, see Fig.~\ref{fig:mode_degeneracies}(a).
If the nanostructures share the symmetries of the pattern, then diffracted waves with wave vectors of the same magnitude have the same complex scattering coefficient. It leads to degenerate modes of order 4 or 8 in a square lattice. For instance, the four modes associated to the diffracted orders $(0,\pm1)$ and $(\pm1,0)$ (blue dots) are degenerate of order 4, and the diffracted orders $(\pm1,\pm2)$ and $(\pm2,\pm1)$ (green dots) are degenerate of order 8.

The coupling between degenerate modes leads to symmetric and antisymmetric coupled modes and removes the degeneracy. However, small coupling leads to a small spectral splitting and a strong overlap of coupled modes, reducing the effective density of modes.

Square patterns or disks in a square lattice are examples of such highly symmetric structures. For large numbers of modes, the absorption spectrum is dominated by modes with degeneracy of order 8. Neglecting the coupling between degenerate modes (strong overlapping between coupled modes), the effective spectral density of modes for each polarization state becomes $\rho=\rho_t/8$, and the minimal light path enhancement factor is: $F_{sq-low} = \pi n^2$.

A pattern with a single mirror plane facilitates the removal of degeneracies by changing the diffraction amplitudes for wave vectors with the same magnitude. An array of L-shaped nanostructures in a square periodical lattice is an example of such pattern with a plane of symmetry along the (1,1) direction. It ensures a single radiative channel for each polarization state, and the maximal optical path enhancement factor is the upper bound $F_{sq-ub}= 4 \pi n^2$ given in Eq.~\ref{eq:Fsquaremax}. \textbf{According to the symmetries and coupling strengths, the maximal $F$ factor obtained with a square lattice is comprised between $F_{sq-low}$ and $F_{sq-ub}$.}

The use of a rectangular lattice is another way to facilitate the spectral splitting between degenerate modes. It is worth mentioning that the degeneracies are also removed at oblique incidence.

\subsubsection{Hexagonal lattice}

Similar arguments apply to hexagonal patterns (Fig.~\ref{fig:mode_degeneracies}(b)). Degeneracies of order 6 (blue and yellow dots) or 12 (green dots)
may decrease the effective spectral density of modes.

In the case of a large density of modes with highly symmetric nanostructures, we have $\rho=\rho_t/12$, and the minimal light path enhancement factor is: $F_{hex-low} = \frac{4\pi}{3\sqrt{3}}n^2 \approx 2.42 n^2$.

\definecolor{shadecolor}{gray}{0.9}
\begin{shaded}
The maximal optical path enhancement factor $F_{ub} = \frac{8 \pi}{\sqrt{3}} n^2$ is obtained for an hexagonal lattice with a single mirror plane.
\end{shaded}

For highly symmetrical patterns with low coupling strengths, the $F$ factor can be lower than the value obtained for square patterns $F_{sq-low}$ and than the optical path enhancement of incoherent scattering $4 n^2$.

\subsubsection{General case with no symmetry}

If the periodical pattern has no symmetry or in the case of incident plane waves at an oblique angle, each resonant mode can be coupled to both polarization states, and polarization rotation can occur in the reflectivity of a linearly polarized incident plane wave. We need to consider two output channels ($N=2$), coupled to the total density of states $\rho_t$ defined in Eq.~\ref{eq:rhoSradSmode}.

If we assume that the coupling to each channel is the same, the multi-channel framework of section~\ref{sec:GeneralFormalism} can be used to derive the maximum absorption enhancement for the light incident through each channel (Eq.~\ref{eq:F_ub_N_channels}) with $N=2$:

\begin{equation}
    F^{N=2}_{ub} = \frac{2\pi \eta^{N=2}_{nr}}{2 \alpha d}.
\end{equation}

\noindent where $\eta^{N=2}_{nr}$ is twice larger than for the single channel case, due to the twice larger available density of states. As a result, the two configurations lead to the same light path enhancement factors: $F^{N=1}_{ub} = F^{N=2}_{ub}$.

%%%%%%%%%%%%%%%%%%%%%%%%
\clearpage
\section{Application to broadband light absorption with isotropic scattering}
\label{sec:OpticalScattering}

In this section, we derive the upper-bound absorption in the case of isotropic optical scattering modeled with an infinite number of diffraction channels (periodical pattern with a period $p \rightarrow +\infty$).

\subsection{Upper bound for isotropic scattering}

We consider an absorbing system with an isotropic optical response.
We use our general theoretical framework developed in Section~\ref{sec:GeneralFormalism}, where absorption is induced by multiple resonances coupled to a large number $N$ of diffraction channels.
Isotropy of the optical response imposes, for each resonant mode, identical coupling rates with all radiative channels.
The optical path enhancement factor (cf Eq.~\ref{eq:F_ub_N_channels}) is given by
\begin{equation}
    F = \frac{2\pi \eta_{nr}}{N \alpha d},
\end{equation}
which can be written (see section~\ref{sec:MultiresonantAbsLight} and Eq.~\ref{eq:F_rho_1}):
\begin{equation}
    F=\frac{2\pi c}{n d} \frac{\rho}{N}.
    \label{eq:F_rho_N}
\end{equation}

Here, in contrast with reference cases studied in section~\ref{sec:MultiresonantAbsLight}, no symmetry restricts the coupling between radiative channels and resonant modes,
so $\rho = \rho_t$ with $\rho_t$ the total density of states in the absorber (Eq.\ref{eq:rhoSradSmode}), leading to:
\begin{equation}
    F= 8 n^2 \frac{S_{rad}}{N S_{mode}}.
    \label{eq:F_Srad_Smode_N}
\end{equation}

The number of radiative channels $N$ is the number of diffraction waves propagating in free space, multiplied by 2 to account for both light polarizations. In the phase space, $N$ is the number of lattice points inside the surface $S_{rad}$ (cf Fig.~\ref{fig:optimal_grating_compacity}(a)).
When $S_{mode} \ll S_{rad}$ (large number of channels), we have, independently of the periodic pattern,
\begin{align}
    N \approx 2 \frac{S_{rad}}{S_{mode}},
    \label{eq:NSradSmode}
\end{align}

\definecolor{shadecolor}{gray}{0.9}
\begin{shaded}
From Eqs.~\ref{eq:F_Srad_Smode_N} and \ref{eq:NSradSmode}, we obtain the light path enhancement factor for isotropic scattering
\begin{equation}
    \boxed{F_{s} = 4 n^2}
    \label{eq:F_ll}
\end{equation}
\noindent and the upper-bound absorption takes the form:
\begin{equation}
    \boxed{\langle A \rangle_{s} = 1-\exp(-F_{s} \ \alpha d).}
    \label{eq:A_ll}
\end{equation}
\end{shaded}

\subsection{Angular restriction of the escape cone}

In the previous derivation, we assumed that light is scattered isotropically into the whole hemisphere above the absorber slab. If we assume a restricted escape cone of half-angle $\theta_e$, the disk that represents the radiative losses (escape cone) in the reciprocal space becomes: $S_{rad} = \pi (k_0 \sin \theta_e)^2$, and the optical path enhancement factor becomes:

\definecolor{shadecolor}{gray}{0.9}
\begin{shaded}
\begin{equation}
    F_{s} (\theta_e) = \frac{4 n^2}{\sin^2 \theta_e}.
    \label{F_4n2theta}
\end{equation}
\end{shaded}

The increase of the optical path enhancement factor is obtained at the cost of a reduced acceptance angle: the illumination cone is restricted to $\theta_i \leq \theta_e$.

%%%%%%%%%%%%%%%%%%%%%%%%
\clearpage
\section{Absorption approximated by the sum of the resonance spectral cross-sections}
\label{sec:AbsCrossSection}

\definecolor{shadecolor}{gray}{0.9}
\begin{shaded}
In this section, we show that the model proposed by Yu et al.~\cite{yu_fundamental_2010,yu_fundamentalgrating_2010}, which is based on the sum of the spectral absorption cross-sections and valid only in the weak absorption regime, can be retrieved by our multi-resonant model based on a pole-zero factorization of the reflectivity.
\end{shaded}

\subsubsection{Derivation of the model proposed by Yu et al.~\cite{yu_fundamental_2010,yu_fundamentalgrating_2010}}

In references~\cite{yu_fundamental_2010,yu_fundamentalgrating_2010}, Yu \textit{et al.} developed a theoretical model based on a series of resonances described by the TCMT that inspired the model presented in this article. However, they made additional assumptions and approximations that strongly limit the domain of validity of their results. In the following, we show that their model, valid only in the weak absorption limit, is a particular case of our theoretical framework.

In references~\cite{yu_fundamental_2010,yu_fundamentalgrating_2010}, absorption is expressed as the sum of the absorption spectral cross-sections $\sigma_m$ of the $M$ resonances located in a given spectral bandwidth $\Delta \omega$ (note that here $\Delta \omega$ is not the resonance width):

\begin{align}
    \sigma_m &= \int_{-\infty}^{+\infty} A_m(\omega) d\omega \label{eq:AbsCrossSect}\\
    A_{\sigma} &= \frac{1}{\Delta \omega} \sum_{m} \sigma_m.
    \label{eq:A_sum_yu}
\end{align}

They further assume that the $N$ radiative channels have the same external coupling rate (isotropic optical response) defined as $\gamma_e = 2 \gamma_{r}/N$, resulting in

\begin{equation}
    A_{\sigma} = \frac{M}{\Delta \omega} \frac{2 \pi \gamma_i \gamma_e}{N \gamma_e + \gamma_i}
    \label{eq:A_yu1}
\end{equation}
with the intrinsic non-radiative losses described by $\gamma_i = 2 \gamma_{nr}$. They observe that the maximum value is reached in the overcoupling regime, when $\gamma_e \gg \gamma_i$, leading to the upper bound:
\begin{equation}
    A_{\sigma} = \frac{2\pi \gamma_i}{\Delta \omega}\frac{M}{N}.
    \label{eq:A_yu2}
\end{equation}

With our definitions, $M/\Delta \omega=1/\delta \omega$ is the spectral density of modes and Eq. \ref{eq:A_yu2} can be rewritten:
\begin{equation}
    A_{\sigma} = \frac{2 \pi \eta_{nr}}{N}.
    \label{eq:A_yu3}
\end{equation}

We observe that Eq.~\ref{eq:A_yu3} can be recovered with a first-order approximation of the upper bound limit derived in our model, by taking $N$ identical radiative channels (Eq.~\ref{eq:upper_bound_N_identical_channels}), for $\langle A \rangle \ll 1$.

This result is a particular case of our model that is only valid for weak absorption, when the overlapping between resonances can be neglected. Starting from Eq.~\ref{eq:A_tot}, if we assume no overlap between modes, then all the coupled terms $A_i * A_j$ in the product are zero and we obtain:
\begin{equation}
    \prod_{m}\left(1-A_m(\omega)\right) \approx 1- \sum_{m} A_m(\omega)
\end{equation}

\noindent leading to
\begin{equation}
    A(\omega) = \sum_{m} A_m(\omega)
    \label{A_sum_cross-sections}
\end{equation}

\noindent and the average absorption
\begin{equation}
    \langle A \rangle = \sum_{m} \langle A_m \rangle
\end{equation}
\noindent is equivalent to Eq.~\ref{eq:A_sum_yu}.

\subsubsection{Comparison with the generalized upper bound}

The theoretical framework developed here provides a more accurate model that takes into account the overlapping between modes (see Section~\ref{sec:GeneralFormalism}, Eq.~\ref{eq:upper_bound_N_identical_channels}):

\begin{equation}
    \langle A \rangle_{ub} = 1 - \exp{\left(-\frac{2\pi  \eta_{nr}}{N} \right)}.
    \label{eq:upper_bound_N_identical_channels2}
\end{equation}
\noindent This expression is valid outside the low absorption regime and ensures $A \leq 1$ for any $\eta_{nr}$.

\begin{figure}[h]
\centering
\includegraphics[width=0.7\textwidth]{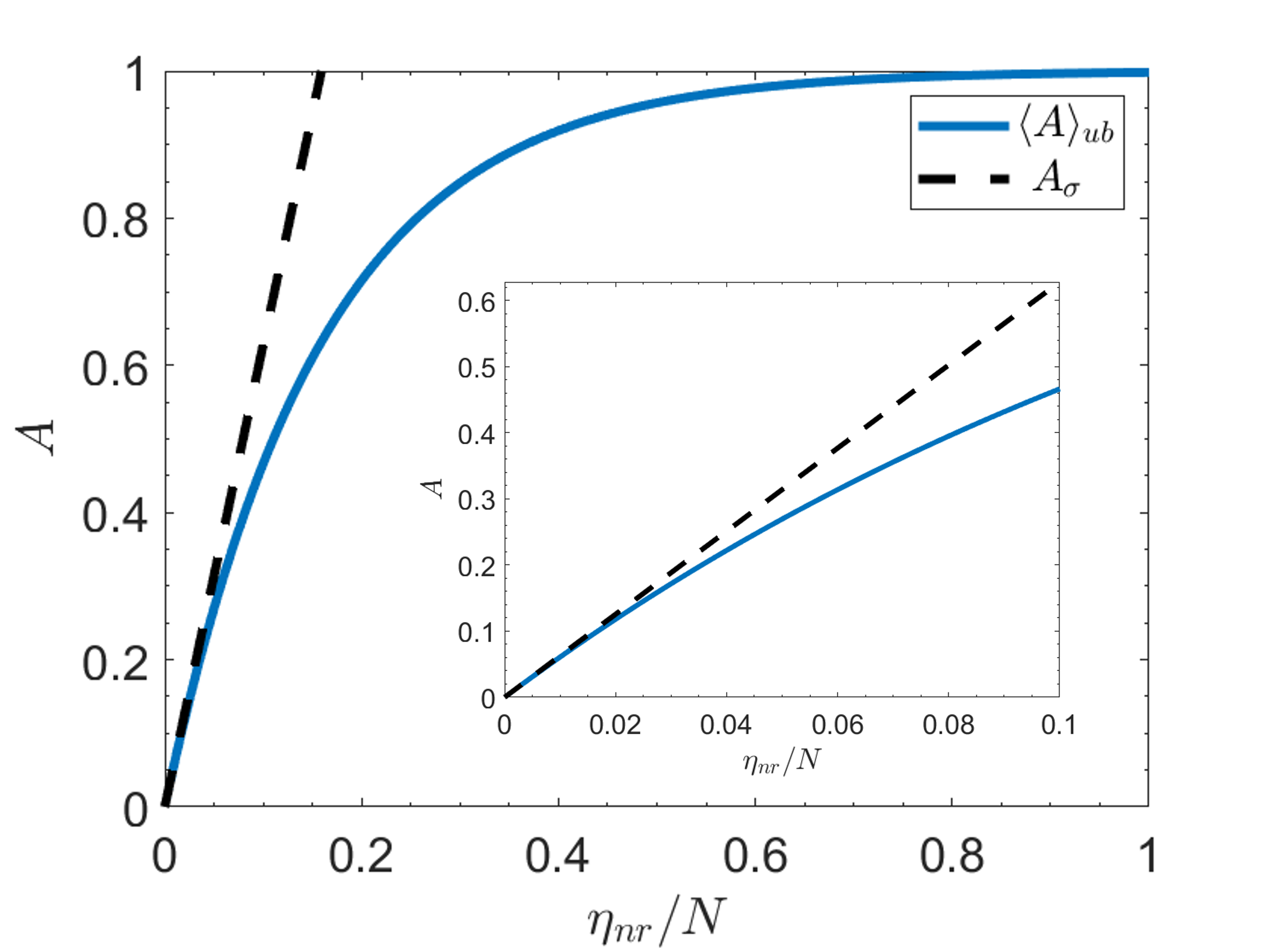}
\caption{Average absorption as a function of the non-radiative overlap factor $\eta_{nr}/N$ for the upper bound absorption ($\langle A \rangle_{ub}$, blue line) and the model derived by Yu \textit{et al.}~\cite{yu_fundamental_2010} and based on the sum of absorption spectral cross-sections ($A_{\sigma}$, dark dashed line). Inset: zoom on small values of $\eta_{nr}/N$. The model $A_{\sigma}$ deviates significantly from the upper bound for $A > 0.3$.}
\label{fig:A_ub_yu}
\end{figure}

Both models are compared in Figure~\ref{fig:A_ub_yu}. The results differ significantly for $A > 0.3$, with a relative error beyond 5\% for $\eta_{nr} > 0.015$ ($A > 0.1$). Moreover, the model developed in references~\cite{yu_fundamental_2010,yu_fundamentalgrating_2010} (Eq.~\ref{eq:A_yu3}), based on the sum of the spectral cross-sections, leads to absorption above 1 for $\eta_{nr}/N \geq 0.16$.

\definecolor{shadecolor}{gray}{0.9}
\begin{shaded}
\textbf{In conclusion, the model based on the sum of the spectral absorption cross-sections ($A_{\sigma}$) developed by Yu \textit{et al.}~\cite{yu_fundamental_2010,yu_fundamentalgrating_2010} appears as a first-order approximation of our multi-resonant absorption model ($\langle A \rangle_{ub}$). It neglects overlapping between resonances and its domain of validity is restricted to weak absorption. This prevents its use in most practical applications, such as broadband sunlight absorption in efficient solar cells.}
\end{shaded}

\subsubsection{Illustration with Fabry-Pérot resonances}

\begin{figure}[h]
\centering
\includegraphics[width=\textwidth]{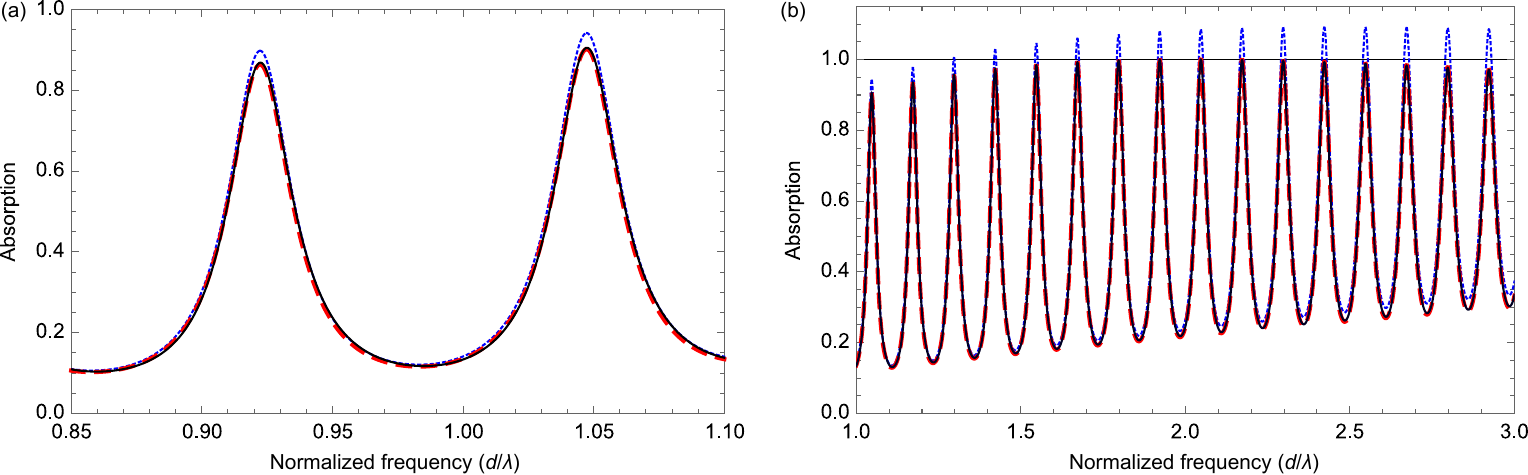}
\caption{Absorption of a slab of absorber of thickness $d$ and refractive index: $4 + 0.02 i$ (silicon at a wavelength of 600~nm), with a metallic non-absorbing back mirror (refractive index: $10 i$). The exact absorption spectrum (dark solid curve) is compared to the multi-resonant model based on the pole-zero factorization (Eq.~\ref{eq:R_tot}, red dashed curve), and to the approximation based on the sum of the resonance spectral cross-sections (Eq.~\ref{A_sum_cross-sections}, blue dashed curve).}
\label{fig:A_Fabry-Perot}
\end{figure}

Figure~\ref{fig:A_Fabry-Perot} shows the absorption spectra of a slab of absorber of thickness $d$ and refractive index $n + i\kappa = 4 + 0.02 i$ (corresponding to silicon at a wavelength of 600~nm), with a metallic non-absorbing back mirror (refractive index: $10 i$). It exhibits well-separated, regularly spaced Fabry-Pérot resonances. The exact optical response at normal incidence (dark solid curve) is calculated using a scattering matrix formalism. 

The frequency and radiative and non-radiative decay rates of each resonance can be calculated using simple formulas valid for Fabry-Pérot resonances when the TCMT formalism applies~\cite{vandamme_ultrathin_2015}.
The non-radiative overlap factor is $\eta_{nr} = 4 \kappa d/\lambda = 0.08 d/\lambda$ while the radiative overlap factor takes the expression $\eta_r = \frac{1}{\pi} \ln(\frac{n+1}{n-1}) \approx 0.163$. Using these overlap factors, we can calculate the absorption using our multi-resonant model and the model based on the sum of Lorentzian functions.
The multi-resonant model based on the pole-zero factorization (Eq.~\ref{eq:R_tot}, red dashed curve) is in very good agreement with the exact calculation.
On the contrary, the model based on the sum of Lorentzian functions (Eq.~\ref{A_sum_cross-sections}) over-estimates the absorption, in particular at resonance frequencies (blue dashed curve). It leads to absorption higher than 1 for $d/\lambda \sim 2$, where the critical coupling regime is achieved ($\eta_{nr} = \eta_{r} \approx 0.163$). The average absorption of about $46.5~\%$ from $d/\lambda = 1.5$ to $2.5$ (8 periods around the critical coupling condition), accurately calculated with the closed-form expression: $\langle A \rangle = \tanh{(\pi \eta_{nr})}$ (see Eq.~\ref{eq:A_av_CC}), is also over-estimated by the sum of Lorentzian functions to $50~\%$. Disagreements would be even more severe in the case of larger resonance overlaps induced either by stronger absorption, or a higher density of modes (sub-wavelength periodical patterning).

%%%%%%%%%%%%%%%%%%%%%%%%%%%%%%%%%%%%%%%%%%%%%%%%%%%%%%%%%%%
\clearpage
\section{Critical comments on the thermodynamic upper bound proposed in Phys. Rev. Lett. \textbf{109}, 173901 (2012)}
\label{YuPRL2012}

In an attempt to generalize the simplified model presented in the previous section (Eqs.~\ref{eq:A_yu1} and \ref{eq:A_yu2}) beyond the weak absorption regime, Yu et al. derived a thermodynamic condition that constrains the external coupling rates and proposed an upper limit for broadband absorption~\cite{yu_thermodynamic_2012}.
In the following, we critically discuss the theoretical derivation and question the validity of the results presented in references \cite{yu_thermodynamic_2012} and \cite{Wang2019a}.

To derive a thermodynamic upper bound, Yu et al. use two different descriptions of the resonant, dissipative system:
\begin{enumerate}

    \item The first model is based on the statistical coupled mode theory formalism for an open and dissipative resonant system. The absorption spectrum is expressed as the sum of Lorentzian functions that describe the absorption cross-section of each resonance (Eqs.~\ref{eq:AbsCrossSect} and \ref{eq:A_sum_yu} of  section~\ref{sec:AbsCrossSection}).

    \item The second model describes the emission of photons when the system is in thermodynamic equilibrium at temperature $T$. According to Eq.~(3) of Ref.~\cite{yu_thermodynamic_2012}, the number of emitted photons $p_m$ from the mode $m$ is expressed as the product of the radiative coupling rate $\gamma_m$ of mode $m$ and the occupation rate given by the Bose-Einstein distribution. Considering a narrow spectral range, the total number of emitted photons $\epsilon_p$ is then expressed as the sum over all modes (Eq.~(3) of Ref.~\cite{yu_thermodynamic_2012}):
\begin{equation}
    \epsilon_p = \sum_m p_m = \sum_m \frac{\gamma_m}{e^{\hbar \omega / k_B T}-1}.
    \label{eq:EmissionYu2012}
\end{equation}
\end{enumerate}

\textbf{In their derivation of the second model (Eq.~\ref{eq:EmissionYu2012}), Yu et al. implicitly assume that the optical density of states can be express as a sum of Dirac delta distributions: $DOS(\omega) = \sum_m \delta (\omega - \omega_m)$. However, such a discrete spectrum is only valid for a lossless system.} When absorption and leakage are included, the optical response of \emph{open and dissipative resonant systems} should be described with quasi-normal modes, and the Dirac terms in the $DOS$ should be replaced by Lorentzian functions.
We invite the interested reader to refer to the comprehensive review by P. Lalanne et al.~\cite{Lalanne2018}.

Most importantly, the incorrect $DOS$ leads to a major inconsistency in reference [Phys. Rev. Lett. \textbf{109}, 173901 (2012)] \cite{yu_thermodynamic_2012}. Eq.~\ref{eq:EmissionYu2012} (Eq.~(3) in \cite{yu_thermodynamic_2012}) results in a number of emitted photons from a resonant mode $p_m$ independent from the non radiative decay rate ($\gamma_{nr}$), and thus independent from internal dissipation. \textbf{This contradicts the Kirchhoff's law which relates emission and absorption.}

This inconsistency can be illustrated by a simple counter-example.
Take an optical cavity made of a semi-infinite slab of semiconductor of complex refractive index $n + i \kappa$ (with $\kappa \ll n$) and thickness $d$, surrounded by air on one side and a perfect mirror on the other side. The spectral spacing between the resonant modes is $\delta \omega = \frac{\pi c}{n d}$, and the radiative losses can be approximated as $\gamma_r = \frac{c}{2 n d} \ln \frac{n+1}{n-1}$~\cite{vandamme_ultrathin_2015}. 
Using our definition of the overlap factor for $N$ modes in a narrow frequency range $\Delta \omega$, Eq.~(7) in Ref.~\cite{yu_thermodynamic_2012} is equivalent to $\eta_r \leq \frac{1}{2 \pi}$, which becomes, for the semiconductor slab:
\begin{equation}
    \ln \frac{n+1}{n-1} \leq \frac{1}{2},
\end{equation}
\noindent or $n \geq 4.083$. This inequality is broken in most semiconductor materials.

\definecolor{shadecolor}{gray}{0.9}
\begin{shaded}
\textbf{In conclusion, the ''thermodynamic upper bound'' presented in Ref.~\cite{yu_thermodynamic_2012} and further discussed in Ref.~\cite{Wang2019a} is derived from an invalid expression of the density of optical states. As a consequence, the main results of both papers are erroneous (Eqs. (1), (7) of Ref.~\cite{yu_thermodynamic_2012} in particular), and the claimed ''thermodynamic upper bound'' does not hold.}

In essence, the guiding idea of Yu et al. was to bound the emission of the system by Planck's law for a blackbody (Eq.~(5) in Ref. \cite{yu_thermodynamic_2012}), which translates into imposing that absorption is equal to or lower than 1. In contrast with the model based on the sum of resonances spectral cross-sections and developed in references~\cite{yu_fundamental_2010,yu_fundamentalgrating_2010} (see section~\ref{sec:AbsCrossSection}), our multi-resonant absorption model fulfills this condition for any dissipation rate.
\end{shaded}

%%%%%%%%%%%%%%%%%%%%%%%%%%%%%%%%%%%%
\clearpage
\section{Application of the multi-resonant absorption model to crystalline silicon}
\label{sec:validity}

In this section, we discuss the application of the multi-resonant absorption model to sunlight absorption in crystalline silicon (c-Si) thin films. We show that the assumptions of the model are valid in the whole visible and near-infrared spectral range of interest, and for thicknesses down to a few hundreds of nanometers, such that the multi-resonant absorption model can be used as a reference to design and evaluate light-trapping structures in silicon solar cells.

%%%%%%%%%%%%%%%%%%%%
\subsection{Absorption as a function of wavelength and thickness}

\begin{figure}[h]
    \centering
    \includegraphics[width=\textwidth]{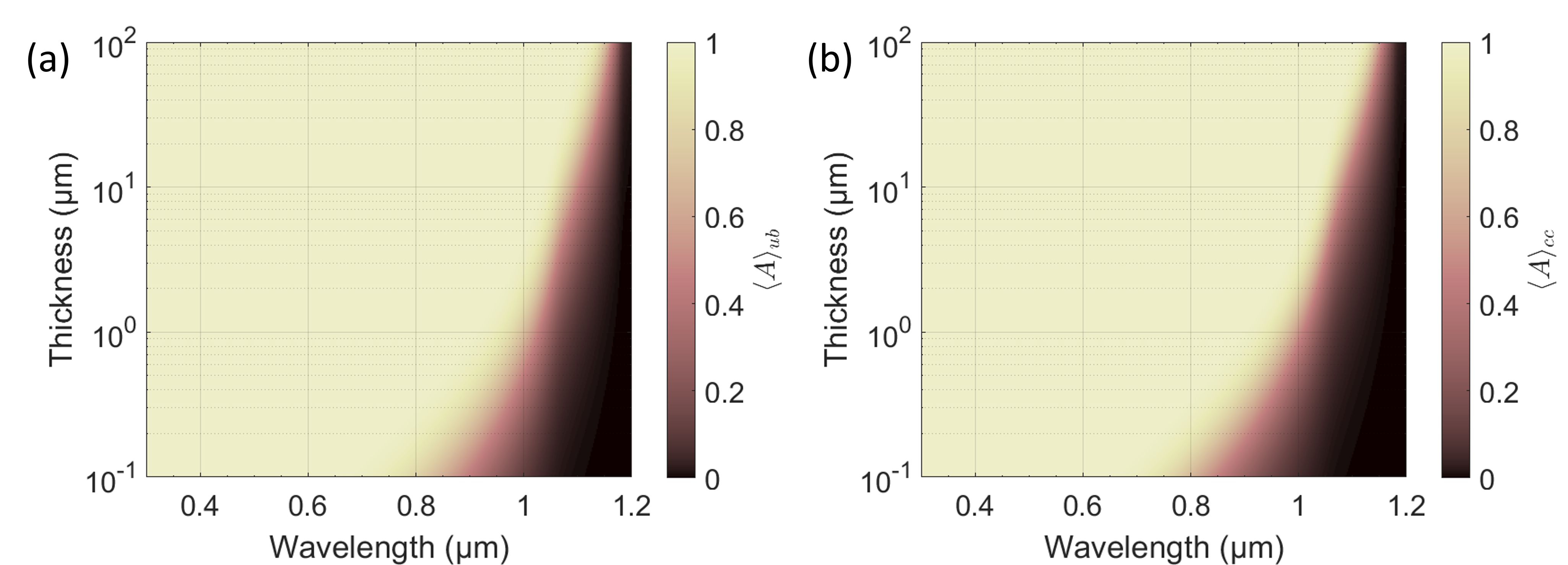}
    \caption{(a) Upper bound (Eq.~\ref{eq:A_ub}) and (b) critical coupling (Eq.~\ref{eq:A_ll}) absorption in c-Si , as a function of the absorber thickness and the wavelength of light. }
    \label{fig:A_lambda_thickness_Si}
\end{figure}

We first look at the absorption that can be achieved in a c-Si thin film as a function of the absorber thickness, for the upper bound (Fig.\ref{fig:A_lambda_thickness_Si}(a)) and for the critical coupling model (Fig.\ref{fig:A_lambda_thickness_Si}(b)). The absorption coefficient of c-Si is taken from reference~\cite{green_improved_2022}.

For any thickness and both light trapping expressions, absorption is close to unity at short wavelengths and smoothly drops to zero as the wavelength increases. As the thickness increases, this transition shifts to longer wavelengths and gets closer to the bandgap of c-Si. The upper bound also systematically extends absorption further into the infrared as compared to critical coupling.

%%%%%%%%%%%%%%%%%%%%

\subsection{Validation of the assumptions}

A number of assumptions need to be checked to ensure the validity of the model in the domain investigated in Fig.~\ref{fig:A_lambda_thickness_Si}: the density of modes is approximated by the formula valid for large volumes, resonances are approximated by Lorentzian functions (TCMT), and their parameters as supposed constant over their spectral width.

\subsubsection{Density of modes approximation}
\label{ssec:DOS_approx}

We have approximated the spectral density of resonances by the density of states in a bulk material, calculated using periodical boundary conditions in a volume $p^2 \times d$. Here, we question its validity in the case of a small number of modes, when the unit cell is of the order of or smaller than $\lambda^3$.
In the following, by calculating the frequency of modes described as plane waves resonating in a slab of thickness $d$ with a two-dimensional pattern, we show that the bulk density of modes is an accurate approximation for thicknesses as low as a few tens of nanometers.

For the sake of simplicity, we assume that normal incident light impinges a slab with a constant real refractive index $n = 3.5$ (typical of c-Si in the near-IR region, as well as other inorganic semiconductors), a shallow square pattern of period $p$, and a perfect back mirror. Resonant modes in the slab are described by their three-dimensional wave vector:

\begin{equation}
    \mathbf{k} = k_x \mathbf{u_x} + k_y \mathbf{u_y}  + k_z \mathbf{u_z}.
    \label{eq:k_decomposed}
\end{equation}

\noindent where $k_x$ and $k_y$ are the in-plane components of the wave vector and $k_z$ the component normal to the plane. 

\begin{itemize}
    \item 
    $k_z$ is constrained by boundary conditions at the top and bottom interfaces of the slab.
The solutions are Fabry-Perot modes as well as TE and TM waveguide modes in the slab, labelled with a positive integer $m_z$. Their dispersion relation $k_{xy}^{m_z} (\lambda)$ are calculated, neglecting the perturbation induced by the shallow pattern, with $k_{xy} = \sqrt{k_x^2 +k_y^2}$, and $|\mathbf{k}| = 2 \pi n / \lambda$.
    
    \item
    The in-plane components $k_x$ and $k_y$ must satisfy the diffraction conditions induced by the periodical patterning:

\begin{equation}
    k_{x,y} = \frac{2\pi}{p} m_{x,y}, \quad m_{x,y} \in \mathbb{Z}. \label{eq:folding} 
\end{equation}
\end{itemize}

The total number of modes is found by summing the number of couples $(m_x,m_y)$ whose in-plane wavevector matches one of the $m_z$ orders in the wavelength range $[\lambda_{min},\lambda_{max}]$.

\begin{figure}[h]
\centering
\includegraphics[width=.7\textwidth]{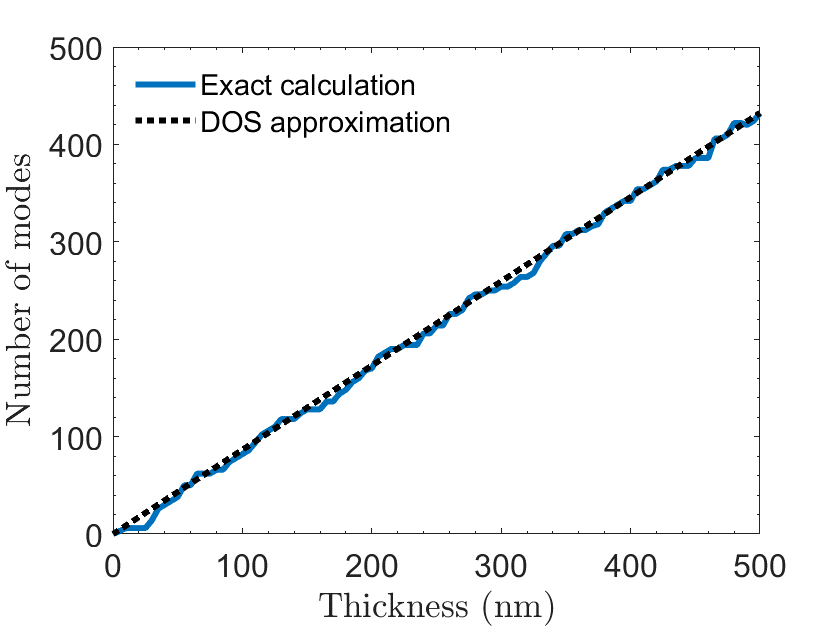}
\caption{Number of modes in a slab of real refractive index $n=3.5$, surrounded by air on the front side and a perfect mirror on the rear side, with a shallow two-dimensional square pattern (period $p = 400 \ \mathrm{nm}$). The wavelength range considered is $[400 \ \mathrm{nm},1200 \ \mathrm{nm}]$. The exact calculation of Fabry-Pérot and TE/TM waveguide modes coupled to free space via the diffracted orders $(m_x,m_y)$ (blue line) is compared to the expression of the density of modes in a bulk material (Eq.~\ref{eq:rho_square}, dashed black line).
}
\label{fig:DOS_comparison}
\end{figure}

In Fig.~\ref{fig:DOS_comparison}, we plot the number of modes $(m_x,m_y,m_z)$ as a function of the slab thickness in the spectral range $[400 \ \mathrm{nm},1200 \ \mathrm{nm}]$, for a sub-wavelength period $p = 400 \ \mathrm{nm}$. The result is compared to the number of modes $N_{DOS}$ derived from the total density of modes in a bulk material with the volume of the unit cell $p^2 \times d$ (Eq.~\ref{eq:rhoSradSmode}):

\begin{equation}
    N_{DOS} = \int_{\lambda_{min}}^{\lambda_{max}}\rho_t(\lambda) d\lambda = \frac{8}{3} \pi n^3 p^2 d \left(\frac{1}{\lambda_{min}^3} - \frac{1}{\lambda_{max}^3} \right)
\label{eq:rho_square}
\end{equation}

For thicknesses above 50 nm, the approximation of the bulk DOS gives a number of modes within 10\% of the exact value. This originates from the relatively large spectral bandwidth required for broadband sunlight absorption.

\definecolor{shadecolor}{gray}{0.9}
\begin{shaded}
\textbf{
The bulk density of states is a good approximation of the real distribution of resonant modes for broadband absorption, even for absorber layers as thin as 50 nm.
}
\end{shaded}

%%%%%%%%%%%%%%%%%%%%
\subsubsection{Lorentzian approximation of resonances (TCMT)}

%%%%
\paragraph{Quality factor.}

The validity of our model fundamentally relies on the assumptions of the TCMT, that approximates the spectral response of each resonant mode by a Lorentzian function (see section \ref{sec:TCMT}).
This requires large quality factors:

\begin{equation}
    Q = \frac{\omega}{2\gamma} \gg \pi.
\end{equation}

From Eq.~(\ref{eq:gamma_nr}) and assuming critical coupling, the quality factor can be expressed as

\begin{equation}
    Q = \frac{n}{4\kappa}.
    \label{eq:Q_n_kappa}
\end{equation}

\begin{figure}[h]
\centering
\includegraphics[width=.6\textwidth]{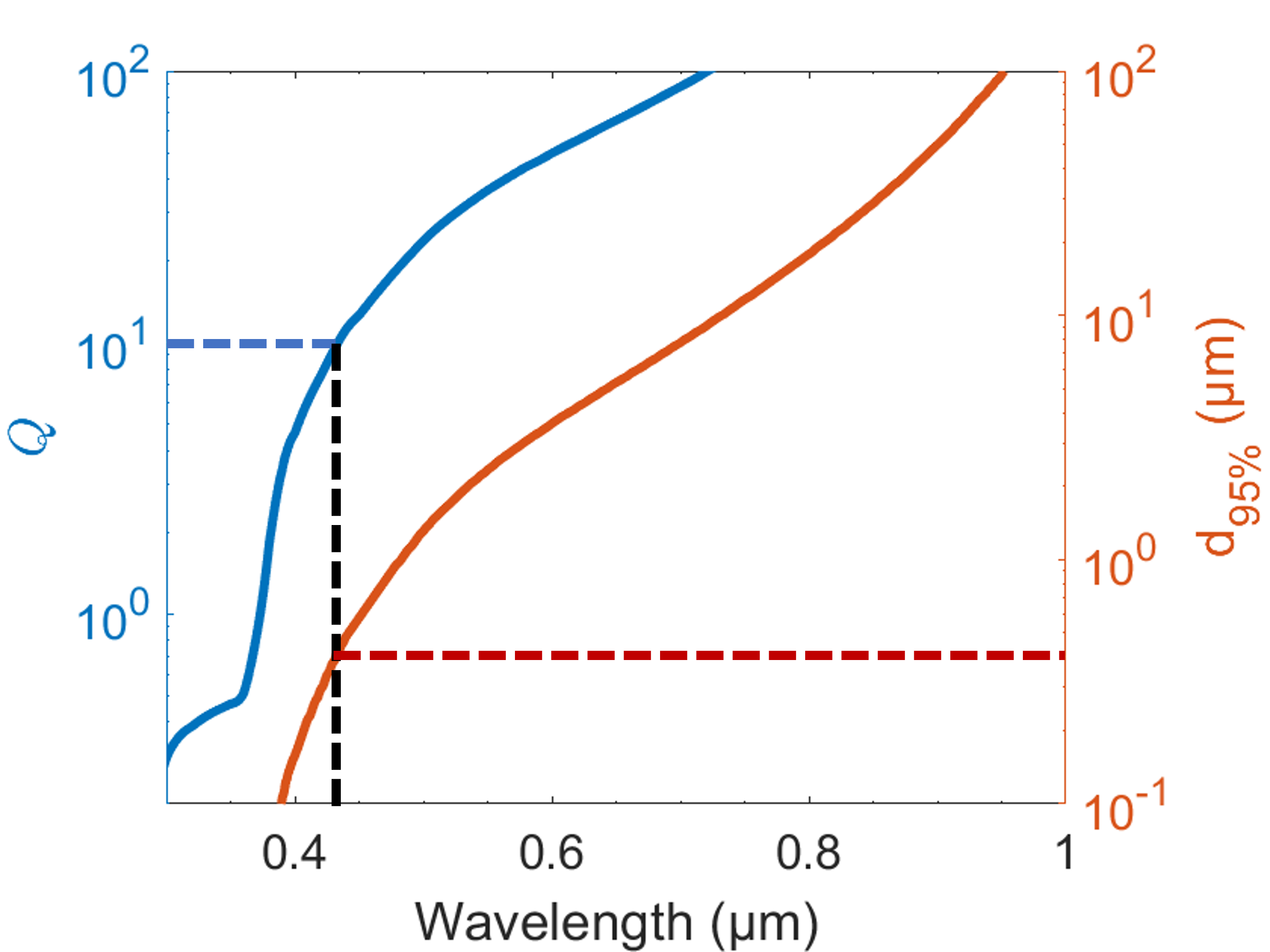}
\caption{Quality factor $Q$ (blue line, left axis) and minimal thickness $d_{95\%}$ required to reach 95\% of absorption with a double-pass in c-Si (red line, right axis). In the region where $Q$ is lower than 10, double pass absorption exceeds $95\%$ for slabs thicker than 400~nm. The spectral region where the theoretical framework is questionable due to low $Q$ corresponds to absorption very close to 1.}
\label{fig:Q_d95_Si}
\end{figure}

$Q$ depends only on the refractive index of the material. We plot $Q$ for c-Si as a function of the wavelength in Fig.~\ref{fig:Q_d95_Si} (blue curve).
The quality factor increases very rapidly to values above 10, indicating that independent modes are very well approximated by Lorentzian functions for wavelengths above 435~nm. 

%%%%
\paragraph{Double-pass absorption.}

Double-pass absorption is obtained with a perfect back mirror and no reflection at the front side, and is simply expressed by:

\begin{equation}
A_{double}=1-e^{-2 \alpha d}.
\end{equation}

$95\%$ of incident light is absorbed in c-Si after a double pass if 

\begin{equation}
    d \geq d_{95\%} = \frac{3}{2 \alpha}.
    \label{eq:double_pass_cond}
\end{equation}

The thickness $d_{95\%}$ is plotted as a function of the wavelength in Fig.~\ref{fig:Q_d95_Si} (red curve).
With thicknesses down to 400 nm, most photons with a wavelength shorter than 435 nm are absorbed after a double pass. This thickness is much lower than the minimal thickness considered for strong absorption in c-Si, even in the upper bound case.

Overall, in the spectral region where the theoretical framework is questionable due to low $Q$, double-pass absorption is sufficient to reach absorption close to 1 even for very thin absorbers, and the different absorption models can still be used as a good estimate for the theoretical limits.

%%%%%%%%%%%%%%%%%%%
\subsubsection{Slow spectral variations of the resonance parameters}

\begin{figure}[h]
\centering
\includegraphics[width=\textwidth]{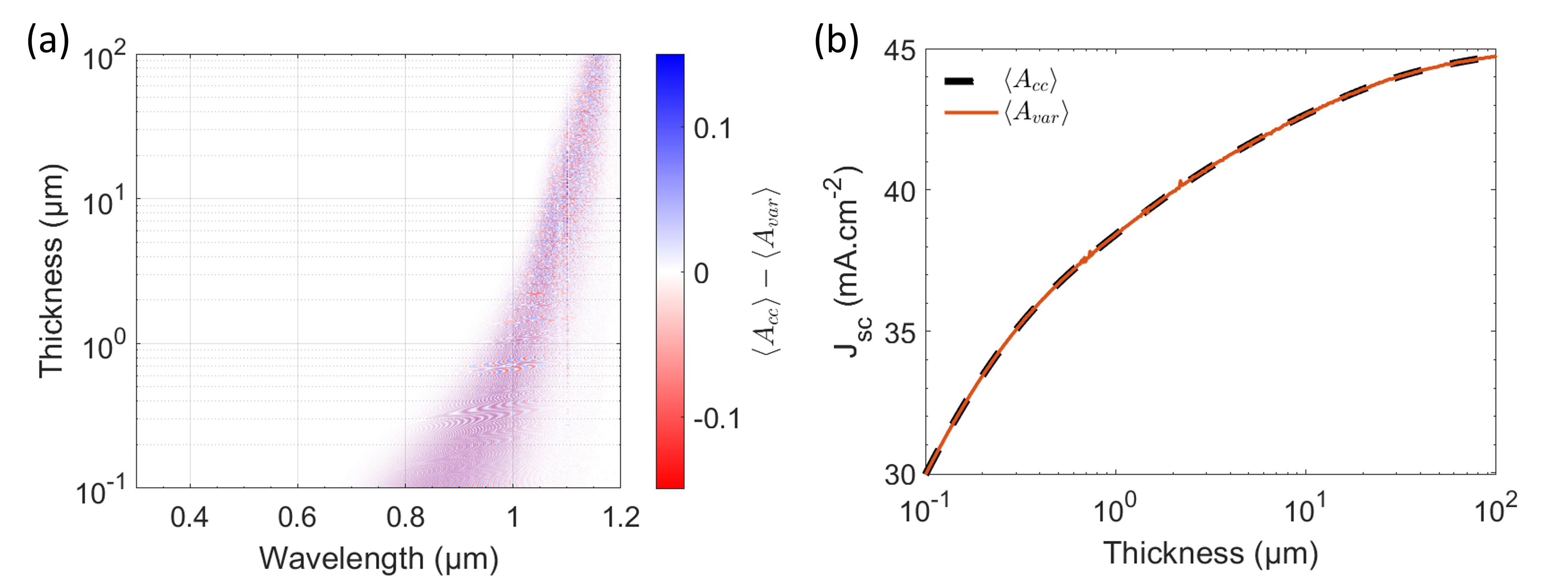}
\caption{
(a) Difference between critical the coupling absorption $\langle A\rangle_{cc}$, and the locally averaged absorption $\langle A \rangle_{var}$ calculated from Eq.~\ref{eq:A_var} in c-Si.
(b) Short-circuit current as a function of the thickness obtained from $\langle A\rangle_{cc}$ and $\langle A\rangle_{var}$, considering the AM1.5G solar spectrum.
}
\label{fig:slow_var}
\end{figure}

A question that was left hanging in section \ref{sec:SlowVaryingEnvelop} is the potential impact of the variations of $\eta_{nr}$ (and in general all quantities) with the wavelength. 
The model requires the variations to be sufficiently small such that these quantities can be considered constant over the spectral width of a resonance.
To check the validity of this assumption in c-Si,
we compare in Fig.~\ref{fig:slow_var}(a) the critical coupling limit $\langle A\rangle_{cc}$ (Eq.~\ref{eq:A_av_CC}) with the locally averaged absorption $\langle A \rangle_{var}$, as a function of wavelength and absorber thickness. 
$\langle A \rangle_{var}$ is obtained by integration of Eq.~\ref{eq:infinite_CC}, taking into account the frequency dependence of $\delta\omega$ and $\gamma$:

\begin{equation}
    \langle A \rangle_{var} (\omega_0) = \int_{\omega_0-\delta \omega(\omega_0)/2}^{\omega_0+\delta \omega(\omega_0)/2} \left[1 - \frac{1}{1+\sinh^2\left(\frac{\pi\gamma(\omega)}{\delta\omega(\omega)}\right)\csc^2\left(\frac{\pi\omega}{\delta\omega(\omega)}\right) }\right] d\omega.
    \label{eq:A_var}
\end{equation}

Differences between both values appear in the region where absorption shifts from 1 to 0 (see Fig.\ref{fig:A_lambda_thickness_Si}(b)). The variations are alternatively positive and negative, but remain within 10\%.
To see whether these variations affect the overall absorption of the device, we calculate the short-circuit current as:

\begin{equation}
    J_{sc} = q \int_{0.3 \ \mu m}^{1.2 \ \mu m} A(\lambda) \ \Phi_{AM1.5G}(\lambda) d\lambda
    \label{eq:J_sc}
\end{equation}

\noindent where $q$ is the electron charge, $A$ is the absorption and $\Phi_{AM1.5G}$ is the incident photon flux of the AM1.5G solar spectrum.
We compare the short-circuit currents obtained with both absorption models in Fig.~\ref{fig:slow_var}(b). The spectral discrepancies between both models compensate each other, leading to essentially identical photocurrents.

\definecolor{shadecolor}{gray}{0.9}
\begin{shaded}
\textbf{
Overall, the assumptions of the multi-resonant model are verified for sunlight absorption in c-Si (and more generally in indirect bandgap semiconductors) over the whole solar spectrum, for thicknesses down to a few hundreds of nanometers.}
\end{shaded}

%%%%%%%%%%%%%%%%%%%%%%%%%%%%%%%%%%%%
\clearpage
\section{Application of the multi-resonant absorption model to GaAs}
\label{sec:GaAs}

This section focuses on absorber slabs made of gallium arsenide (GaAs), which is the reference material for direct-bandgap solar cells and holds the efficiency record for single-junction solar cells. We present the results of the different absorption models and upper bounds, and we check the validity of the main assumptions of the theoretical framework.

%%%%%%%%%%%%%%%%%%%%
\subsection{Absorption and short-circuit current}

We first study the absorption in GaAs thin films as a function of the absorber thickness, for the upper bound (Fig.\ref{fig:A_lambda_thickness_Si}(a)) and for the critical coupling model (Fig.\ref{fig:A_lambda_thickness_Si}(b)). The absorption coefficient of GaAs is taken from reference~\cite{palik_-_1997}, and it is convoluted with a 10~meV Urbach tail~\cite{Chen2021}.

\begin{figure}[h]
\centering
\includegraphics[width=\textwidth]{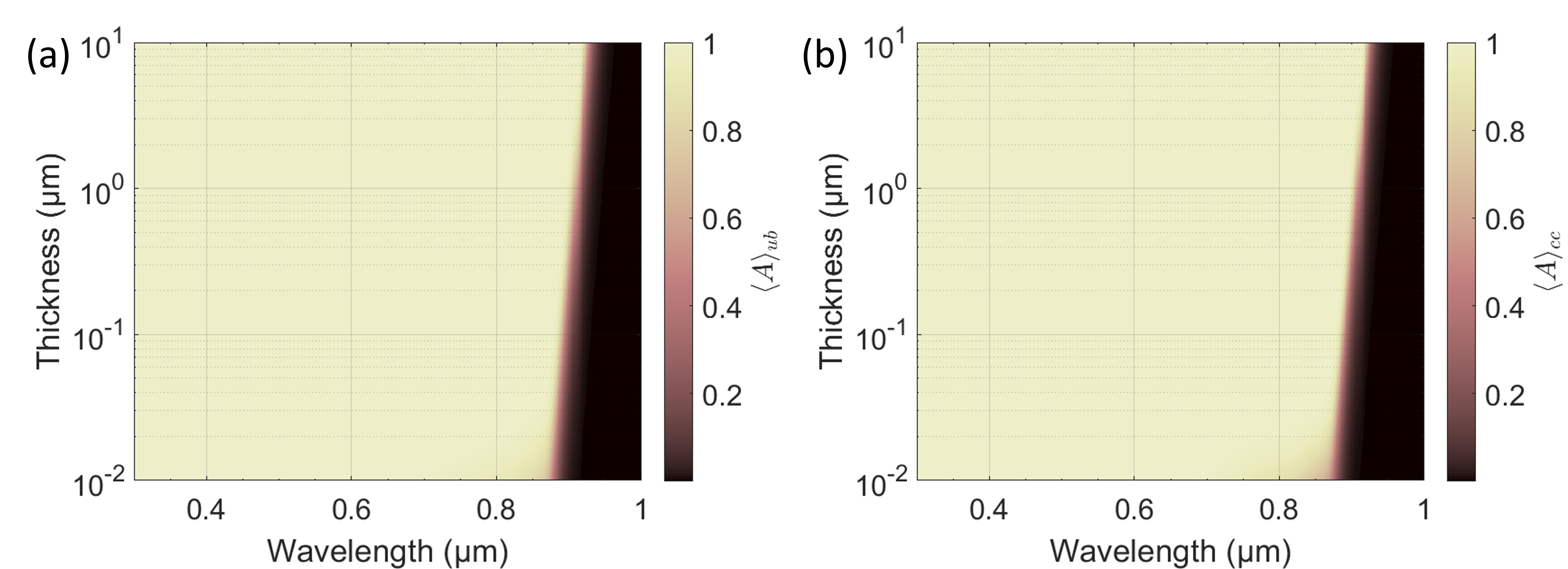}
\caption{(a) Upper bound (Eq.~\ref{eq:A_ub}) and (b) critical coupling (Eq.~\ref{eq:A_ll}) absorption in 
GaAs, as a function of the absorber thickness and the wavelength of light.}
\label{fig:A_lambda_thickness_GaAs}
\end{figure}

\begin{figure}[h]
\centering
\includegraphics[width=\textwidth]{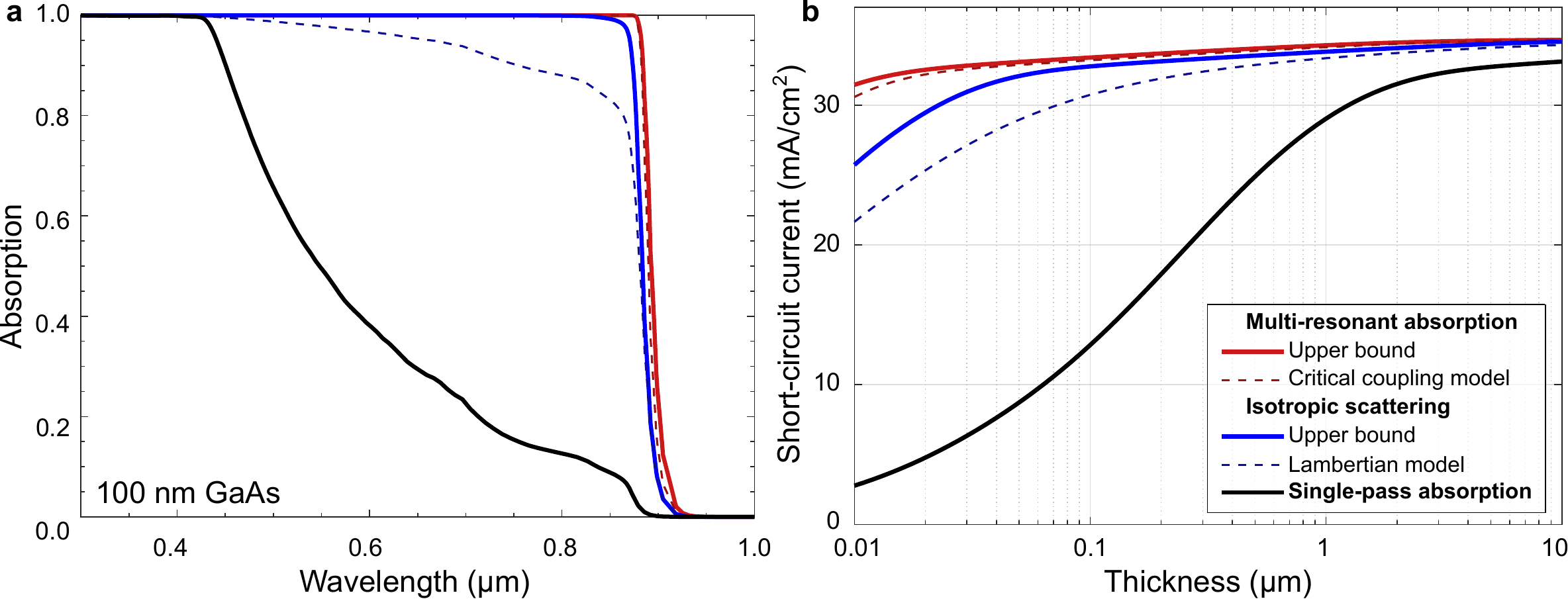}
\caption{(a) Absorption spectra for a 100~nm-thick GaAs slab for different absorption models and upper bounds. (b) Short-circuit current density $J_{sc}$ calculated for the same models, for AM1.5G solar irradiance.
}
\label{fig:Abs_Jsc_GaAs}
\end{figure}

The absorption drops steeply from 1 to 0 in a narrow wavelength range close to the bandgap.
From these color maps, we extract the absorption spectra for a 100~nm-thick GaAs absorber. In Fig.~\ref{fig:Abs_Jsc_GaAs}(a), they are compared to single-pass absorption, to the upper bound absorption with isotropic scattering (Eqs.~\ref{eq:F_ll}-\ref{eq:A_ll}) and to the Lambertian model (Eq.~\ref{eq:ray_optics_lamb}).
In Fig.~\ref{fig:Abs_Jsc_GaAs}(b), the short-circuit current density is plotted as a function of the absorber thickness for the same absorption models.
Without light trapping, thicknesses around 2~µm are required to achieve current densities above $31.5~\mathrm{mA.cm^{-2}}$. According to the multiresonant and scattering upper bounds and the critical coupling model, a thickness of 100~nm is sufficient for near-perfect absorption.

\subsection{Validation of the assumptions}

We investigate the validity of the model assumptions for GaAs in the domain of interest.
We have shown in Sec.~\ref{ssec:DOS_approx} that the approximated density of modes is valid for slabs as thin as 50~nm. In the following, we focus on the Lorentzian description of resonances.

\subsubsection{Lorentzian approximation of resonances (TCMT)}

\begin{figure}[h]
\centering
\includegraphics[width=.6\textwidth]{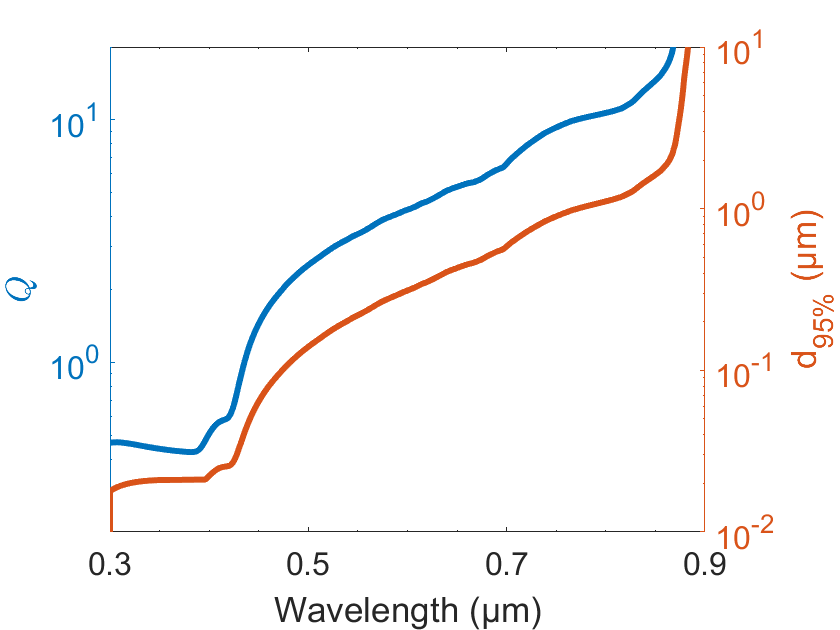}
\caption{Quality factor and minimal thickness required to reach 95\% absorption with a double pass in GaAs, as a function of the wavelength.}
\label{fig:Q_d95_GaAs}
\end{figure}

The quality factor $Q$ (Eq.\ref{eq:Q_n_kappa}) and the thickness required for 95\% double-pass absorption $d_{95\%}$ (Eq.\ref{eq:double_pass_cond}) are plotted as a function of the wavelength in Fig.~\ref{fig:Q_d95_GaAs}.

At wavelengths shorter than 600~nm, the quality factor drops below 5, raising questions on the applicability of the model in this spectral range. However, the models predicts absorption close to 1 even for GaAs slabs as thin as a few tens of nanometers at short wavelengths, and numerical simulations go in the same direction~\cite{Grandidier2012a,Massiot2014,camarillo_abad_transparent_2022}. In most practical cases, challenges for absorption enhancement lie in the 600-900~nm wavelength range and the absorption models still provide a good estimate of the theoretical limits.

\subsubsection{Slow spectral variations of the resonance parameters}

\begin{figure}[ht!]
\centering
\includegraphics[width=\textwidth]{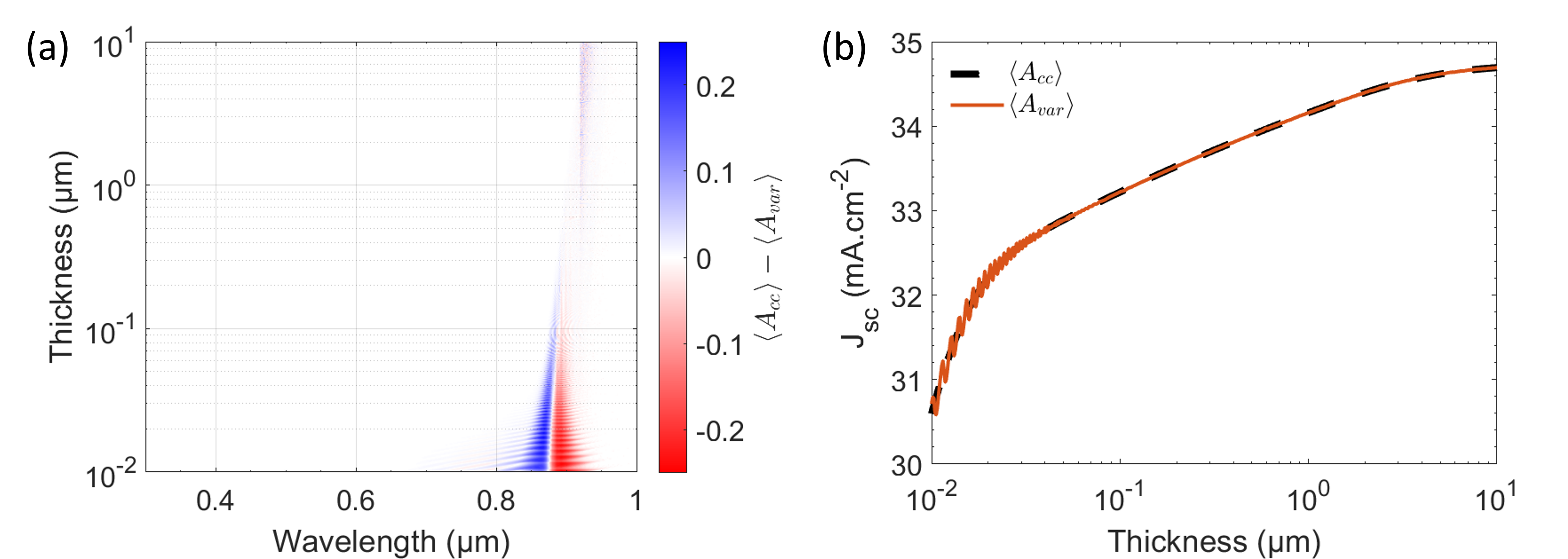}
\caption{
(a) Difference between the critical coupling absorption $\langle A\rangle_{cc}$, and the locally-averaged absorption $\langle A \rangle_{var}$ calculated from Eq.~\ref{eq:A_var} for GaAs.
(b) Short-circuit current as a function of thickness obtained from $\langle A\rangle_{cc}$ and $\langle A\rangle_{var}$, considering AM1.5G solar irradiance.
}
\label{fig:slow_var_GaAs}
\end{figure}

We compare the absorption calculated with the critical coupling absorption model (Eq.~\ref{eq:A_av_CC}) with the results obtained by accounting for the spectral dispersion of the resonance parameters (Eq.~\ref{eq:A_var}). The difference in absorption and the resulting short-circuit current are shown in Fig.~\ref{fig:slow_var_GaAs}.

For thicknesses below 100 nm, significant differences between the two models can be observed close to the bandgap (Fig.~\ref{fig:slow_var_GaAs}(a): the CC model overestimates absorption below the bandgap and underestimates it above the bandgap (sub-bandgap absorption tail). This originates from the very sharp change in the absorption coefficient of direct-bandgap semiconductors. When integrated spectrally to calculate $J_{SC}$ (Fig.~\ref{fig:slow_var_GaAs}(b)), the differences lead to very small oscillations with the thickness.
For thicknesses above 100~nm, $\langle A_{CC} \rangle$ and $\langle A_{var} \rangle$ are essentially the same.

\definecolor{shadecolor}{gray}{0.9}
\begin{shaded}
\textbf{
Overall, the multiresonant absorption framework still holds for direct-bandgap semiconductors. At short wavelength ($\lambda < 600~\rm{nm}$, the low quality factor questions the validity of the models, but absorption is anyway very close to 1 in most practical cases.
}
\end{shaded}

%%%%%%%%%%%%%%%%%%%%%%%%%%%%%%%%%%%
\clearpage
\bibliography{MRABib.bib}